\documentclass{amsart}

\usepackage{amsthm,amssymb,amsmath}
\usepackage{bm,url}
\usepackage{bbm}
\usepackage{graphicx}
\usepackage{subfigure}
\usepackage{tikz}
\usepackage{array}
\usepackage{mathptmx}
\usepackage{multirow, makecell}
\usetikzlibrary{arrows,positioning,shapes.geometric}
\usetikzlibrary{matrix,chains,decorations.pathreplacing}

\usepackage{xcolor}
\usepackage{microtype}
\usepackage{booktabs}
\usepackage{url,diagbox}
\usepackage{setspace}
\usepackage{algorithm}
\usepackage{algorithmic}

\newtheorem{theorem}[subsubsection]{Theorem}
\newtheorem{def.}{Definition}
\newtheorem{exam.}[subsubsection]{Example}
\newtheorem{prop.}[subsubsection]{Proposition}
\newtheorem{cor}[subsubsection]{Corollary}
\newcommand{\bbR}{\mathbb{R}}
\newcommand{\bbZ}{\mathbb{Z}}

\newcommand{\im}{{\rm im}}

\newcommand{\VR}{{\rm VR}}
\newcommand{\diam}{{\rm diam}}
\newcommand{\lleft}{{\rm left}}
\newcommand{\rright}{{\rm right}}

\begin{document}

\title[Persistent homology and heart rate variability]{A persistent homology approach to heart rate variability analysis with an application to sleep-wake classification}

		\author{Yu-Min Chung}
		\address{Department of Mathematics and Statistics, University of North Carolina at Greensboro, NC, USA}
		
		\author{Chuan-Shen Hu}
		\address{Department of Mathematics, National Taiwan Normal University, Taipei, Taiwan}
		 
		\author{Yu-Lun Lo}
		\address{Department of Thoracic Medicine, Chang Gung Memorial Hospital, Chang Gung University, School of Medicine, Taipei, Taiwan}
		
		\author[Chung {\it et al.}]{Hau-Tieng Wu}
\address{Department of Mathematics and Department of Statistical Science, Duke University, Durham, NC, USA; Mathematics Division, National Center for Theoretical Sciences, Taipei, Taiwan}
\email{hauwu@math.duke.edu}

\begin{abstract}
Persistent homology (PH) is a recently developed theory in the field of algebraic topology {to study shapes of datasets}. It is an effective {data analysis} tool {that is robust to noise} and has been widely applied. 
We demonstrate a general pipeline to apply PH to study time series; particularly the {instantaneous heart rate time series for the} heart rate variability (HRV) {analysis}. {The first step is capturing} the shapes of time series {from} two different {aspects} -- {the PH's and hence persistence diagrams of its} sub-level set and Taken's lag map.  
Second, we propose a systematic {and computationally efficient} approach to summarize persistence diagrams, {which we coined {\em persistence statistics}.}  
To demonstrate our proposed method, we apply these tools to the HRV analysis and the sleep-wake, REM-NREM (rapid eyeball movement and non rapid eyeball movement) and sleep-REM-NREM classification problems. The proposed algorithm is evaluated on three different datasets via the cross-database validation scheme. The performance of our approach is { better than} the state-of-the-art algorithms, and {the result is} consistent throughout different datasets.  
\end{abstract}

	\maketitle

\section{Introduction}
\label{sec:introduction}

Heart rate variability (HRV) is the physiological phenomenon of variation in the lengths of consecutive cardiac cycles, or the rhythm of heart rate \cite{Draghici2016}. 
Interest in HRV has a long history \cite{Billman2011}, and there have been several theories describing how the heart rate rhythm, {including,} for example, the polyvagal theory \cite{porges2009polyvagal} and the model of neurovisceral integration \cite{thayer2006beyond}. In short, HRV {results from an integration of} complicated interactions between various physiological systems and external stimuli \cite{Vanderlei_Pastre_Hoshi_Carvalho_Godoy:2009,Draghici2016,Shaffer2014} on various scales, and the autonomic nervous system (ANS) plays a critical role \cite{porges2009polyvagal,thayer2006beyond}.  
A correct quantification of HRV yields dynamical information of various physiological systems and has various clinical applications \cite{stys1998current}, including improving diagnostic accuracy and treatment quality \cite{Vanderlei_Pastre_Hoshi_Carvalho_Godoy:2009}.

In practice, the heart rhythm is {quantified} by the time series called {\em instantaneous heart rate} (IHR) coming from intervals between consecutive pairs of heart beats, which is usually determined from the {R peak to R peak} interval (RRI) by reading the electrocardiogram (ECG). To quantify HRV, a common approach is studying various {statistics} of IHR. 
There have been a lot of efforts trying to quantify HRV, and proposed {statistics} could be briefly classified into four major categories -- time domain approach, frequency domain approach \cite{Electrophysiology1043}, nonlinear geometric approach \cite{voss2008methods,marwan2002recurrence}, and information theory based approach \cite{costa2002multiscale}.
It is worth mentioning that while there have been a lot of researches in this direction with several proposed {statistics}, there is limited consensus and { it is still an active research field} due to the non-stationarity nature of the {IHR} time series \cite{Pincus1994,Glass2009}.

{Topological data analysis} (TDA) is a data analysis {framework based on tools} from {algebraic} topology \cite{epstein2011topological,carlsson2009topology}. In the past decades, its theoretical foundation has been actively established, and {various algorithms have been proposed to study datasets from} different fields. {The basic idea underlying TDA is that the data organization can be well captured by {\em counting holes}. Theoretically, the number of holes of different dimensions characterizes how the data is organized. {Thus, researchers} design useful statistics based on { the information of holes.} 
This simple {yet} powerful idea has been applied to different fields. Specifically,}
there {have} been several efforts applying TDA to analyze time series. 
For example, the {Vietoris-Rips  (VR)} complex filtration and the bottleneck or Wasserstein distances among {persistence diagrams} (PDs) are applied to study voices and body motions \cite{LeeMSeversky-CVPR2016,Venkataraman-ICIP-2016}. A transformation of the PD, called persistence landscapes \cite{bubenik2015statistical}, has been applied to study trading records \cite{Gidea-Katz-PhysicaA-2018}, EEG signals \cite{wang2018} and cryptocurrency trend forecasting \cite{kim2018time}. {However, to the best of our knowledge, its application to HRV is not yet considered. Moreover, existing TDA approaches usually suffer from computational issues,} {which limits its application to large scale database. Finding a computationally efficient TDA algorithm is thus critical.}

In this article, motivated by the complicated interaction among different physiological systems over various scales and inter-individual variability, { the need for} a useful tool for the HRV analysis, and the {numerical limitation} of the recently developed TDA tools, we hypothesize that topological information could be useful to quantify the HRV{, and propose a computationally efficient approach to analyze time series via TDA}.

%
%While these approaches have been widely considered, they have limitations. Some may be complicated to construct and/or to implement, some may require prior knowledge from different disciplines, some may need to tune parameters, and some may require more computational efforts. 

\subsection{Our contribution}

Based on {the} flexibility {of TDA tools}, and due to the non-stationarity of {complicated time series we commonly encounter in real life, like the IHR}, we propose a systematic, {principled, and computationally efficient} approach to {study complicated time series by the TDA tools.}

\begin{figure}[h]
	\centering 
	\includegraphics[width=\linewidth]{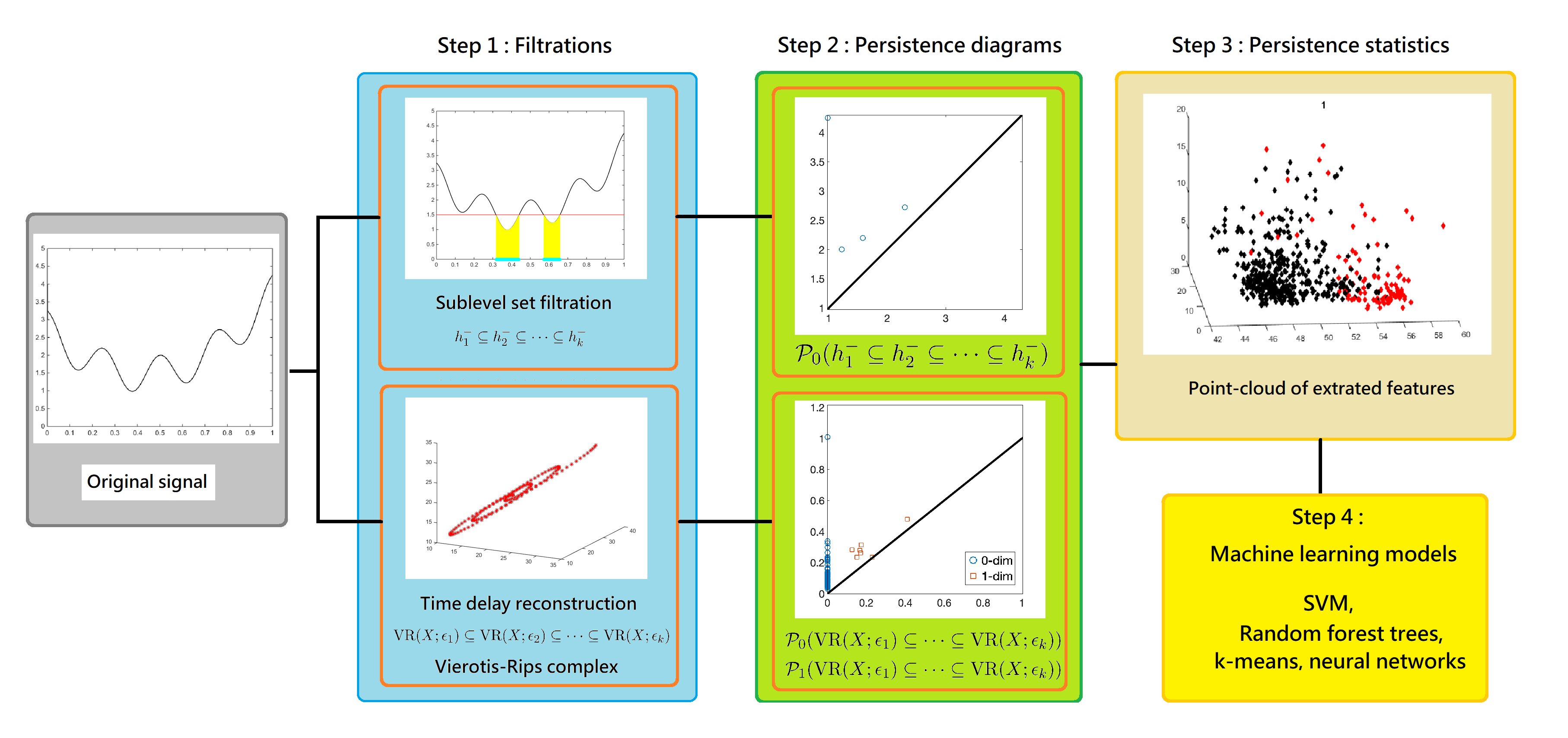}
	\caption{The scheme of our proposed time series analysis can be separated by three steps: constructing filtrations, computing PD's and extracting PS as features. The features are applied to train a machine learning model for the classification purpose.}
	\label{Figure : main_scheme}
\end{figure}

Our main scheme for {studying a complicated time series} is shown in Fig. \ref{Figure : main_scheme}, which is divided into three steps that we will detail later. First, consider two {\em filtrations}, the Vietoris-Rips (VR) complex filtration {of the Takens' lag map \cite{Takens:1981}} and the sub-level set filtration {of the time series}, and {\em persistent homology} (PH). Second, compute corresponding {PDs}. Finally, calculate {\em persistence statistics} (PS) as a novel {statistic of the time series of interest}. 
{We mention that compared with existing TDA approach for time series analysis}, our proposed PS features based on both sub-level set and VR complexes filtrations are intuitive, straightforward to implement, and also {computationally efficient}. 
%
%The proposed PS features will be used for the automatic sleep stage annotation problem.

\subsection{Application -- Sleep dynamics}

To demonstrate the usefulness of the proposed PS{, we apply it to study IHR time series recorded during sleep, and use obtained statistics to classify} sleep stages. 
Sleep is a universal recurrent physiological phenomenon. %{In recent} decades, a growing body of evidence shows that sleep is not only intimately related to {personal} health \cite{Karni1994,Kang2009} but also has a direct impact on public health \cite{Colten_Altevogt:2006}. 
%
%In clinics, sleep experts score sleep stage by reading the electroencephalogram (EEG), electrooculogram (EOG) and electromyogram (EMG) based on the American Academy of Sleep Medicine (AASM) criteria \cite{Iber_Ancoli-Isreal_Chesson_Quan:2007,berry2012aasm}. 
% 
Sleep impacts the whole body, so we can read sleep via reading different physiological signals.
Taking ECG into account is specifically attractive, since {the ECG sensor is} easy to install, and {it is now widely available} in mobile health devices. 
HRV {of a subject is usually quantified by analyzing} ECG, {and it} has {been} shown to be related to sleep dynamics \cite{Zemaityte1984,Vaughn1995,Toscani1996,Bonnet1997,SigridElsenbruchMichaelJHarnish1999,Chouchou2014,Penzel2016}. In other words, the heart rate rhythm provides a non-invasive window for researchers to study sleep. While there have been several studies trying to classify sleep stages based on HRV \cite{Lewicke2008,Mendez2010,Long2012,Xiao2013,Aktaruzzaman2015,Ye2016,malik2018sleep}, {it still remains a challenging problem in the field.} The challenge and difficulty of this mission can be appreciated from the reported results. 
In this article, we apply the proposed PS to quantify HRV during sleep, and propose a new prediction algorithm for the sleep stage; for example, an automatic classification of wake and sleep,  REM and NREM, and wake, REM and NREM. {We remark that w}hile we focus on the HRV and sleep stage classification, the result indicates the potential of applying TDA-based approaches to study other complicated time series.
%
%{Specifically, w}hen a subject is awake, since the sympathetic tone of the ANS is dominant, he/she has a higher heart rate and a less stable heart rhythm due to external stimuli \cite{Somers:1993}.   
%
%When a subject is asleep, the heart rate is relatively lower, and it reaches its lowest value during deep (slow wave) sleep \cite{Snyder:1964}. During NREM (non-rapid eye movement) sleep, the parasympathetic nervous system dominates the sympathetic tone and the energy restoration and metabolic rates reach their lowest levels, so the heart rate decreases and the rhythm of the heart stabilizes \cite{Somers:1993}. 
%

%The above physiological facts indicate that the heart rate rhythm provides a non-invasive window for researchers to study sleep. There have been several studies trying to classify sleep stages based {\em solely} on HRV \cite{Lewicke2008,Mendez2010,Long2012,Xiao2013,Aktaruzzaman2015,Ye2016,vicente2016drowsiness,malik2018sleep}. Most of them focus on classifying wake and sleep \cite{Lewicke2008,Long2012,Aktaruzzaman2015,Ye2016,malik2018sleep}, some focus on detecting drowsiness \cite{vicente2016drowsiness}, and some focus on classifying rapid eye movement (REM) and NREM \cite{Mendez2010}, or wake, REM and NREM \cite{Xiao2013}. 

\subsection{Organization}

In Section \ref{sec:background}, we review the mathematical background of the PH and PD. In Section~\ref{sec:applications to time series}, we demonstrate two ways to use the PH to study time series,
 {and} propose a new approach to summarize the PD, called the PS. 
The classification model based on the PS for the sleep stage classification will be discussed in detail in Section \ref{sec:method}.  The discussion of our classification performance and a comparison with the state-of-arts results will be included in Section~\ref{sec:discussion and conclusion}. {More technical details and numerical results are postponed to the Online Supplementary.} %XXX {Make sure we come back!}

\section{Mathematical Background}
\label{sec:background}

%Our main focus in this study is quantifying HRV by analyzing the IHR time series based on the TDA tool.  
%
In this section, we describe the mathematical background, including simplicial complex, homology, filtration of sets and the PH. %and then present two PH approaches to study time series -- the sub-level set approach and the delay map approach. Via the proposed PS, we obtain different information about time series.  
Although these topics can be studied in an abstract and general way (see e.g. \cite{Munkres}), to enhance the readability, we present {them} in a relatively concrete way without losing critical information.  %{Readers who are familiar with these concepts might skip this section.}

\subsection{Simplicial Complexes}

To investigate the complicated structure of an object, an intuitive way is to use simple objects as building blocks to approximate the original object. 
In TDA, the main building block is the {\it simplicial complex}, which we briefly recall now.   
See Section \ref{section:simplicialcomplexes} for more detailed mathematical background and illustrative examples.

We start with the {\em simplex}.
Intuitively, a simplex is a ``triangle'' of different dimension. 
{ Let $x_0, x_1, \ldots, x_q$ be affinely independent points in $\mathbb{R}^d$, where $d, q \in \mathbb{N}$ and $d \geq q$}. The $q$-\textbf{simplex}, denoted by $\sigma := \langle x_0, x_1, \ldots, x_q \rangle$, is defined to be the convex hull of $x_0, x_1, \ldots, x_q$. 
Denote ${\rm Vert}(\sigma):=\{x_0, x_1, \ldots, x_q\}$.
Any $q$-simplex is a $q$-dimensional object {consisting} of lower degree simplexes. We are interested in the relation among simplexes of different dimensions.
Since any $V\subset{\rm Vert}(\sigma)$ is also affinely independent, the convex hull of $V$, called a {\em face} of $\sigma$, forms a simplex of dimension $|V| \leq q$, where $|V|$ is the cardinality of $V$.  
If $|V| = k$, the face $\tau = \langle V \rangle$ is called a $k$-face of $\sigma$.
A {\em simplicial complex} $\mathcal{K}$ in $\bbR^d$ is a collection of finite simplexes $\sigma$ in $\bbR^d$ so that any intersection of two arbitrary simplexes is { a face to each of them}; that is,
	\begin{itemize}
		\itemsep = -1pt
		\item If $\sigma \in \mathcal{K}$ and $\tau$ is a face of $\sigma$, then $\tau \in \mathcal{K}$;
		\item If $\sigma_1, \sigma_2 \in \mathcal{K}$, then $\sigma_1 \cap \sigma_2$ is a face of $\sigma_1$ and $\sigma_2$. In particular, $\sigma_1 \cap \sigma_2 \in \mathcal{K}$.
	\end{itemize}

\subsection{Homology and Betti numbers}

In order to study the topological information of a given simplicial complex, we study relations among simplexes of different dimensions, and hence the ``holes''. \textit{Homology} and { \textit{Betti numbers}} are classic subjects in the algebraic topology \cite{Munkres}, which capture ``holes'' of geometric objects of different dimensions. While we can discuss these topics in a more general setup, in this work, we mainly consider simplicial complexes as our target object.  
See Section \ref{section:MoreBettiHomology} for more information and illustrative examples.

%In order to count $q$-dimensional holes in $\mathcal{K}$, 
We need an algebraic structure of simplexes. 
Given $q$-simplexes $\sigma_1, \sigma_2, \ldots, \sigma_n$ in a simplicial complex $\mathcal{K}$, define the sum over $\bbZ_2$ as $c = \sum_{i = 1}^n \nu_i \sigma_i$, where $\nu_i\in \bbZ_2$.  This formal sum is commonly known as a $q$-{\it chain}. One could also define an addition operator as $\sum_{i = 1}^n \nu_i \sigma_i + \sum_{i = 1}^n \mu_i \sigma_i := \sum_{i = 1}^n (\nu_i+\mu_i) \sigma_i$.  We consider the collection of all $q$-chains, denoted as 
\begin{align}
C_q(\mathcal{K}) := \bigg\{ \sum_{i = 1}^n \nu_i \sigma_i~\Big|~ \nu_i \in \bbZ_2,~\sigma_i\in \mathcal{K},~\dim(\sigma_i)=q \bigg\}.
\end{align}
One could prove that $C_q(\mathcal{K})$ is actually a vector space over $\bbZ_2$ with the above addition.  
There is a natural relation between $C_q(\mathcal{K})$ and $C_{q-1}(\mathcal{K})$, called the {\em boundary map} \cite[Sec. 1.5, p. 30]{Munkres}.
%
%Let $\sigma = \langle x_0, x_1, \cdots, x_q \rangle \in C_q(\mathcal{K})$. 
The $q^{\rm th}$ {\em boundary map} $\partial_q : C_q(\mathcal{K}) \rightarrow C_{q-1}(\mathcal{K})$ over $\bbZ_2$ is defined by
	\begin{equation}
	\partial_q(\langle x_0, x_1, \cdots, x_q \rangle) := \sum_{i = 0}^q \langle x_0, \cdots, \widehat{x_i}, \cdots x_q \rangle,
	\end{equation}
	where $\langle x_0, x_1, \cdots, x_q \rangle\in \mathcal{K}$ and the $\widehat{\bullet}$ denotes the drop-out operation. 
With the boundary maps, there is a nested relation among chains
\begin{equation}
\cdots\xrightarrow[]{\partial_{n+1}} C_n(\mathcal{K}) \xrightarrow[]{\partial_{n}} C_{n-1}(\mathcal{K}) \xrightarrow[]{\partial_{n-1}} \cdots C_1(\mathcal{K}) \xrightarrow[]{\partial_{1}} C_0(\mathcal{K}).
\end{equation}
This nested relation among chains is known as the {\em chain complex}, which is denoted as $\mathcal{C} = \{ C_q, \partial_q \}_{q \in \bbZ}$.

A fundamental result in the homology theory (\cite{Munkres} Lemma 5.3 Sec. 1.5, p. 30) is that the composition of any two consecutive boundary maps is trivial, i.e.  $\partial_{q-1} \circ \partial_q = 0$. This result allows one to define the following quotient space.  Denote \textit{cycles} and \textit{boundaries} by $Z_q$ and $B_q$, respectively, which are defined as
\begin{align}
Z_q &:= \ker(\partial_q) = \{ c \in C_q \ | \ \partial_q(c) = 0 \}, \,
B_q:= {\rm im}(\partial_{q+1}) = \{ \partial_{q+1}(z) \in C_{q} \ | \ z \in C_{q+1} \}.\nonumber
\end{align}
Note that each $B_q$ is a subspace of $Z_q$ since $\partial_{q-1} \circ \partial_q = 0$. Therefore, we can define the $q^{\rm th}$ \textit{homology} to be the quotient space
\begin{equation}
\label{Equation : Quotient space as homology}
H_q(\mathcal{K}) := \frac{Z_q}{B_q} = \frac{\ker(\partial_q)}{{\rm im}(\partial_{q+1})},
\end{equation}
which is again a vector space.
The $q^{\rm th}$ { \textit{Betti number}} is defined to be the dimension of the $q^{\rm th}$ homology; that is,
\begin{align}
\beta_q(\mathcal{K}) = \dim(H_q(\mathcal{K}))\,,
\end{align}
which measures the number of $q$-dimensional holes.
As a result, given { a simplicial} complex $\mathcal{K}$, the {\em homology of $\mathcal{K}$} is a collection of vector spaces $\{ H_q(\mathcal{K}) \}_{q = 0}^{\infty}$, and its {\it Betti numbers} is denoted as $\beta(\mathcal{K}):=\{ \beta_q(\mathcal{K}) \}_{q = 0}^{\infty}$.  

\subsection{Persistent Homology}
\label{subsec:persistent homology}

\begin{figure} 
	\centering 
	\includegraphics[width =0.7 \textwidth]{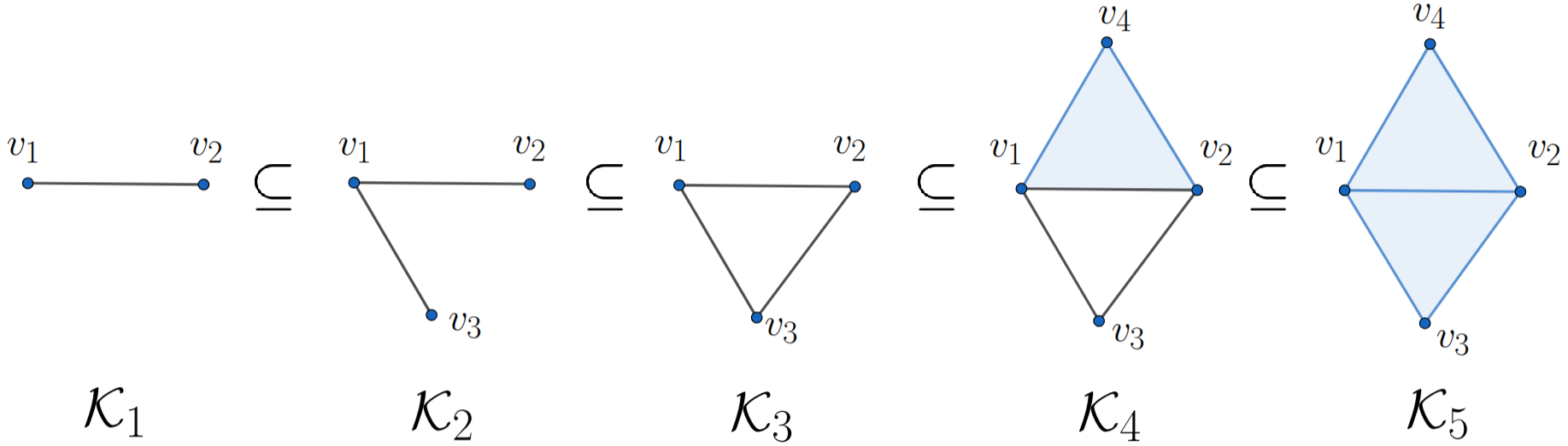}
	\caption{Illustration of a filtration of simplicial complexes $\mathcal{K}_1, \mathcal{K}_2, \mathcal{K}_3, \mathcal{K}_4$ and $\mathcal{K}_5$.\label{fig:filtation of simplicial complexes}}
\end{figure}

We now introduce a natural generalization of homology, the PH, that is suitable for data analysis.
PH {is more suitable for data analysis than homology due to this capability of dealing with inevitable noise in real world dataset}.  It depends on the {notion of} {\it filtration} {to handle noise}. {In general, filtration} is a sequence of simplicial complexes (see Fig.~\ref{fig:filtation of simplicial complexes} for an example).  We are interested in how the ``holes'' vary in the filtration. Intuitively, if certain holes are ``robust'', they will survive in the filtration.  

\begin{def.}
[\cite{Edelsbrunner2010} Sec. III.4, p. 70]
	For an index set  $I$, a {\em filtration} is a sequence of simplicial complexes, $\{\mathcal{K}_{t}\}_{t\in I}$, satisfying 
	\begin{align}
	\mathcal{K}_{t_1} \subseteq \mathcal{K}_{t_2},\text{ whenever } t_1 \leq t_2.
	\end{align}
\end{def.}

{ From the previous discussion}, for each $\mathcal{K}_t$ in a filtration, one could compute its homology group and Betti number.
Because of the nested subset relation in a filtration, { there exist relations among simplicial complexes}. This allows one to track and record the changes of the homology group and the Betti numbers, which we detail now. 
Given a fixed $q \geq 0$, each $\mathcal{K}_i$ induces homology $H_q(\mathcal{K}_i)$. Denote $\iota_i : \mathcal{K}_i \rightarrow \mathcal{K}_{i+1}$ to be the inclusion map. Then $\iota_i(Z_q(\mathcal{K}_i)) \subseteq Z_q(\mathcal{K}_{i+1})$ and $\iota_i(B_q(\mathcal{K}_i)) \subseteq B_q(\mathcal{K}_{i+1})$ (\cite{Edelsbrunner2010} Sec. IV.4, p. 95). Therefore, the mapping
\begin{equation}
\iota_{i,H_q} : H_q(\mathcal{K}_i) \rightarrow H_q(\mathcal{K}_{i+1}), \ \ \overline{c} \longmapsto \overline{\iota_i(c)}
\end{equation}
induced by $\iota_i$ is a well-defined linear transformation over $\bbZ_2$. We also define a linear transformation
\begin{equation}
\label{Equation : Functoriality of maps}
\rho_q^{i,i+k} :=  \iota_{i+k-1,H_q} \circ \cdots \circ \iota_{i+1,H_q} \circ \iota_{i,H_q},
\end{equation}
which maps $H_q(\mathcal{K}_i)$ to $H_q(\mathcal{K}_{i + k})$. The following definition is crucial for defining {\em lifespans} of connected components or holes in homology theory. 

\begin{def.}
[\cite{Edelsbrunner2010} Sec. VII.1, p. 151]
\label{def:birth-death}
	Let $\{ \mathcal{K}_i \}_{i=0}^n$
	be a filtration of simplicial complexes. For $q \in \bbZ_{\geq 0}$ and $i, j \in \bbZ_{\geq 0}$ with $i \leq j$, we define the PH as 
	\begin{equation}\
	\label{equ:Hqij}
	H_q^{i,j} := \frac{Z_q(\mathcal{K}_i)}{B_q(\mathcal{K}_j) \cap Z_q(\mathcal{K}_i)}.
	\end{equation}
\end{def.}
Since $\mathcal{K}_0 \subseteq \mathcal{K}_1 \subseteq \cdots \subseteq \mathcal{K}_n$, we have inclusions of $q$-chains: $C_q(\mathcal{K}_0) \subseteq C_q(\mathcal{K}_1) \subseteq \cdots \subseteq C_q(\mathcal{K}_n)$ for all $q \geq 0$. Hence, the intersection $B_q(K_j) \cap Z_q(\mathcal{K}_i)$ is a well-defined subspace of $Z_q(\mathcal{K}_i)$. Moreover, for $i \leq j$, the kernel of the linear transformation
\begin{equation*}
\phi : Z_q(\mathcal{K}_i) \longrightarrow \frac{Z_q(\mathcal{K}_j)}{B_q(\mathcal{K}_j)}, \ \ c \longmapsto \overline{c} = c + B_q(\mathcal{K}_j)
\end{equation*}
induced by the inclusion map is $B_q(\mathcal{K}_j) \cap Z_q(\mathcal{K}_i)$. By the first isomorphism theorem, we obtain an injective linear transformation
\begin{equation*}
\overline{\phi} : \frac{Z_q(\mathcal{K}_i)}{B_q(\mathcal{K}_j) \cap Z_q(\mathcal{K}_i)} \longrightarrow \frac{Z_q(\mathcal{K}_j)}{B_q(\mathcal{K}_j)} \,.
\end{equation*}
Via the one-to-one linear mapping $\overline{\phi}$, the vector space $H_q^{i,j}$ may be viewed as a subspace of $H_{q}(\mathcal{K}_j)$. In particular, if $i = j$, then $H_q^{i,j} = H_q(\mathcal{K}_i) = H_q(\mathcal{K}_j)$, which means that the PH is a generalization of the homology. 
With the inclusion $H_q^{i,j} \hookrightarrow H_q(\mathcal{K}_j)$, we define the moments of {\em birth} and {\em death} of a ``hole'' in the filtration. 

\begin{def.}
[\cite{Edelsbrunner2010} Sec. VII.1, p. 151]
	Let $\{ \mathcal{K}_i \}_{i=0}^n$
	be a filtration of simplicial complexes and $q \in \bbZ_{\geq 0}$. 
	\begin{itemize}
		\itemsep = -1pt
		\item {
		A $q$-th class $\overline{c}$ ($c \in Z_q(\mathcal{K}_i)$) is \textbf{born} at $\mathcal{K}_i$ if $\overline{c} \in H_q(\mathcal{K}_i) \setminus \{ 0 \}$, but $\overline{c} \notin {\rm \im}(\rho_q^{i-1,i})$};
		\item A $q$-th class $\overline{c}$ ($c \in Z_q(\mathcal{K}_i)$) \textbf{dies} at $\mathcal{K}_j$ if $\rho_q^{i,j-1}(\overline{c}) \notin H_q^{i-1,j-1}$, but $\rho_q^{i,j}(\overline{c}) \in H_q^{i-1,j}$. 
	\end{itemize}
\end{def.}
{

We use Fig.~\ref{fig:filtation of simplicial complexes} to explain the relation between these two abstract definitions.
For instance, the non-trivial class $\overline{c}$ represented by $1$-chain
\begin{equation*}
\begin{split}
	c &= \langle v_1,v_2 \rangle + \langle v_2,v_3 \rangle + \langle v_3,v_1 \rangle
\end{split}
\end{equation*}
in $H_1(\mathcal{K}_{3})$ is born at $\mathcal{K}_{3}$ i.e. $\overline{c} \notin {\rm im}(\rho_1^{2,3})$ because $H_1(\mathcal{K}_2) = \{0 \}$ and $\bbZ_2 = H_1(\mathcal{K}_3) = \langle \overline{c} \rangle$. On the other hand, the fact $\{ 0 \} \subseteq H_1^{2,5} \subseteq H_1(\mathcal{K}_5) = \{ 0 \}$ shows that $\rho_1^{3,5}(\overline{c}) \in H_1(\mathcal{K}_5) = H_1^{2,5}$ and $\rho_1^{3,4}(\overline{c}) \notin H_1^{2,4}$ because $H_1^{2,4} = \{ 0 \}$ (since $Z_1(\mathcal{K}_2) = \{ 0 \}$) and $\bbZ_2 = H_1(\mathcal{K}_{4}) = \langle \rho_{1}^{3,4}(\overline{c}) \rangle$, thus $\overline{c}$ dies at $\mathcal{K}_5$.
}
We refer readers with interest to \cite{Edelsbrunner2010} for more details in PH.

\subsection{Persistence Diagram}\label{Subsection persistence diagram}
{\em Persistence diagram} (PD) proposed in \cite{edelsbrunner2000topological} {or equivalently {\em persistence barcodes} proposed in \cite{carlsson2005persistence}} is a tool {to {visualize} the complicated} lifespans {of holes in} a given filtration for data analysis.  {We use PD in this paper. }

%{\em Persistence diagram} (PD) {or {\em persistence barcodes}} is a tool proposed in \cite{edelsbrunner2000topological} {to summarize the complicated} lifespans {of holes in} a given filtration for data analysis. %The PD summarizes lifespans of holes in a filtration. 
%``Change of structures'' means the {\it lifespans} of holes in a filtration. 
%
{ The PD possesses} the desired {\em stability property} \cite{Steiner-Edelsbrunner-Harer2007} -- a {bounded} perturbation of a given filtration leads to a {bounded} perturbation of the corresponding PD. {Due to the inevitable noise in real data, this stability property renders PD-based approaches suitable for data analysis.} The bottleneck and Wasserstein distances \cite{Steiner-Edelsbrunner-Harer2007} are typical ways to measure differences among PDs. The formal statements {of the stability property based on these two distances} are provided in Section~\ref{subsection : sub-level set Filtration} and Section~\ref{subsec : Vietoris-Rips Complexes Filtration}.
We refer readers with interest { to \cite{Edelsbrunner2010} for details} in PD.

\begin{def.}
[\cite{Edelsbrunner2010} Sec. VII.1, p. 152]
	Let $\{ \mathcal{K}_i \}_{i=0}^n$ be a filtration of simplicial complexes and $q \in \bbZ_{\geq 0}$. The $q^{\text{th}}$ PD, denoted as $\mathcal{P}_q(\{ \mathcal{K}_i \}_{i=0}^n)$, of the filtration is the multiset of $q$-dimension holes in the filtration.  More precisely, $\mathcal{P}_q(\{ \mathcal{K}_i \}_{i=0}^n)$ is the multiset of all tuples $(b,d)$ corresponding to $q$-dimensional holes $\overline{c}$ { which satisfy} $\rho_q^{b,d-1}(\overline{c}) \notin H_q^{b-1,d-1}$ and $\rho_q^{b,d}(\overline{c}) \in H_q^{b-1,d}$. 
	\end{def.}
In other words, a $q$-dimensional hole in a filtration is recorded by a pair $(b,d)$ of integers where $b$ and $d$ are called the {\it birth} and {\it death} of the hole respectively \cite{Edelsbrunner2010}.
Although the above definition of PD seems technical, its interpretation is intuitive.  For instance, consider the filtration shown in Fig.~\ref{fig:filtation of simplicial complexes}. We look for the ``changes'' of topological structure (holes). Note that since a connected component is born at $\mathcal{K}_1$ (specifically, $\langle v_1, v_2 \rangle$), its birth value is $b=1$; since it lives throughout the filtration, its death value is $\infty$.  We now turn our focus to the 1-dimensional hole.  Note that a 1-dimensional hole (specifically, $\langle v_1, v_2 \rangle + \langle v_2, v_3 \rangle + \langle v_1, v_3 \rangle$) is formed at $\mathcal{K}_3$, so its birth value is $3$; note also that this hole is filled at $\mathcal{K}_5$, so its death value is $5$.   Since there is no more change of holes, we have the persistence diagrams $\mathcal{P}_0(\{ \mathcal{K}_i \}_{i=1}^5)= \{ (1, \infty)\}$ and  $\mathcal{P}_1(\{ \mathcal{K}_i \}_{i=1}^5)= \{ (3, 5)\}$. 

\begin{figure}[hbt!]
	\centering
	\subfigure[Point cloud data $X$.]{
		\includegraphics[width = 0.48\textwidth]{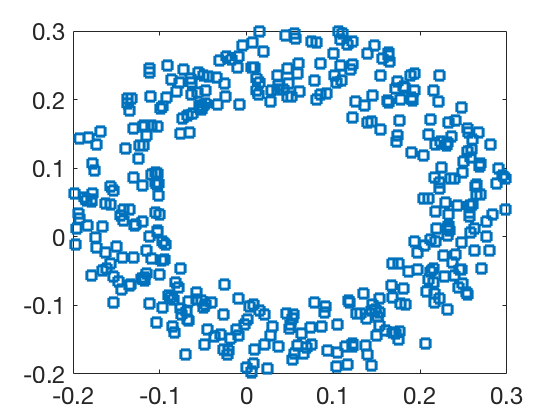}}
	\subfigure[$\mathcal{P}_1(X)$.]{
		\includegraphics[width = 0.48\textwidth]{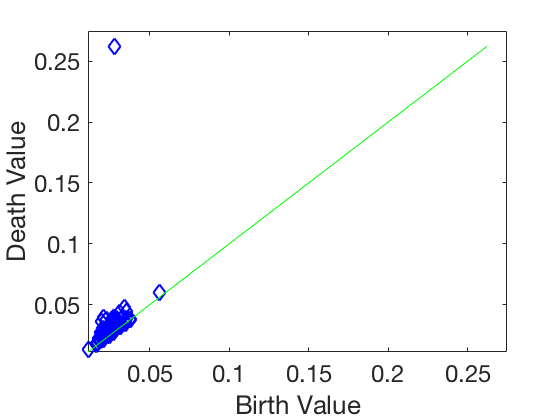}}
	\caption{Toy example of the persistent homology.  (a) Data points are sampled from a circle with the Gaussian noise. (b) 1st dimensional persistence diagram.  }
	\label{fig:toy example}
\end{figure}

{{Before closing this subsection, we illustrate how PH and PD work by taking a noisy point cloud} sampled from a circle {contaminated by} Gaussian noise shown in Fig.~\ref{fig:toy example}(a). If there is no noise, the 1st Betti {number} of {the} circle is $\beta_1=1$. In the {noisy case}, the Betti number information is contained in the form of the PD as shown in Fig.~\ref{fig:toy example}(b), where each point represents {one} 1-dimensional hole associated with its birth and death value.  In Fig.~\ref{fig:toy example}(b), we observe that there is {an} outstanding point {with long lifespan} (located around birth value 0.05 and death value 0.25), while lifespans for other points are very small.  This suggests that the {noisy} point cloud has a strong/robust 1-dimensional hole. {This captures the main topology information, $\beta_1=1$,} about this data.  %In this subsection, we will give an overview of PH.
%Intuitively, the data seems like a circle, and there is a hole that is born before 0.05 and dies around 0.25.  This means that the original data contains a prominent circle structure.
}

\subsection{Data analysis with PD and commonly considered TDA statistics}
\label{subsec:da with pd}
{Usually, researchers design statistics on the PD of a given dataset via the chosen filtration. One basic result supporting this approach is} 
\cite{mileyko2011probability}, {where} authors showed that the space of PDs with certain metric is complete and separable. {This result} forms a theoretic foundation for any statistical methods. In \cite{fasy2014confidence,Blumberg2014}, authors derived confidence sets of PDs in order to separate the long lifespan holes from noisy ones, and also proposed four ways to estimated them. 
{While these theoretical results shed light on applying TDA to analyze complex data,} however, any operation in the space of PDs is complicated and difficult to compute.  
{For example, c}omputing bottleneck or Wasserstein distances among PDs is a difficult task and can be time consuming, even for the state-of-art algorithm \cite{kerber2017geometry}.  {Another result indicates that} the mean in the space of PDs may not be unique \cite{Turner2014}. 
{This computational burden} renders it less applicable to data analysis. 

\begin{figure}[hbt!]
	\centering
		\includegraphics[width = 0.75\textwidth]{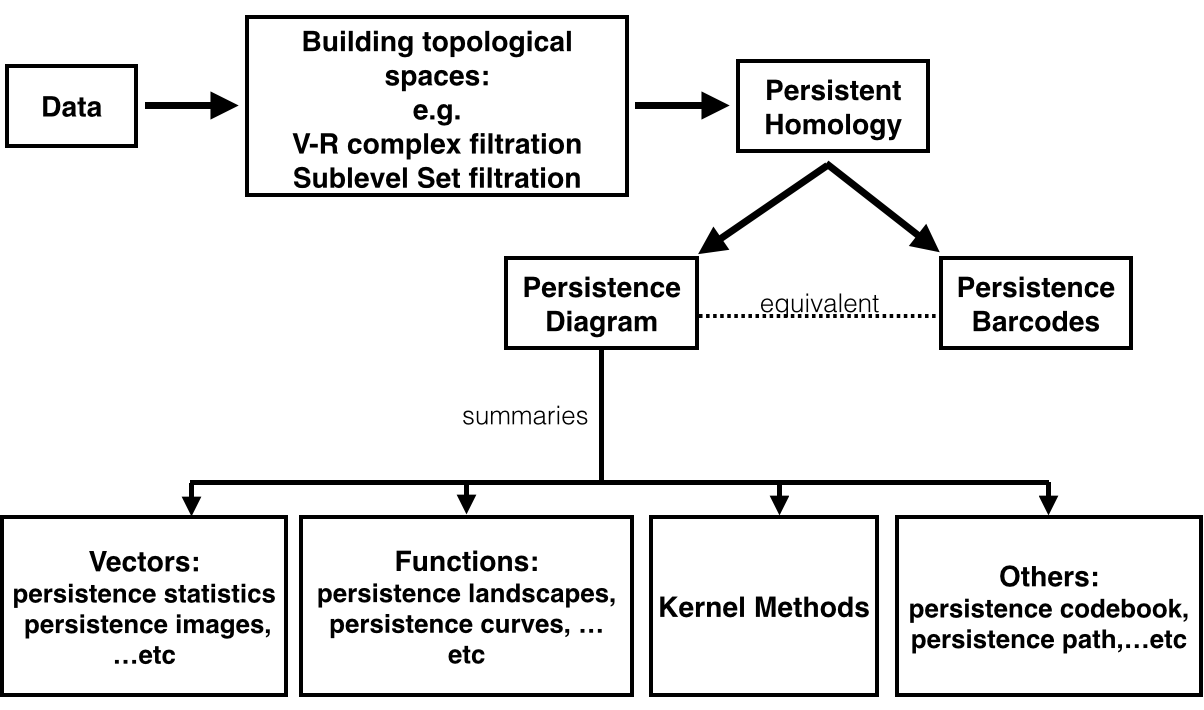}
	\caption{A chart depicting relationship among existing TDA tools.}
	\label{fig:all TDA tools}
\end{figure}

To get around the {computational issue when working with} those distances, one {major approach} is to ``vectorize'' PDs; that is, researchers {map} the space of PDs into another space. For example, persistence landscapes \cite{bubenik2015statistical} {map} PDs into a Banach space, specifically $L^p$ space.  
More examples include persistence image \cite{adams2017persistence}, generalized persistence landscapes \cite{berry2018functional}, persistence path \cite{chevyrev2018persistence}, persistence codebook \cite{zielinski2018persistence}, persistence curves \cite{chung2019persistence}, kernel based methods \cite{reininghaus2015stable,kusano2016persistence}, and persistent entropy \cite{atienza2017persistent,chintakunta2015entropy}. These methods have been studied and {applied to} different applications. In Fig.~\ref{fig:all TDA tools}, we provide a chart depicting the relationship among existing TDA tools. {We mention that} the proposed PS in Section \ref{sec:applications to time series} could be viewed as a computationally efficient vectorization of PD's. % (we will give more references in Section~\ref{sec:ps})

\section{TDA for time series analysis {and features extraction}} 
\label{sec:applications to time series}

%We now apply the PH theory to analyze complicated time series. 
{Armed with the theoretical background in Section \ref{sec:background}, we are ready to describe how to apply TDA for time series analysis.} To apply the PH to analyze complicated time series, { we introduce two useful filtrations, the sub-level set filtration and the VR complexes filtration. With these two filtrations, we introduce a novel features extraction methods, {coined PS,} based on the PDs of the sub-level set filtration and the VR complexes filtration.}

\subsection{{First useful filtration --} Sub-level set Filtration}
\label{subsection : sub-level set Filtration}
To simplify the discussion and illustrate the idea, we identify a time series as a discretization of a continuous function $f : [0, T] \rightarrow \bbR$, where $T$ is some fixed constant. For each $h \in \bbR$, the {\it sub-level set} of $f$ is defined as
\begin{equation}
\label{equ:sub-level set def}
f_h := f^{-1}((-\infty, h]) = \{ t \in [0,T] \ | \ f(t) \leq h \}.
\end{equation}
Clearly, $f_{h_1} \subseteq f_{h_2}$ whenever $h_1 \leq h_2$. 
Therefore, for any increasing sequence $\{h_i\}_i$, the collection of sub-level sets, $\{ f_{h_i} \}$, forms a filtration.  
Intuitively, the sub-level set filtration reveals the {\em oscillating information} of the functions.  
Since each $f_h$ is a subset of $[0,T] \subseteq \bbR$, it only contains $0$-dimensional { structures, i.e.,} connected components.  Hence, the only non-trivial PD in this case is $\mathcal{P}_0$.  
For simplicity, when there is no danger of confusion, for a given function $f$, we use $\mathcal{P}_0(f)$ to denote $\mathcal{P}_0(\{ f \}_{h_i})$, the PD associated with the sub-level sets filtration of $f$.
As discussed in \cite{Edelsbrunner2010}, each element in $\mathcal{P}_0$ is a min-max pair in the original function $f(t)$.  
The concept of this filtration is closely related to the size function theory (see \cite{biasotti2008describing} and references therein) and is commonly used as a shape descriptor \cite{biasotti2008describing}.  %In this essence, we treat signals as 1D shapes, and use PS as shape descriptors.  
%
%As we mentioned in Section~\ref{sec:background}, PDs are stable measurements, and 
{In practice, PD is robust to noise under the {\em bottleneck distance}. This fact renders PD an useful data analysis quantity. A precise} statement of this {\em robustness} is below.

\begin{theorem}
	\label{Theorem : Stability of PD with continuous tame functions}
	Let $X$ be an $n$-dimensional rectangle in $\bbR^n$. {Take two} continuous functions $f, g: X \rightarrow \bbR$ {with} finitely many local extremums (minimums or maximums). Then, we have for $q\in\mathbb{N}$, 
	\begin{equation*}
	d_B(\mathcal{P}_q(f), \mathcal{P}_q(g)) \leq \Vert f - g \Vert_{\infty},
	\end{equation*}
	where $d_B$ is the {\em bottleneck distance} defined as
	$d_B(M,N) = \inf_{\gamma} \sup_{m \in M} \Vert m - \gamma(m) \Vert_{\infty}$,
	where $\gamma$ ranges over all bijections from multisets $M$ to $N$. Here we interpret each point with multiplicity $k$ as $k$ individual points and the bijection is between the resulting sets. 
\end{theorem}
In fact, Theorem~\ref{Theorem : Stability of PD with continuous tame functions} is a special form of a stability theorem (Main Theorem in \cite{Steiner-Edelsbrunner-Harer2007}, p. 109). 
See Section \ref{Section:MoresublevelVRfiltrations} for an illustrative example of the sub-level sets filtration.

\subsection{{Second useful filtration --} Vietoris-Rips Complexes Filtration}
\label{subsec : Vietoris-Rips Complexes Filtration}

To introduce Vietoris-Rips (VR) complexes filtration for a given time series, we first embed the time series into a high dimension point cloud via {\em Taken's lag map} \cite{Takens:1981}, which is constructed in the following way. Take $p\in \mathbb{N}$ to be the dimension of the embedding, and $\tau\in \mathbb{N}$ to be the lag step. 
For a given time series $H:\mathbb{Z}\to \mathbb{R}$, the lag map with lag $\tau$ and dimension $p$ is defined as 
\begin{align}
R_{p, \tau}(H) = \{ (H(t),  H(t-\tau), H(t-2\tau), \ldots,  H(t-(p-1)\tau))^\top|\,t\in\mathbb{Z}\}\,, 
\end{align}
%{ Should we change this notation to $R_{p, \tau}(H)$???}
which is a subset of $\mathbb{R}^p$. We postpone details of Taken's lag map to Section \ref{SI : Taken's Lag Map}.
With the point cloud $R_{p, \tau}\subset \mathbb{R}^p$, we are ready to introduce the VR complex. 

In general, given a point cloud $\mathcal X = \{ x_1, \ldots, x_N \}\subset \mathbb{R}^p$, the main idea of VR complex is to build simplicial complexes from $\mathcal X$ if points in $\mathcal X$ are closed enough.  A formal definition is given below. 

\begin{def.}
[\cite{Edelsbrunner2010} Sec. III.2, p. 61]
	Let $X = \{ x_1, x_2, \ldots, x_N \} \subseteq \bbR^p$ be a point cloud and take $\epsilon > 0$. The VR complex is a collection of all $q$-simplexes $\sigma$ with vertices in $X$ with $\diam(\sigma) \leq 2\epsilon$, where $\diam(\sigma)$ is the diameter of $\sigma$; that is,
	\begin{equation}
	\VR(X;\epsilon) := \bigcup_{q = 0}^p \left \{ q{\rm -simplex} \ \sigma \ | \ \diam(\sigma) \leq 2\epsilon, {\rm Vert}(\sigma) \subseteq X \right \}.
	\end{equation}
\end{def.}
Clearly, for an increasing sequence $\epsilon_1 < \epsilon_2 < \cdots < \epsilon_N$, the corresponding sequence of VR complexes forms a filtration:
\begin{equation}
\VR(X;\epsilon_1) \subseteq \VR(X;\epsilon_2) \subseteq \cdots \subseteq \VR(X;\epsilon_N).
\end{equation}
After determining the representation rules of connected components, the lifespan of holes of different dimensions can be computed easily. 
 See Section \ref{Section:MoresublevelVRfiltrations} for an illustrative example of the VR filtration.

For simplicity, we denote the $q$-th PD associated the VR filtration as $\mathcal{P}_q(R_{p,\tau}):=\mathcal{P}_q(\{ \VR(R_{p,\tau};~\epsilon) \}_{\epsilon})$.  
In parallel with Theorem~\ref{Theorem : Stability of PD with continuous tame functions}, the stability of PDs extracted from a VR filtration has been discussed in \cite{Chazal-Silva-Oudot2014}. 
\begin{theorem}[\cite{Chazal-Silva-Oudot2014}, Theorem 5.2]
\label{Theorem : Stability of PD of VR complexes}
	For finite metric spaces $(X, d_X)$ and $(Y, d_Y)$, {then} for $q\in \mathbb{N}$,
	\begin{equation*}
	d_B(\mathcal{P}_q({\rm VR}(X)), \mathcal{P}_q({\rm VR}(Y))) \leq  2d_{GH}(X,Y),
	\end{equation*}
	where $d_B$ is the {\em bottleneck distance} and $d_{GH}$ is the Gromov-Hausdorff distance.
\end{theorem}

%\section{Persistence Statistics}
\subsection{{Persistence Statistics}}
\label{sec:ps}

We {now} introduce a { set of} { new} features to represent PDs. {It is {computationally} efficient and straightforward to implement.} Motivated by features considered in \cite{ChungSkin2018} to classify different types of skin lesions, and those considered in \cite{Mittal-Gupta2017} to study bifurcations and chaos in complex dynamic systems,
we propose to explore distributions of the birth $b$ and the death $d$ of all possible holes.  
{ To be more specific}, given a PD $\mathcal{P}$, we transform it into two {multi-sets} of numbers, $M$ and $L$, defined as
\begin{equation}
M = \left\{ \frac{d+b}{2} ~\Big|~ (b,d)\in \mathcal{P} \right\} \text{ and } L = \left\{ d-b ~\Big|~ (b,d)\in \mathcal{P} \right\}\,.
\end{equation}
Note that {{for the} VR complex filtration,} $\frac{d+b}{2}$ captures the ``size'' of the associated hole, and $d-b$ captures the robustness of the associated hole.  {On the other hand, {for} the sublevel set filtration, $\frac{d+b}{2}$ {reveals} the locations of holes, and $d-b$ captures the differences between low and high peaks in a time series.}  Note that since the hole $(0, \infty)$ always exists in the PD as is shown in the previous section, it is omitted.

In this paper, {for each persistence diagram,} we consider eight summary statistics to represent the multi-set $M$, including mean, standard deviation, skewness, kurtosis, 25th, 50th, 75th percentile and  
the {\it persistent entropy} \cite{chintakunta2015entropy}. We {number them} from $1$ to $8$. We consider the same summary statistics for the multi-set $L$, and {number them} from $9$ to $16$.
\begin{def.}[{Persistence Statistics}]
	\label{def:persistence stat}
Given a PD, the PS is defined as a map, $\Phi^{(\texttt{PS})}$, that transforms the PD to $\mathbb{R}^{16}$.
\end{def.}

The persistent entropy of $L$ and $M$, denoted as $E(L)$ and $E(M)$ respectively, describes the complexity of $L$ and $M$.  
It has been used to study the cell arrangements \cite{atienza2019persistent} and emotion recognition \cite{gonzalez2019towards}. From the theoretical perspective, it has been shown that the $E(L)$ is a stability measurement \cite{persistentEntropyStability}. On the other hand, $E(M)$ is the new quantity that we propose. It would be interesting to investigate theoretic properties of $E(M)$ for the future work.

Note that while {\em intuitively}, holes with long lifespans are considered important features and those with short lifespans are considered noises,  in our proposed features, we do not discriminate holes with long or short lifespans. In other words, we take all holes into consideration. This approach is supported by a recent discovery that those considered as noisy holes might actually contain important information. For example, in the drivers' behavior classification \cite{bendich2016topological}, authors transformed the space of PDs into ``binned'' diagrams, and found that the main differences occurred in those short lifespan holes.  Another work on the leave classification \cite{Patrangenaru2018} also suggested that holes with short lifespans could better distinguish different {types} of leaves. 

\section{Application to sleep stage classification}
\label{sec:method}

{In recent} decades, a growing body of evidence shows that sleep is not only intimately related to {personal} health \cite{Karni1994,Kang2009} but also has a direct impact on public health \cite{Colten_Altevogt:2006}. 
In clinics, sleep experts score sleep stage by reading the electroencephalogram (EEG), electrooculogram (EOG) and electromyogram (EMG) based on the American Academy of Sleep Medicine (AASM) criteria \cite{Iber_Ancoli-Isreal_Chesson_Quan:2007,berry2012aasm}. 
Sleep, however, impacts the whole body, and we can read sleep via reading physiological signals other than EEG, particularly ECG and HRV mentioned in Introduction.
The relationship between HRV and sleep dynamics has been widely studied in the physiology society \cite{Zemaityte1984,Vaughn1995,Toscani1996,Bonnet1997,SigridElsenbruchMichaelJHarnish1999,Chouchou2014,Penzel2016}. 
{Specifically, w}hen a subject is awake, since the sympathetic tone of the ANS is dominant, he/she has a higher heart rate and a less stable heart rhythm due to external stimuli \cite{Somers:1993}.   
When a subject is asleep, the heart rate is lower, and it reaches its lowest value during deep (slow wave) sleep \cite{Snyder:1964}. During NREM (non-rapid eye movement) sleep, the parasympathetic nervous system dominates the sympathetic tone and the energy restoration and metabolic rates reach their lowest levels, so the heart rate decreases and the rhythm of the heart stabilizes \cite{Somers:1993}. 

The above physiological facts indicate that the heart rate rhythm provides a non-invasive window for researchers to study sleep. There have been several studies trying to classify sleep stages based {\em solely} on HRV. Most of them focus on classifying wake and sleep \cite{Lewicke2008,Long2012,Aktaruzzaman2015,Ye2016,malik2018sleep}, some focus on detecting drowsiness \cite{vicente2016drowsiness}, and some focus on classifying rapid eye movement (REM) and NREM \cite{Mendez2010}, or wake, REM and NREM \cite{Xiao2013}. The challenge and difficulty of this mission can be appreciated from the reported results. {In this section, we apply the TDA tool and the proposed PS to study this problem.}

\subsection{Datasets}

The databases we use here are the same as those used in \cite{malik2018sleep}. Here we summarize them and refer readers with interest in the database details to \cite{malik2018sleep}.
{The {\em CGMH-training} database consists of standard polysomnogram (PSG) signals on patients suspicious of sleep apnea syndrome} at the sleep center in Chang Gung Memorial Hospital (CGMH), Linkou, Taoyuan, Taiwan. The Institutional Review Board of CGMH approved the study protocol (No. 101-4968A3).  
All recordings were acquired on the Alice 5 data acquisition system (Philips Respironics, Murrysville, PA). Each recording {lasts for} at least 5 hours. 
The sleep stages, including wake, REM and NREM (REM and NREM constitute the {\em sleep} stage), were annotated by two experienced sleep specialists according to the AASM 2007 guidelines \cite{Iber_Ancoli-Isreal_Chesson_Quan:2007}, and a consensus was reached. According to the protocol, the sleep specialists provide annotation for {non-overlapping 30 seconds} long epochs.  
{In this study,} we focus on { the second lead of the ECG recording, which was} sampled at $200$ Hz. There are 90 participants without sleep apnea (each with apnea-hypopnea index (AHI) less than $5$) in this database, among which we consider only $56$ participants {who have} at least $10\%$ epochs {labeled as} wake to avoid the imbalanced data issue. 

We consider three validation databases. The first one is the {\em CGMH-validation database}. {This database} consists of $27$ participants acquired independently of CGMH-training from the same sleep laboratory in CGMH {under the same Institutional Review Board}. 
The other two validation databases are publicly available. The DREAMS Subjects Database\footnote{DOI: 10.5281/zenodo.2650142} % The extracted IHR can be found in } 
(DREAMS), consists of $20$ recordings from healthy participants, {where the ECG recordings were acquired by} the Brainnet\texttrademark{} system (Medatec, Brussels, Belgium). The sampling rate is $200$ $\mathrm{Hz}$, and the minimum recording duration is $7$ hours. Although the race information is not provided, we may assume that its population constitution is different from that of the CGMH databases since it is collected from Belgium. This database is chosen to assess the model's performance on participants of a different race {recorded from different} recording machine. 
The third database is the St. Vincent's University Hospital/University College Dublin Sleep Apnea Database (UCDSADB) from Physionet \cite{Goldberger_Amaral_Glass_Hausdorff_Ivanov_Mark_Mietus_Moody_Peng_Stanley:2000} \footnote{\url{https://archive.physionet.org/pn3/ucddb/}}. %\footnote{\url{https://physionet.org/pn3/ucddb}}. 
It consists of $25$ participants with sleep apnea of various severities. {The ECG signal was recorded by} Holter monitor at the sampling { rate of $128$ Hz}. The minimum recording is $6$ hours long. We focus on the first ECG lead in this study. 
The UCDSADB is chosen to assess the model's performance on recordings which come from participants with sleep disorders.
{We remark the these} validation databases are not used to tune the model's parameters, and {no subject is rejected}.

\subsection{Time series to analyze -- Instantaneous Heart Rate}

The data preprocessing steps are the same as those shown in \cite{malik2018sleep}. Here we summarize those steps and refer readers to \cite{malik2018sleep} for more details. First, apply a standard automatic R peak detection algorithm \cite{elgendi:qrs}. Suppose there are $n_k$ R peaks in the $k$-th subject's ECG recording. Denote $\{r_{k,i}\}_{i=1}^{n_k}$ the location in time (sec) of the detected R peaks of the $k$-th subject. 
{We apply the $5$-beat median filter} to remove artifacts in the detected R peaks; that is, if a detected beat is too close or too far from their preceding beats, it is removed or interpolated. Then, the IHR of the $k$-th recording, denoted as $H_k$, is determined by {the shape-preserving piecewise cubic} interpolation \cite{Electrophysiology1043} over the nonuniform sampling
\begin{gather}
H_k( r_{k,i} ) = 60(r_{k,i} - r_{k,i-1})^{-1}.
\end{gather}
{$H_k$ describes the IHR at each time in beats-per-minute.
The IHR is sampled} at a sampling rate of $4$ Hz. 
We break the IHR signal into $30$-second epochs {following} the {same epoch segmentation in the experts' annotations. We discard all epochs with fewer than five detected R peaks. This step is adjusted by physiological knowledge.} 
For each labeled epoch, we build { a time} series of 90 seconds in length { by} concatenating the epoch with the preceding 2 epochs. {For the sake of handling} the inter-individual variance, each 90 seconds time series is normalized by subtracting its median value. 
Thus, for the $j$-th epoch of the $k$-th recording, the associated time series we consider is 
\begin{align}
H^{(k,j)}:=\,&\big[H_k(t_j-359/4),H_k(t_j-358/4),\ldots,H_k(t_j-1/4),H_k(t_j)\big]^\top\label{equ:IHR}\\
&\,{-\texttt{median}\{H_k(t_j-(q-1)/4) |\,q=1,\ldots,360\}\in \mathbb{R}^{360},} \nonumber
\end{align}
where $t_j$ indicates the ending time of the $j$-th epoch.

%\subsection{Illustration of PD and PS during different sleep stages/{IHR time series and their PDs}}
\subsection{{IHR time series and their PDs}}
\label{subsec:ihr time series}
%{\color{teal}YM: which title would be better?}

{
Following the discussion in Section~\ref{sec:applications to time series}, 	we apply {TDA} to IHR time series defined in \eqref{equ:IHR}, $H^{(k,j)}$.  More precisely, we consider $\mathcal{P}_0(H^{(k,j)})$ via the sub-level set filtration, and $\mathcal{P}_i(R_{120,1}(H^{(k,j)}))$ for $i=0,~1$, via the VR complex filtration.  We extract PS from both $\mathcal{P}_0(H^{(k,j)})$ and $\mathcal{P}_i(R_{120,1}(H^{(k,j)}))$, where $i=0,~1$.  We {summarize} Section~\ref{sec:applications to time series} and highlight our approach in the following {pseudocode. See} also Fig.~\ref{Figure : main_scheme} {for an illustration}.
}
	% % % %new algorithm env % % % % % % % % % % %
	\floatname{algorithm}{Algorithm:}
	\renewcommand{\algorithmicrequire}{\textbf{Input:}}
	\renewcommand{\algorithmicensure}{\textbf{Output:}}
	\renewcommand{\thealgorithm}{}
	\begin{algorithm} % enter the algorithm environment
		\setstretch{1.2}
		\caption{Feature Extraction Scheme} % give the algorithm a caption
		\label{alg1} % and a label for \ref{} commands later in the document
		\begin{algorithmic} % enter the algorithmic environment
			\REQUIRE A time series, $H(t)$.
			\ENSURE Topological features used in this article.
			\STATE 1. Calculate $\mathcal{P}_0(H)$ via sublevel set filtration (as in Section~\ref{subsection : sub-level set Filtration}).
			\STATE 2. Calculate $\mathcal{P}_i(R_{120,1}(H))$ via VR complex filtration (as in Section~\ref{subsec : Vietoris-Rips Complexes Filtration}).
			\STATE 3. Calculate PS features:\\ $\Big[ \Phi^{(\texttt{PS})}(\mathcal{P}_0(H)),~\Phi^{(\texttt{PS})}(\mathcal{P}_0(R_{120,1}(H))),~\Phi^{(\texttt{PS})}
		(\mathcal{P}_1(R_{120,1}(H)))\Big]$ (as in Section~\ref{sec:ps}).
		\end{algorithmic}
	\end{algorithm}
	% % % % % % % % % % % % % % % % % % % % % % % % %

\begin{figure}[hbt!]
\centering
	\subfigure[Waking epoch]{\includegraphics[width = 0.45\textwidth]{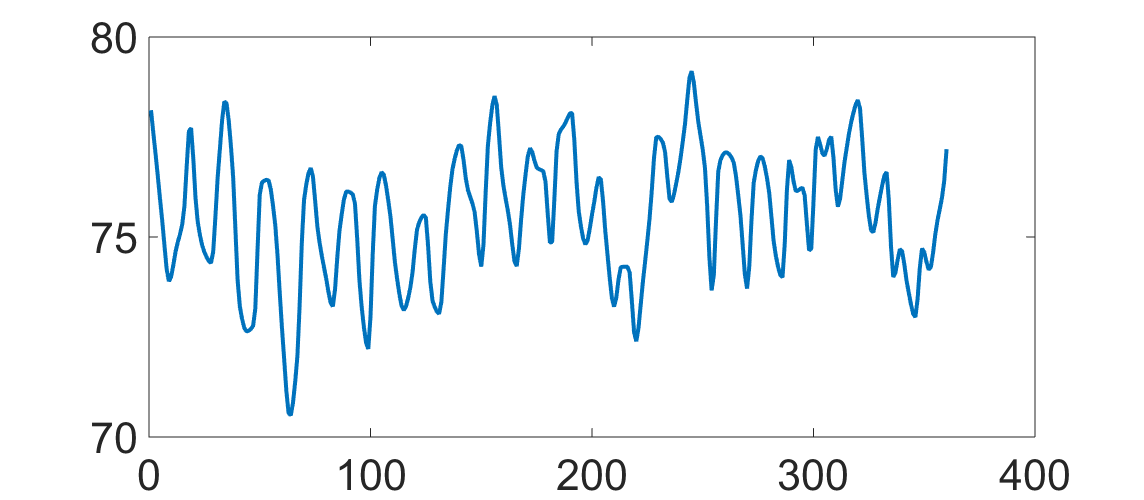}}
	\subfigure[$\mathcal{P}_0(H^{(k,j)})$]{\includegraphics[width = 0.45\textwidth]{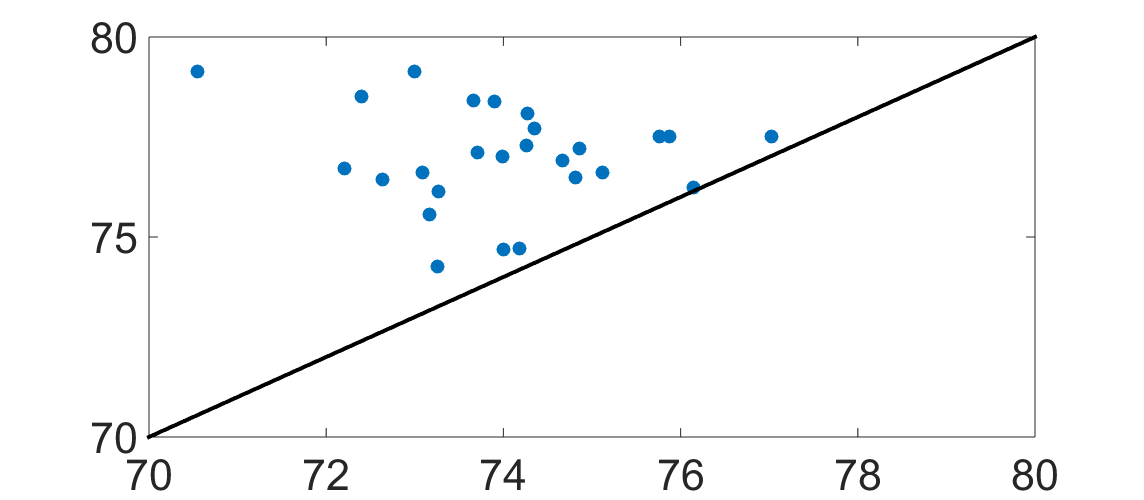}} \\
	\subfigure[$R_{120,1}(H^{k,j})$]{\includegraphics[width = 0.45\textwidth]{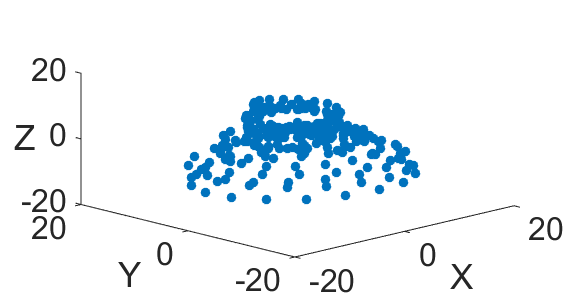}}
	\subfigure[$\mathcal{P}_q(R_{120,1}(H^{k,j}))$]{\includegraphics[width = 0.45\textwidth]{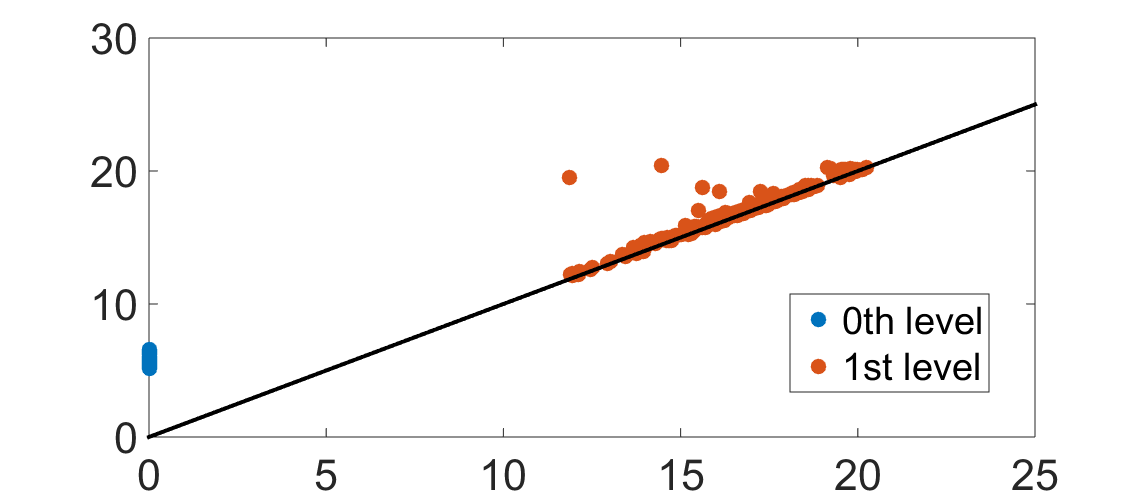}}
	\caption{{ An illustration} of the IHR during the wake stage. (a) the IHR signal $H^{(k,j)}$; (b) the PD of the sub-level set filtration, $\mathcal{P}_0(H^{(k,j)})$; (c) The first three { principal components} of the point cloud $R_{120,1}(H^{k,j})$; (d) the PD's of the VR filtration, $\mathcal{P}_0(R_{120,1}(H^{k,j}))$ and $\mathcal{P}_1(R_{120,1}(H^{k,j}))$ are superimposed, where blue and red points represent $q=0$ and $q=1$ respectively.}
	\label{fig. Waking sample}
\end{figure}

\begin{figure}[hbt!]
\centering
	\subfigure[Sleep epoch]{\includegraphics[width = 0.45\textwidth]{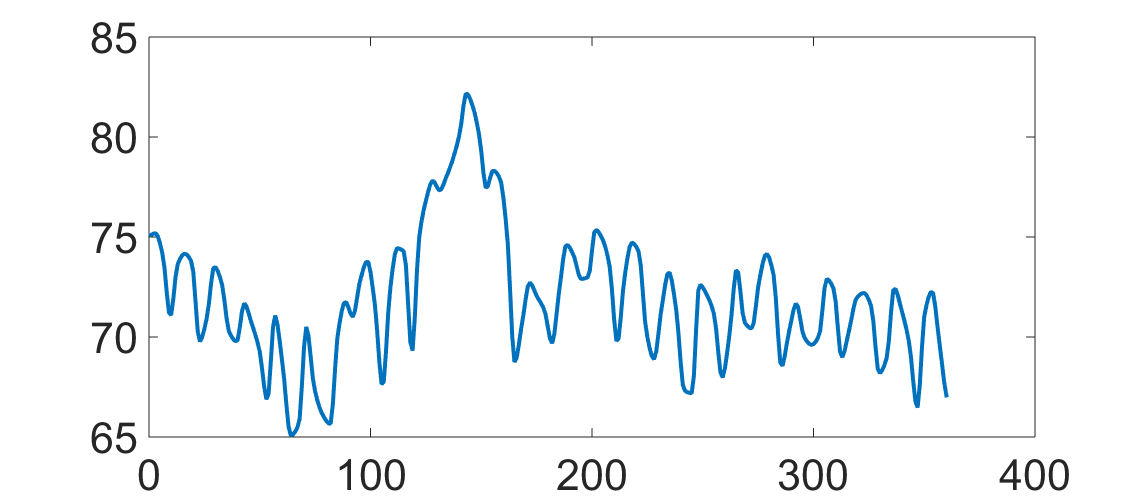}}
	\subfigure[$\mathcal{P}_0(H^{(k,j)})$]{\includegraphics[width = 0.45\textwidth]{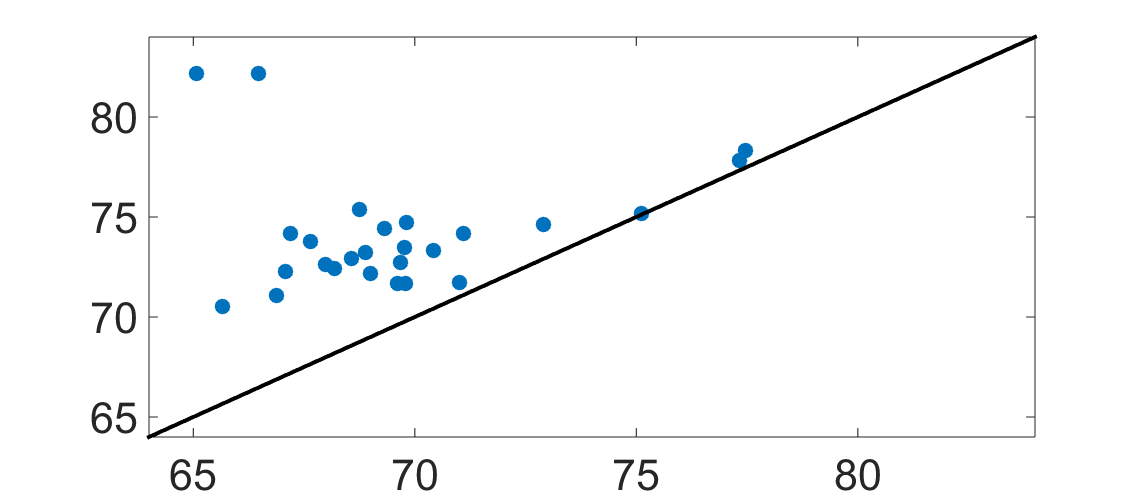}} \\		
	\subfigure[$R_{120,1}(H^{k,j})$.]{\includegraphics[width = 0.45\textwidth]{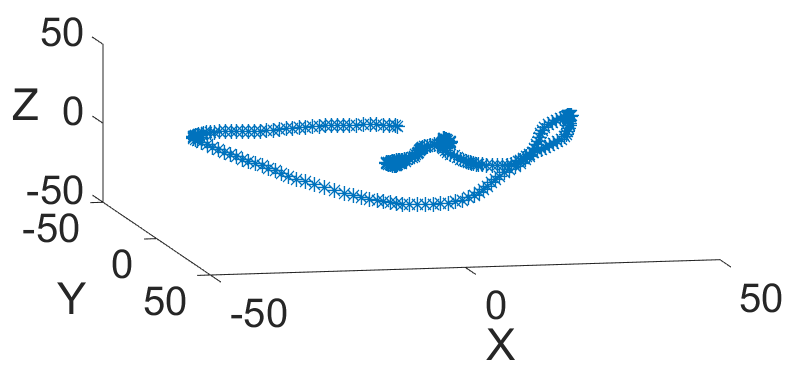}}
	\subfigure[$\mathcal{P}_q(R_{120,1}(H^{k,j}))$]{\includegraphics[width = 0.45\textwidth]{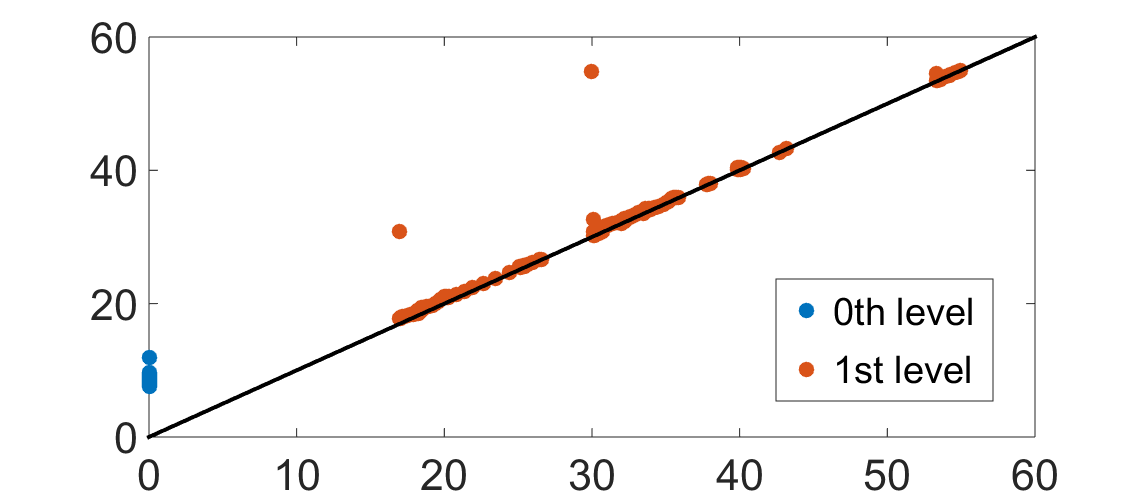}}
	\caption{{ An illustration} of the IHR during the sleep stage. (a) the IHR signal $H^{(k,j)}$; (b) The PD of the sub-level set filtration, $\mathcal{P}_0(H^{(k,j)})$; (c) The first three { principal components} of the point cloud $R_{120,1}(H^{k,j})$; (d) the PD's of the VR filtration $\mathcal{P}_0(R_{120,1}(H^{k,j}))$ and $\mathcal{P}_1(R_{120,1}(H^{k,j}))$ are superimposed, where blue and red points represent $q=0$ and $q=1$ respectively.}
	\label{fig. Sleeping sample}
\end{figure}

%Let $\{ H^{(k,j)}_{h_i} \}_i$ be the sub-level set filtration of $H^{(k,j)}$ associated with the $j$-th epoch of the $k$-th subject, where $h_1<h_2<\ldots$.   
%For simplicity, we denote its PD by $\mathcal{P}_0(H^{(k,j)})$. 
{  
We illustrate the IHR time series and their PD's with different filtrations {in Fig.~\ref{fig. Waking sample} and Fig.~\ref{fig. Sleeping sample}. From a IHR time series during a wake (resp. sleep) epoch shown in Fig.~\ref{fig. Waking sample}(a) (resp. Fig.~\ref{fig. Sleeping sample}),}} we observe that these IHR's seem to be different: {wake epoch seems to have more variability than sleep one does.  Sub-level set filtration captures such variability in the form of the PD.  As shown in Fig.~\ref{fig. Waking sample}(b) and Fig.~\ref{fig. Sleeping sample}(b), their PD's of sub-level set filtration are different.  Points in Fig.~\ref{fig. Waking sample}(b) spread widely while most points in Fig.~\ref{fig. Sleeping sample}(b) are clustered around lower left portion of the diagram. Moreover, Fig.~\ref{fig. Waking sample}(b) seems to have more long-lived points than Fig.~\ref{fig. Sleeping sample} (b) does.}
Next, we examine the {PD's of VR complex filtration}.  In this work, we take $(p, \tau)=(120, 1)$, where $p=120$ is equivalent to a 30 seconds long time series. Therefore,  $H^{(k,j)}$ {denotes} its lag map by $R_{120,1}(H^{(k,j)}) \subseteq \mathbb{R}^{120}$.  
Fig.~\ref{fig. Waking sample}(c) and Fig.~\ref{fig. Sleeping sample}(c) show examples of $R_{120,1}(H^{(k,j)})$ projected onto their first three { principal components}. Visually, the point clouds of Fig.~\ref{fig. Waking sample}(c) and Fig.~\ref{fig. Sleeping sample}(c) have different shapes (former seems to have a ``lamp'' shape while the latter does not), and their PDs shown in Fig.~\ref{fig. Waking sample}(d) and Fig.~\ref{fig. Sleeping sample}(d) are also different.   For instance, the red points {in} Fig.~\ref{fig. Waking sample}(d) cluster around birth values $10\sim20$, the red points in Fig.~\ref{fig. Sleeping sample}(d) have three clusters around the birth values $15\sim25$, $30\sim 40$, and $55$.  {It is important to note that the computations on $\mathcal{P}_0(R_{120,1}(H^{k,j}))$ {are} done on the $\mathbb{R}^{120}$ space, and {projection onto their first three} principal components {is} merely for the visualization purpose.}

\begin{figure}[hbt!]
	\centering
	\subfigure[Boxplot.]{
	\includegraphics[width = 0.48\textwidth]{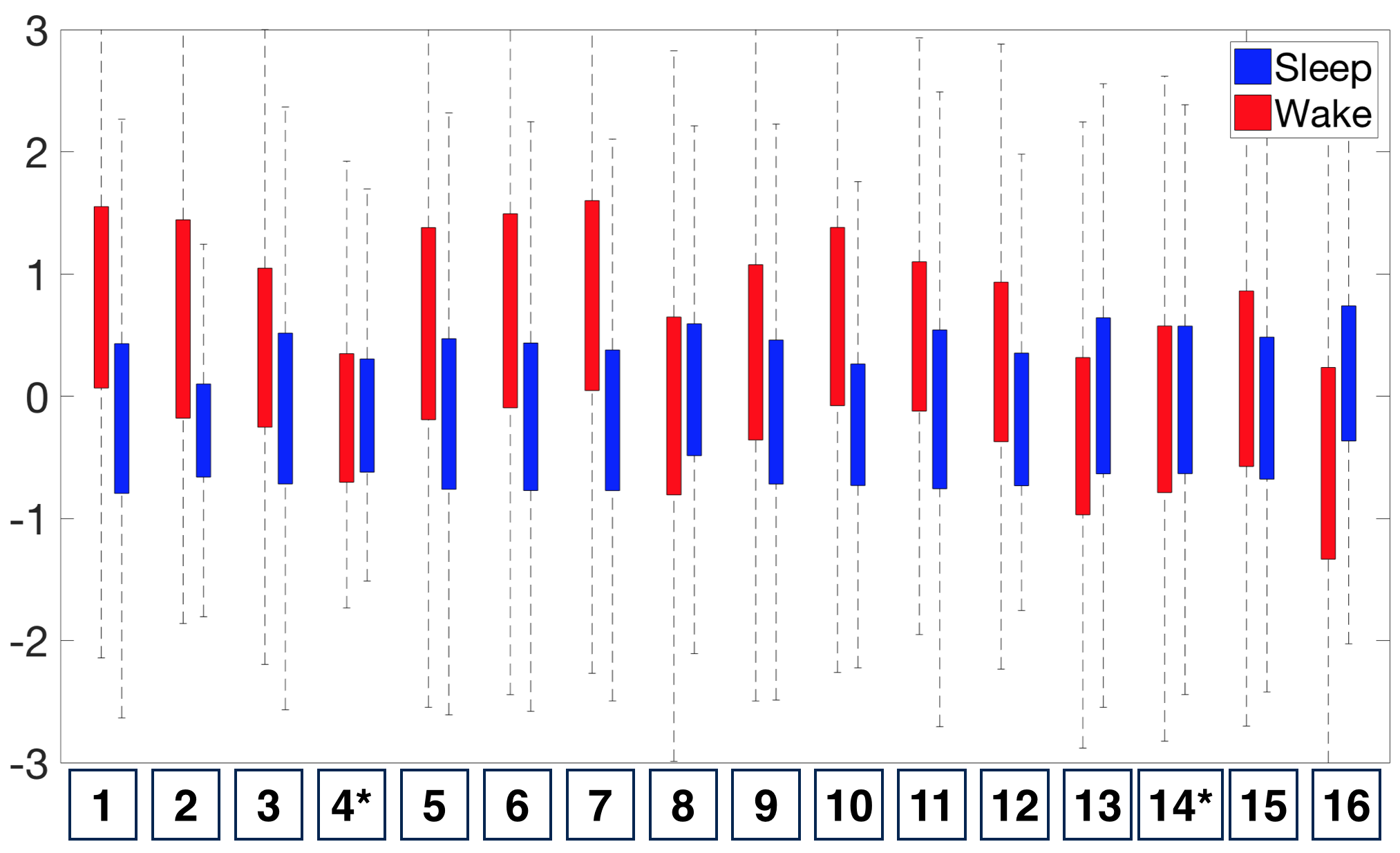}}
	\subfigure[Scatter plot.]{
	\includegraphics[width = 0.48\textwidth]{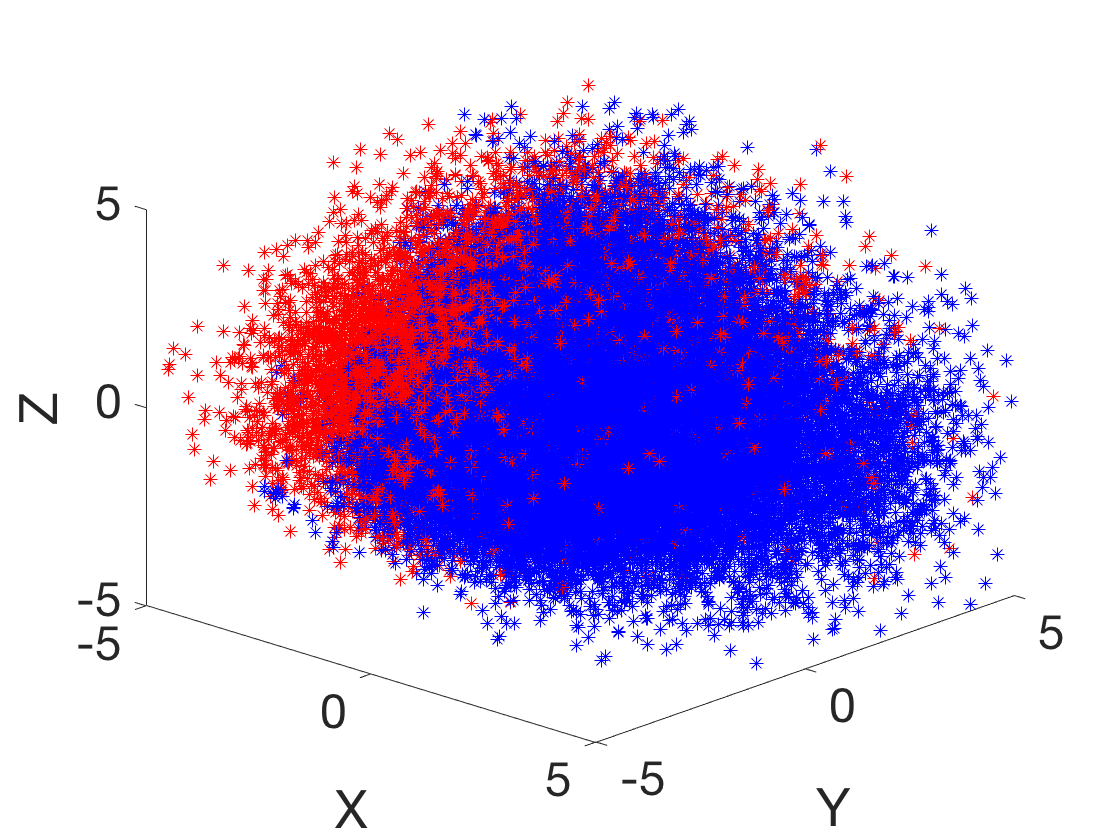}}
	\caption{Distribution of normalized PS features, $\Phi^{(\texttt{PS})}(\mathcal{P}_0(H^{(k,j)}))$.  (a) Boxplot of the $\Phi^{(\texttt{PS})}(\mathcal{P}_0(H^{(k,j)})).$  The numbers listed on the horizontal axis indicates the number of PS. * indicates that the feature {\em fails} to reject the null hypothesis ({that two samples have equal medians}) of the significance level of $0.05$ on the rank sum test with Bonferroni correction. (b) Visualization of $\cup_k\{\Phi^{(\texttt{PS})}(\mathcal{P}_0(H^{(k,j)}))\}_{j=1}^{n_k}$ by the first three { principal components}.  }
	\label{fig:H90CGMH features}
\end{figure}

{As discussed in Section \ref{Subsection persistence diagram}, while it is possible to analyze the data via PD's, it is usually computationally challenging. The proposed PS} allows us to {further summarize the PD's and} quantify the above observations.  To examine the PS features, take $\cup_k\{\Phi^{(\texttt{PS})}(\mathcal{P}_0(H^{(k,j)}))\}_{j=1}^{n_k}$ as an example. In order to compare them on the same scale, we perform the standard $z$-score normalization for {each subject}. %each parameter. % {Specifically, $z$-normalization was performed on each subject's features.} 
We abuse the notation and use $\Phi^{(\texttt{PS})}(\mathcal{P}_0(H^{(k,j)}))$ to denote the normalized parameters.
In Fig.~\ref{fig:H90CGMH features}(a), we show the boxplot of each normalized PS parameter, where blue (red) bars represent the PS associated with an IHR time series associated with the sleep (wake) stage.   
We performed a rank sum test {with the null hypothesis that two samples have equal medians} with a significance level of $0.05$ with the Bonferroni correction. We found that there are significant differences between waking and sleeping features for all PS parameters, {except for the kurtosis of $M$ (labeled as 4 in Fig.~\ref{fig:H90CGMH features}(a)), and the median of $L$ (labeled as 14 in Fig.~\ref{fig:H90CGMH features}(a)).} % except the 75-th percentile of $L$. 
{The boxplot as shown in Fig.~\ref{fig:H90CGMH features}(a) shows that the mean and standard deviation of $M$ are the most distinguishable PS parameters between sleep and wake epochs.}
%The boxplot shows that the standard deviation of $M$ and the skewness of $L$ are the most distinguishable PS parameters between sleep and wake epochs. 
%
To further visualize these features, we apply the principle component analysis (PCA) to $\cup_k\{\Phi^{(\texttt{PS})}(\mathcal{P}_0(H^{(k,j)}))\}_{j=1}^{n_k}$, and plot the first three { principal components} in $\mathbb{R}^3$ as shown Fig.~\ref{fig:H90CGMH features}(b).  We observe a separation between sleep and wake features. The visualization of $\Phi^{(\texttt{PS})}(\mathcal{P}_i(R_{120,1}(H^{k,j})))$, where $i=0,1$, is shown in Fig. \ref{fig:R120CGMH features}.

Motivated by the above observation and discussion, we consider the following features for $H^{(k,j)}$ to distinguish sleep and wake epochs: 
\begin{equation}
\label{equ:main features}
[\Phi^{(\texttt{PS})}(\mathcal{P}_0(H^{(k,j)})),~\Phi^{(\texttt{PS})}(\mathcal{P}_0(R_{120,1}(H^{k,j}))),~\Phi^{(\texttt{PS})}(\mathcal{P}_1(R_{120,1}(H^{k,j})))].
\end{equation}

\subsection{Automatic sleep stage annotation system}
\label{subsec:auto sleep stage}

\subsubsection{Support Vector Machine as the learning model}
We consider the widely applied classifier with { a solid theoretical foundation}, the support vector machine (SVM), to establish the heartbeat classification model. This is Step 4 (machine learning step) shown in Fig.~\ref{Figure : main_scheme}.
The linear function kernel is considered in this work and we use the Matlab built-in function {\tt fitcsvm} with default parameters. 
The input data are features shown in \eqref{equ:main features}, which are calculated by the publicly available libraries DIPHA (https://github.com/DIPHA/dipha) and Ripser (https://github.com/Ripser/ripser).
When there are more than 2 classes, we apply the Error-Correcting Output Codes (ECOC) \cite{dietterich1994solving} with one-versus-one design.  Specifically, we use the Matlab built-in function {\tt fitcecoc}  with default parameters. The Matlab version is 2014b.  

\subsubsection{Statistics}
\label{subsubsec:stat}
We carry out the {\em cross-database validation}. Specifically, we train the SVM model on one database, and evaluate the performance on the other databases. 
One of the main challenges in this automatic annotation problem is that the datasets are usually imbalanced; for example, the number of wake epochs is usually much smaller than that of sleep epochs (e.g., in the CGMH-training, the total number of wake epochs is $9,150$, while the total number of sleep epochs is $54,547$).  
In order to account for the imbalanced dataset, we adopt a {down-sampling} process. Let $E_s$ and $E_w$ be the collection of all sleep and wake epochs respectively across all subjects in the training set, and denote their cardinality by $|E_s|$ and $|E_w|$ respectively. We take all epochs in $E_w$, and randomly select $|E_w|$ epochs from $E_s$. The SVM model will then be built on these balanced epochs.
Once the model is built on the training dataset, we test it on the entire testing dataset. %To account for this subsampling process, we repeat the training process $20$ times. 

We report the following performance measurement indices.
When there are $m$ labels, denote $M\in \mathbb{R}^{m\times m}$ to be the confusion matrix of the automatic classification model, where $M_{kl}$      represents the count of epochs whose known group labels are $k$ and whose predicted group labels are $l$. The sensitivity (SE), positive predictivity (+P) and F1 for the $k$-th class, the Cohen kappa, and the overall accuracy (Acc) are defined as
\begin{align}
  &SE_{k}=\frac{M_{kk}}{\sum_{l=1}^{m}M_{kl}},\,\, {  +P_{k}=\frac{M_{kk}}{\sum_{l=1}^{m}M_{lk}}},\,\,{F1}_{k}=\frac{2 (+P_{k})\cdot SE_{k}}{(+P_{k})+SE_{k}}
  \end{align}
  \begin{align}
  &Acc=\frac{\sum_{k=1}^{m}M_{kk}}{\sum_{k=1}^{m}\sum_{l=1}^{m}M_{kl}},\,\,Kappa= \frac{Acc-{EA}}{1-{EA}}
\end{align}
respectively, where EA means the expected accuracy and is defined by
\begin{equation}
{EA}= \frac{\sum_{p=1}^m\left(   \sum_{q=1}^m M_{pq}\right)\times \left(\sum_{q=1}^mM_{qp}\right)}{\left(\sum_{p,q=1}^m M_{pq}\right)^{2}}.
\end{equation}
When $m=3$, $k=1$ means wake, $k=2$ means REM, and $k=3$ means NREM. When classifying wake and sleep stages, $k=1$ means wake, and $k=2$ means sleep; when classifying REM and NREM stages, $k=1$ means REM, and $k=2$ means NREM. When { $m=2$}, $SE_1$ { is reduced to the usual sensitivity} (SE), $SE_2$ is reduced to the usual specificity (SP), and $+P_1$ is reduced to the precision (PR). 
For each database and each performance measurement, we report the mean $\pm$ standard deviation of all subjects.

All experiments in this and next sections { were done using Windows 7 operating system environment} equipped with i5-4570 CPU and 32 GB RAM. 
Under this computational environment, given a random seed, the whole training process of an SVM model takes $5 \sim 7$ minutes { on average}. {For the reproducibility purpose, the Matlab code is available in the GitHub repository website\footnote{\url{https://github.com/peterbillhu/TDA_for_SleepWake_Classifications}}.}

\subsubsection{Automatic sleep stage classification result}
We performed three classification tasks---sleep v.s. wake, REM v.s. NREM, and finally wake v.s. REM v.s. NREM. The random seed is fixed to $1$ in all cases when we ran the subsampling scheme.  { The {results are shown in Tables}~\ref{tab: main results-subject SD 1 seed},~\ref{tab:main results for REM classification (SVM)-subject SD 1 seed}, and~\ref{tab:main results for 3 classes classification (SVM)-subject SD 1 seed}, where {the} SVM model was trained on the CGMH-training dataset and tested on CGMH-validation, DREAMS, and UCDSADB{, respectively}.  For the interested readers, we also include extensive experimental results with different settings in \ref{sec:more automatic annotation results}, such as results of training on different datasets, %higher dimension features $\mathcal{P}_2$, 
and different random seeds.  All {results are similar} to those {reported} in the main article.}

\begin{table}
\caption{\footnotesize SVM cross-database performance of subjects for Wake and Sleep classification with a single random seed. %{ Each feature vector used here is normalized via standard $z$-score among features of the corresponding subject}. 
The training database is CGMH-training. For each database and each performance measurement, we report the mean $\pm$ standard deviation of all subjects.\label{tab: main results-subject SD 1 seed}}
		\fbox{%
		\footnotesize 
		\begin{tabular}{c|cccc}
			& CGMH-training  & CGMH-validation & DREAMS & UCDSADB \\ \hline\hline
			TP & 76 $\pm$ 43 & 76 $\pm$ 44 & 101 $\pm$ 55 & 85 $\pm$ 44 \\
			FP & 151 $\pm$ 49 & 126 $\pm$ 48 & 175 $\pm$ 61 & 149 $\pm$ 59 \\
			TN & 462 $\pm$ 68 & 449 $\pm$ 102 & 592 $\pm$ 112 & 448 $\pm$ 110 \\
			FN & 27 $\pm$ 33 & 42 $\pm$ 43 & 56 $\pm$ 45 & 73 $\pm$ 52 \\
			\hline
			SE $(\%)$ & 78.3 $\pm$ 14.7 & 70.9 $\pm$ 16.0 & 66.9 $\pm$ 16.1 & 57.6 $\pm$ 15.5 \\ 
			SP $(\%)$ & 76.0 $\pm$ 6.1 & 78.9 $\pm$ 5.4 & 77.6 $\pm$ 5.8 & 75.3 $\pm$ 5.5 \\
			Acc $(\%)$ & 75.2 $\pm$ 5.4 & 75.8 $\pm$ 4.4 & 74.7 $\pm$ 5.0 & 70.6 $\pm$ 5.4\\
			\hline 
			PR $(\%)$ & 34.0 $\pm$ 17.0 & 38.1 $\pm$ 19.6 & 37.0 $\pm$ 18.8 & 35.6 $\pm$ 17.3 \\
			F1  & 0.438 $\pm$ 0.161 & 0.452 $\pm$ 0.140 & 0.445 $\pm$ 0.146 & 0.407 $\pm$ 0.140 \\
			AUC  & 0.839 $\pm$ 0.084 & 0.824 $\pm$ 0.090 & 0.789 $\pm$ 0.090  & 0.702 $\pm$ 0.094 \\
			Kappa & 0.320 $\pm$ 0.146 & 0.322 $\pm$ 0.123 & 0.308 $\pm$ 0.148 & 0.238 $\pm$ 0.133 \\ \hline
		\end{tabular}}
\end{table}

Table \ref{tab: main results-subject SD 1 seed} lists the result of classifying wake and sleep stages with { different testing sets}. For each testing database, we show the mean$\pm$standard deviation of each prediction outcome measurement of all subjects in that database. Table~\ref{tab: main results-subject SD 1 seed} shows the performances of training the model on CGMH-training and testing it on CGMH-validation, DREAMS, and UCDSADB.  When considering the CGMH database, the (SE, SP) pair for CGMH-training and CGMH-validation are { $(78.3\pm 14.7\%, 76.0\pm6.1\%)$ and $(70.9\pm 16.0\%, 78.9\pm5.4\%)$} respectively.
When testing on DREAMS, the (SE, SP) pair becomes {$(66.9\pm 16.1\%, 77.6\pm 5.8\%)$}. SP remains in the range of $70\%$, although SE falls below $70\%$.  {This result of the cross database testing is similar to that of validation result.}
When tested on UCDSADB, the pair of (SE, SP) becomes {$(57.6\pm 15.5\%, 75.3\pm 5.5\%)$}.  The overall performance on UCDSADB drops as expected since it contains sleep apnea subjects, and their sleep dynamics is disturbed by the sleep apnea. %Note that the standard deviation of SE's and SP's on UCDSADB seems to be larger than other testing databases, but a two sample F-test cannot reject the null hypothesis on the statistical significance level $0.05$.
Overall, the cross-database validation results suggest that our model does not overfit. Moreover, we found that the down-sampling  scheme alleviates the imbalance database issue. 
%

%%%

Table~\ref{tab:main results for REM classification (SVM)-subject SD 1 seed} shows the performance for the REM and NREM classification. In this task, since the number of NREM epochs is much more than that of REM epochs, we apply the same down-sampling process to NREM as discussed in Section~\ref{subsubsec:stat}.  
Table~\ref{tab:main results for REM classification (SVM)-subject SD 1 seed} and Table~\ref{tab: main results-subject SD 1 seed} have several similarities. 

%The SP can reach above $70\%$ and the pair of (SE, SP) is overall balanced. 
 
\begin{table}
	\caption{SVM cross-database performance for REM and NREM classification with a single random seed in the training procedure. %{ Each feature vector used here is normalized via standard $z$-score among features of the corresponding subject}. 
	The training database is CGMH-training. The subject $\# 24$ in CGMH-validation and subject $\# 9$ in UCDSADB were dropped because they do not have REM epochs. For each database and each performance measurement, we report the mean $\pm$ standard deviation of all subjects.
	\label{tab:main results for REM classification (SVM)-subject SD 1 seed}}
\centering
\fbox{\footnotesize 
		\begin{tabular}{c|cccc}
			& CGMH-training  & CGMH-validation & DREAMS & UCDSADB \\ \hline\hline
			TP & 75 $\pm$ 31 & 68 $\pm$ 30 & 93 $\pm$ 34 & 64 $\pm$ 34 \\
			FP & 113 $\pm$ 32 & 106 $\pm$ 43 & 138 $\pm$ 51 & 133 $\pm$ 35 \\
			TN & 400 $\pm$ 68 & 391 $\pm$ 94 & 490 $\pm$ 97 & 373 $\pm$ 68 \\
			FN & 25 $\pm$ 23 & 20 $\pm$ 17 & 46 $\pm$ 27 & 50 $\pm$ 37 \\
			\hline
			SE $(\%)$ & 76.3 $\pm$ 16.0 & 78.1 $\pm$ 17.4 & 67.5 $\pm$ 16.7 & 58.0 $\pm$ 18.0 \\ 
			SP $(\%)$ & 78.0 $\pm$ 4.8 & 79.6 $\pm$ 6.5 & 78.4 $\pm$ 5.2 & 73.7 $\pm$ 5.8 \\
			Acc $(\%)$ & 77.4 $\pm$ 5.6 & 77.8 $\pm$ 8.3 & 76.3 $\pm$ 6.4 & 70.4 $\pm$ 6.4\\
			\hline 
			PR $(\%)$ & 39.4 $\pm$ 13.6 & 41.4 $\pm$ 19.0 & 41.0 $\pm$ 15.4 & 31.9 $\pm$ 16.0 \\
			F1  & 0.505 $\pm$ 0.138 & 0.510 $\pm$ 0.160 & 0.503 $\pm$ 0.144 & 0.390 $\pm$ 0.156 \\
			AUC  & 0.842 $\pm$ 0.094 & 0.849 $\pm$ 0.108 & 0.796 $\pm$ 0.115 & 0.711 $\pm$ 0.120 \\
			Kappa & 0.382 $\pm$ 0.150 & 0.393 $\pm$ 0.175 & 0.312 $\pm$ 0.175 & 0.227 $\pm$ 0.162 \\ \hline
		\end{tabular}
		
	}
\end{table}

Finally, Table \ref{tab:main results for 3 classes classification (SVM)-subject SD 1 seed} shows the performance for the wake, REM, and NREM classification.  In this experiment, since the number of NREM is much more than those of wake and REM, the down-sampling scheme is applied to NREM.
The Acc's in all cases are about $60\%$, except the UCDSADB. The SE's of wake, REM, and NREM are balanced and consistent across databases, except UCDSADB. 
Again, this result might be due to the fact that UCDSADB contains subjects with sleep apnea. On the other hand, note that the +P of NREM is higher than other classes, which is expected due to the dependence of +P on the database prevalence. In the Online Supplementary, we provide more cross-database validation results.

\section{Discussion and Conclusion}
\label{sec:discussion and conclusion}

In this work, the TDA tools are considered to analyze time series. Specifically, we propose a set of novel PS features to {quantify HRV by analyzing IHR time series by TDA tools}. The proposed HRV features are applied to predict sleep stages, ranging from wake, REM and NREM. In addition to being computationally efficient, the algorithm is theoretically sound supported by mathematical and statistical results. 
Note that while we focus on the HRV analysis for the sleep stage annotation, the proposed algorithm has a potential to be applied to analyze other time series and study the HRV for other clinical problems.
 
\begin{table}
	\caption{SVM cross-database performance for Wake, REM and NREM classification with a single random seed in the training procedure. %{ Each feature vector used here is normalized via standard $z$-normalization among features of corresponding subject}. 
	The training database is CGMH-training. The subject $\# 24$ in CGMH-validation and the subject $\# 9$ in UCDSADB were dropped because they do not have REM epochs. For each database and each performance measurement, we report the mean $\pm$ standard deviation of all subjects.
	\label{tab:main results for 3 classes classification (SVM)-subject SD 1 seed}}
	\centering
	\fbox{	
	\small 	
	\begin{tabular}{c|cccc}
			& CGMH-training  & CGMH-validation & DREAMS & UCDSADB \\ \hline\hline
			SE $(\%)$ (Wake) &  63.7 $\pm$ 15.3 & 61.1 $\pm$ 19.0 & 56.2 $\pm$ 14.5 & 39.5 $\pm$ 11.9 \\					
			SE $(\%)$ (REM) & 62.8 $\pm$ 17.4 & 67.1 $\pm$ 20.9 & 57.0 $\pm$ 18.0 & 48.4 $\pm$ 19.6 \\
			SE $(\%)$ (NREM) & 71.9 $\pm$ 6.7 & 72.6 $\pm$ 6.4 & 72.5 $\pm$ 6.8 & 66.0 $\pm$ 7.6 \\
			+P $(\%)$ (Wake) & 40.6 $\pm$ 19.1 & 43.7 $\pm$ 16.6 & 44.1 $\pm$ 18.3 & 39.3 $\pm$ 17.9 \\
			+P $(\%)$ (REM) & 39.1 $\pm$ 14.9 & 40.0 $\pm$ 17.7 & 39.4 $\pm$ 15.5 & 28.5 $\pm$ 15.8 \\
			+P $(\%)$ (NREM) & 89.3 $\pm$ 8.8 & 85.6 $\pm$ 16.0 & 83.7 $\pm$ 8.6 & 76.6 $\pm$ 9.2 \\
			\hline
			Acc $(\%)$ & 68.3 $\pm$ 6.4 & 67.6 $\pm$ 9.2 & 66.3 $\pm$ 6.4 & 57.1 $\pm$ 7.1 \\
			Kappa & 0.401 $\pm$ 10.2 & 0.390 $\pm$ 0.117 & 0.372 $\pm$ 0.116 & 0.244 $\pm$ 0.108 \\
			\hline
		\end{tabular}		
	}
\end{table}

\subsection{Theoretical Supports and Open Problems}

We find that empirically,  $M$ and $L$ are simple yet effective representations of the PD and reveal signatures about the underlying object. It would be interesting to investigate, in theory, the probability distribution of $M$ and $L$ for a given simplicial complex. However, to the best of our knowledge, while there have been several works in this direction, it is still a relatively open problem. 
Recently, { there has been some theoretical work} towards this direction, namely the theory of random complexes (see e.g. survey papers \cite{bobrowski2018topology,kahle2014topology} and references therein).  In order to understand the role of noises in the PD, there have been studies on the topology of the noise. 
In the theory of random fields, authors in \cite{mischaikow2010topology} used sub-level sets as filtration to study the number of components ($\beta_0$) with various random processes; 
in \cite{adler2010persistent}, authors studied the relation between random fields and the PH in general.  
In particular, as mentioned in \cite{adler2010persistent}: ``It would be interesting to know more about the real distributions lying behind the PD, but at this point we know very little.''  { There is also a result in} random cubical complexes \cite{hiraoka2018limit}, and a few work on the limiting theorem of total sum persistence \cite{owada2016limit} {and persistence diagrams \cite{hiraoka2018limit}}.  
It would also be interesting to study the stability of each PS.  As of now, only the sum of $L$, the max of $L$, and the entropy of $L$ have been shown to be stable \cite{cohen2010lipschitz,persistentEntropyStability}.  However, the rest of PS is still unknown. 
Another interesting direction, instead of focusing on each PS, is to study the probability distributions of $M$ and $L$.  For instance, let $\rho_M$ and $\rho_L$ be the empirical probability density function of the sets $M$ and $L$, respectively.  For $S=M$ or $S=L$, could one establish $\| \rho_S(f) - \rho_S(g) \|_{D} \leq d_B ( \mathcal{P}_q(f), \mathcal{P}_q(g) )$,
where $\|\cdot\|_D$ is some suitable statistical distance?
We leave those interesting theoretical problems to future work.

\subsection{Comparison with existing automatic sleep stage annotation results}
%{In this subsection, we compare our results with state-of-art ones.  Our results outperform the state-of-arts methods.  However, it is important to note that these comparisons are not fair/comparable.  The main reasons are all experimental settings are different .  The only common thing is that they all use IHR signals.}
There have been several results in automatic sleep stage annotation by taking {\em solely} the HRV into account. A common conclusion is that classifying sleep-wake by quantifying HRV is a challenging job. 
In general, due to the heterogeneity of the data sets, various evaluation criteria and different features and models used in these publications, it is difficult to have a direct comparison. 
But to be fair, below we summarize some related existing literatures for a discussion. 
To the best of our knowledge, except \cite{malik2018sleep}, there is no result reporting a cross-database validation. 
For those running validation on a single database, we shall distinguish two common cross validation (CV) schemes -- {\em leave-one-subject-out} CV (LOSOCV) and {\em non-LOSOCV}. When the validation set and the training set come from different subjects, we call it the LOSOCV scheme; otherwise we call it the non-LOSOCV scheme. The LOSOCV scheme is in general challenging due to the uncontrollable inter-individual variability, while the non-LOSOCV scheme tends to over-estimate. %
Therefore, for a fair comparison, below we only summarize papers considering {\em only} the IHR features and carrying out the LOSOCV scheme. 

%{{{In this study, we find that in addition to metrics discussed in Section~\ref{subsubsec:stat}, Youden's index, which is SE+SP-1, would also be a good indicator for measuring the performance.  What we observed empirically was that when tuning different experimental settings, SE might be increased, but at the same time SP will be decreased or vice versa.  Thus, in order to diagnose performances of models, we seek those with SE$+$SP as high as possible, and SE$-$SP is as close to 0 as possible.}}}
%In order to improve the overall performance, both SE and SP need to be improved simultaneously. Also, in this study, since we deal with the imbalanced datasets, SE would be more important than SP
	
%	the more balanced SE and SP indicates the better performance.}

%
In \cite{Xiao2013}, the database was composed of healthy participants aged $16-61$ years. A random forest model was established to differentiate between the wake, REM, and NREM stages for those epochs labeled as ``stationary''. Based on the confusion matrix provided in \cite{Xiao2013}, the SE, SP, Acc, and F1 for detecting the wake stage are $51.2$\%, $90.2$\%, $84.0$\%, and $0.50$. 
The authors also provided the SE of wake, REM, and NREM, which are $53.72\% \pm 20.15\%$, $59.01\% \pm 19.72\%$, and $79.50\% \pm 7.82\%$ respectively. In Table~\ref{tab:main results for 3 classes classification (SVM)-subject SD 1 seed}, our validation on CGMH-validation is {$61.1\pm 19.0\%$, $67.1\pm 20.9\%$, and $72.6\pm 6.4\%$}, respectively.  {  Observe that our SE of wake and REM are better, and SE of NREM is on the similar level.}  %Moreover, if we consider the Youden's index, we find that their score is $53.72+59.01+79.50 = 192.23 $ and our score is $61.1+67.1+72.6=200.8$.
%We see that the SE's of wake and REM are similar, while our SE of NREM is a bit worse than theirs. 
In addition to the balance of all classes due to the sub-sampling scheme in our result, note that we focus on all epochs but not ``stationary epochs'', and the subjects in CGMH-validation are not healthy but simply without sleep apnea.

In \cite{Mendez2010}, the database was composed of 24 participants aged $40-50$ years with $0$ AHI. The authors took the temporal information and the phase and magnitude of the ``sleepy pole'' as features to train a hidden Markov model to differentiate REM and NREM stages. The reported SE, SP, and Acc were $70.2\%$, $85.1\%$, and $79.3\%$, respectively.  Our results outperform theirs.  Our SE, SP and Acc of the REM and NREM classification in CGMH-validation shown in Table~\ref{tab:main results for REM classification (SVM)-subject SD 1 seed} are { $78.1\pm 17.4\%$, $79.6\pm 6.5\%$, and $77.8\pm8.3\%$}, respectively. { Observe that both Accs are similar which means portion of correct predictions are similar.  Not only our SE is better, but SE and SP are also balanced.   
%{{Furthermore, if we consider the Youden's index, their index is $70.2\%+85.1\% - 1 = 0.552$ while ours is $78.1\%+79.6\%-1 = 0.577$. }}
} %While { our algorithm} has a lower Acc, our SE and SP are more balanced with the cross-database validation.

In \cite{Lewicke2008}, the database is composed of 190 infants. A variety of features and classification algorithms were considered and the wake and sleep classes were balanced for the analysis. The SE and SP of their multi-layer perceptron model without rejection was $79.0 \%$ and $77.5 \%$, respectively. In Table~\ref{tab: main results-subject SD 1 seed}, the SE and SP of our result on CGMH-validation is {$70.9\pm 16.0\%$ and $78.9\pm 5.4\%$}. %At the first glance, our SP and SE are worse than theirs. 
{Our performance is comparable to theirs.}
However, { there is a fundamental difference} between their experiments and ours -- the sleep dynamics of infants and adults are different. 

In \cite{Aktaruzzaman2015}, the database is composed of $20$ participants aged $49$-$68$ years with varying degrees of sleep apnea. Detrended fluctuation analysis and a feed-forward neural network were applied to differentiate the wake and sleep stages. Various epoch lengths were considered, and the highest performance was recorded on an epoch length of $5$ minutes. The Acc, SE, SP, and Cohen's kappa were $71.9 \pm 18.2 \%$, $43.7 \pm 27.3 \%$, $89.0 \pm 7.8 \%$, and $0.29 \pm 0.24$, respectively. We consider UCDSADB for a comparison. In Table~\ref{tab: main results-subject SD 1 seed},  the Acc, SE, SP, and Cohen's kappa of our testing result on UCDSADB is {$70.6\pm 5.4\%$, $57.6\pm 15.5\%$, $75.3\pm 5.5\%$, and $0.238\pm 0.133$}.  { Their Acc and ours are on the same level, our SE is better than theirs, while their SP is better than ours. However, our SE and SP are balanced compared with theirs. 
%{{However, if we consider Youden's index, their index is 0.32 and our index is 0.329, which may suggest performances are on the similar level.  }}
A major difference is that our standard deviations for Acc, SE, SP are much smaller. Thus, our performance is comparable to theirs.} %We observe that our result is balanced and hence a lower Acc and Cohen's kappa. 

In \cite{Long2012}, fifteen participants aged $31.0 \pm 10.4$ years with the Pittsburgh Sleep Quality Index less than $6$ were considered. 
The linear discriminant-based classifier was trained with spectral HRV features. The SE, SP, Cohen's kappa and AUC were $49.7 \pm 19.2 \%$, $96 \pm 3.3\%$, $0.48 \pm 0.24$ and $0.54$ respectively.
As shown in Table~\ref{tab: main results-subject SD 1 seed},  the SE, SP, Cohen's kappa and AUC of our result on CGMH-validation is {$70.9\pm 16.0\%$, $78.9\pm 5.4\%$, $0.322\pm0.123\%$, and $0.824\pm 0.090$}, respectively.   
%{{If we consider the Youden's index, we find out their index is 0.457, and ours is 0.498.  Moreover, our AUC is much better than their.}} 
{Again, compared with their results, our SE and SE are more balanced.}
%We observe that our AUC outperforms theirs, and the overall performance is again balanced, while it leads to a lower Cohen's kappa.  

{To make a conclusion, we emphasize that all those results under comparison are not carried out in the cross-database scheme. Also, usually the SE and SP are not balanced with high SP, which leads to the high accuracy. Therefore, the results suggest that the proposed PS features and chosen learning model lead a better, or at least similar, performance compared with the state-of-the-art results. The cross-database validation further suggests the usability of the PS features and the proposed learning scheme in clinical setups. Last but not the least, due to the numerical efficiency of the proposed PS features, it is potential to apply it to analyze large scale time series.}
%To sum up, while we cannot conclude that our selected features and chosen learning model outperform the state-of-the-art results, we have shown the potential of this combination. . 

{
\subsection{Technical issues}
\label{subsec:technical issues}
We remark that although it is possible to {include} $\mathcal{P}_i(R_{120,1}(H^{k,j}))$ for $i\geq 2$ {in \eqref{equ:main features}}, in practice, it is a challenging task due to its computational complexity.  Its computation is known to be poorly scalable in dimension and memory-intensive. We refer readers to \cite{otter2017roadmap} for more details and comparisons among state-of-arts TDA packages and extensive benchmark.  To get an idea of the computational cost, for {any epoch}, the computational time by the state-of-art package {\tt{Ripser}} for $\mathcal{P}_1(R_{120,1}(H^{k,j}))$, $\mathcal{P}_2(R_{120,1}(H^{k,j}))$, and $\mathcal{P}_3(R_{120,1}(H^{k,j}))$ are $0.06$, $1.7$, and $106$ seconds {in a standard laptop}, respectively. {This echos the fact} that the computation of $\mathcal{P}_i(R_{120,1}(H^{k,j}))$ does not scale well in $i$ \cite{otter2017roadmap}.  %In {the automatic sleep stage annotation problem}, there are more than $120,000$ epochs in total to process.  
{Thus,} it would be inefficient to obtain the higher dimensional persistence features.  %{We refer the interested readers to Table~\ref{tab: ADD H2 classification training on DREAMS (Subjects)} and \ref{tab: ADD H2 classification training on UCD (Subjects)} in the Online Supplementary for classification results using the features $\Phi^{\text{PS}}(\mathcal{P}_2(R_{120,1}(H^{k,j})))$.}

%Moreover, those higher dimensional features might not be helpful in this study.  We discovered that sub-level set filtration is more effective for this work than VR complex filtration.  In Table \ref{tab: H Features Performance} and Table \ref{tab: VR Features Performance} in \ref{sec:more automatic annotation results}, we found that the classification result of $\mathcal{P}_i(R_{120,1}(H^{k,j}))$ is worse than that of $\mathcal{P}_0(H^{(k,j)})$, in particular the performance of SEs.  Furthermore, even if we concatenate $\Phi^{PS}(\mathcal{P}_2(R_{120,1}(H^{k,j})))$ to the main features \eqref{equ:main features}, when comparing Table~\ref{tab:SW classification training on DREAMS (Subjects)} and Table~\ref{tab: ADD H2 classification training on DREAMS (Subjects)}, and comparing Table~\ref{tab:SW classification training on UCDSB (Subjects)} and Table~\ref{tab: ADD H2 classification training on UCD (Subjects)} in \ref{sec:more automatic annotation results} the improvement is too marginal to justify the additional computational efforts of $\mathcal{P}_2(R_{120,1}(H^{k,j}))$.  
}

\subsection{Limitations and future directions}

In addition to the theoretical development discussed above, there are several interesting practical problems left untouched.  
While we systematically consider the inter-individual variance, the race, the machine, and the sleep disorder by taking three different databases into account, we acknowledge the fact that the data is collected from the sleep lab. When the data is collected from the real-world mobile device, it is not clear if the algorithm could perform as well and run in real-time. Moreover, its performance for the home-based screening needs to be further evaluated.
Yet, in the current mobile health market, the photoplethysmography (PPG) sensor has been widely applied, and its applicability for the sleep-wake classification has been reported in \cite{malik2018sleep}. It is interesting to see how the TDA approach could be applied to analyze the HRV from the PPG for the sleep stage classification mission.
From the data analysis perspective, it would be interesting to perform a more sophisticated analysis and take other features from the PD. For instance, the PH Transformation (PHT) \cite{turner2014persistent} was recently developed and proven to be a sufficient statistic, and had been successfully applied to the shape analysis. It would be interesting to combine the PHT and PS.
IHR is a well-known non-stationary time series. Based on the encouraging results of applying the TDA, we suspect that the PS features could be applied to study other clinical problems related to HRV, and furthermore, analyze other physiological time series. We will explore those limitations/directions in our future work.

\section*{Acknowledgement}
The { authors acknowledge} the hospitality of National Center for Theoretical Sciences (NCTS), Taipei, Taiwan during summer, 2019, when finishing this manuscript. The authors would like to thank Mr. Dominic Tanzillo for his help of proofreading. { Chuan-Shen Hu want to thank Prof. Jung-Kai Chen (NCTS), Prof. Chun-Chi Lin (NTNU) and Prof. Mao-Pei Tsui (NTU) for the kindly financial support for the work. Chuan-Shen Hu is funded by the project MOST108-2119-M002-031 hosted by 
the Ministry of Science and Technology in Taiwan.}

\bibliographystyle{plain}
\bibliography{sleepWake,BibSleep}

\appendix

\setcounter{equation}{0}
\setcounter{section}{0}
\setcounter{page}{1}
\renewcommand{\theequation}{SI.\arabic{equation}}
\renewcommand{\thesection}{SI.\arabic{section}}
\renewcommand{\thetheorem}{SI.\arabic{theorem}}
\renewcommand{\thepage}{SI.\arabic{page}}
\renewcommand{\thefigure}{SI.\arabic{figure}}

\section{More Mathematical Background}

\subsection{More Simplicial Complexes}\label{section:simplicialcomplexes}

To investigate the complicated structure of an object, an intuitive way is to use simple objects as building blocks to approximate the original object. 
For instance, in computer graphics, curves and surfaces in Euclidean spaces are approximated by line segments and triangles, {\it e.g.} Figure \ref{fig:example of Betti numbers} (D).  
In TDA, the main building blocks are {{\it simplicial complexes}}.   
Simplicial complexes are important tools to approximate continuous objects via combinatorial objects, such as vertices, edges, faces, and so on. 
Although simplicial complexes and simplicial homology can be studied in an abstract and general way (see e.g. \cite{Edelsbrunner2010, Munkres}), to enhance the readability, we present the notion in a relatively concrete way without losing critical information.

\begin{figure} 
	\centering 
	\includegraphics[width = \textwidth]{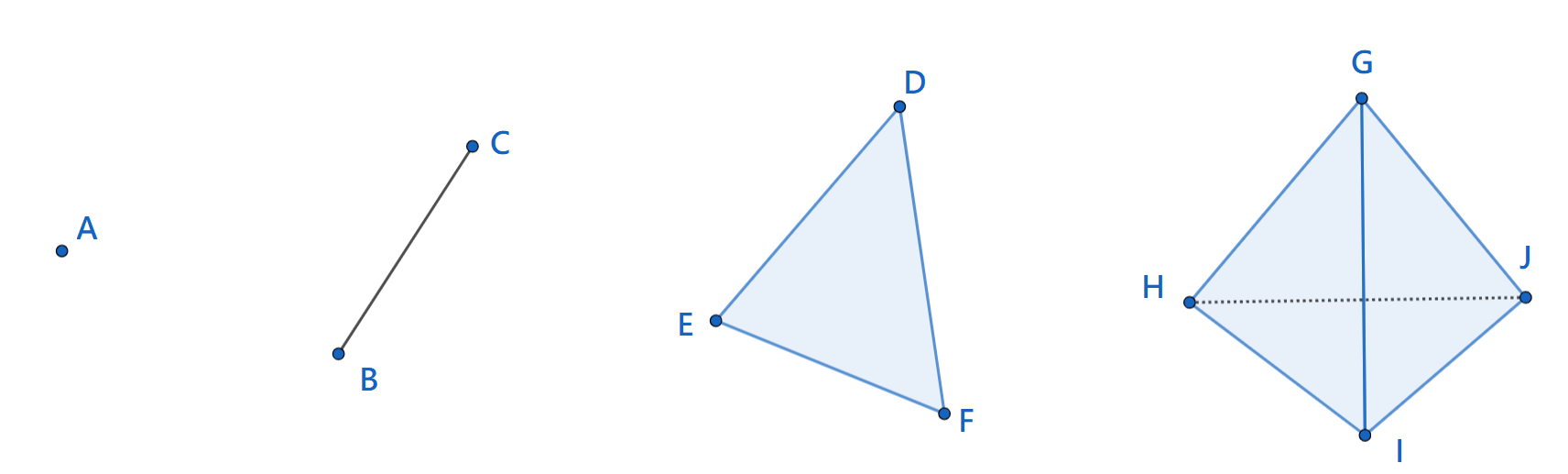}
	\caption{Illustration of simplices in dimension $0,1,2,3$. The singleton point $\{ A \}$ forms a $0$-simplex and convex hulls $\langle B,C \rangle$, $\langle D,E,F \rangle$ and $\langle G,H,I,J \rangle$ are $1$-simplex, $2$-simplex, $3$-simplex in { an Euclidean} space respectively.
	\label{fig:illustration of simplices}}
\end{figure}

We start with introducing simplex.
Intuitively, a simplex is a ``triangle'' in different dimension. As shown in Figure \ref{fig:illustration of simplices}, vertices, line segments, triangles and tetrahedron in $\bbR^d$ are $0,1,2,3$-dimensional triangles respectively, and they are called $0$-simplexes, $1$-simplexes, $2$-simplexes and $3$-simplexes in $\bbR^d$ respectively.  The formal definitions can be found below.

\begin{def.}
[\cite{Munkres} Sec. 1.1, p. 3]
	Let $x_0, x_1, \ldots, x_q$ be affinely independent points in $\mathbb{R}^d$. The $q$-\textbf{simplex}, denoted by $\sigma := \langle x_0, x_1, \ldots, x_q \rangle$, is defined to be the convex hull of $x_0, x_1, \ldots, x_q$. In other words,
	\begin{equation*}
	\sigma = \left \lbrace \lambda_0x_0 + \lambda_1x_1 + \cdots + \lambda_{q}x_{q} \ \left | \ \lambda_0, \lambda_1, \ldots, \lambda_{q} \in [0,1], \sum_{i=0}^q \lambda_i = 1 \right. \right \rbrace.
	\end{equation*}
	We denote ${\rm Vert}(\sigma):=\{x_0, x_1, \ldots, x_q\}$.
\end{def.}
Recall that $x_0, x_1, \ldots, x_q$ are affinely independent points in $\bbR^d$ if and only if $x_1 - x_0, \ldots, x_q - x_0$ are linearly independent vectors in $\bbR^d$.
Any $q$-simplex is a $q$ dimensional object, and it consists of lower degree simplexes. We are interested in the relation among simplexes of different dimensions.
Since any subset $V$ of ${\rm Vert}(\sigma)$ is also affinely independent, the convex hull of $V$ also forms a simplex with dimensional $|V| \leq q$ where $|V|$ denotes the cardinality of $V$.  This lower dimensional simplex is called a {\em face} of $\sigma$.

\begin{def.}
[{ \cite{Munkres} Sec. 1.1, p. 5}]
Let $\sigma$ be a $q$-simplex, and $V\subset{\rm Vert}(\sigma)$. The convex hull of $V$ is called a {\em face} of $\sigma$.  Moreover, if $|V| = k$, then the face $\tau = \langle V \rangle$ is called a $k$-face of $\sigma$.
\end{def.}

For example, in Figure \ref{fig:illustration of simplices}, the $3$-simplex (tetrahedron) $\langle G,H,I,J \rangle$ has the following faces:
\begin{itemize}
\item  $0$-face: $\langle G \rangle$, $\langle H \rangle$, $\langle I \rangle$ and $\langle J \rangle$;
\item $1$-face: $\langle G,H \rangle$, $\langle G,I \rangle$, $\langle G,J \rangle$, $\langle H,I \rangle$, $\langle H,J \rangle$ and $\langle I,J \rangle$;	
\item $2$-face: $\langle G,H,I\rangle$, $\langle G,I,J \rangle$, $\langle G,H,J \rangle$ and $\langle H,I,J \rangle$.
\end{itemize}
A {\em simplicial complex} $\mathcal{K}$ in $\bbR^d$ is a collection of finite simplexes $\sigma$ in $\bbR^d$ so that any intersection of two arbitrary simplexes is a face { to each of them}. It is formalized in the following definition.

\begin{def.}
[\cite{Munkres} Sec. 1.2, p. 7]
	A collection $\mathcal{K}$ of simplexes in $\bbR^d$ is said to be a simplicial complex if it satisfies the following two properties:
	\begin{itemize}
		\itemsep = -1pt
		\item If $\sigma \in \mathcal{K}$ and $\tau$ is a face of $\sigma$, then $\tau \in \mathcal{K}$;
		\item If $\sigma_1, \sigma_2 \in \mathcal{K}$, then $\sigma_1 \cap \sigma_2$ is a face of $\sigma_1$ and $\sigma_2$. In particular, $\sigma_1 \cap \sigma_2 \in \mathcal{K}$.
	\end{itemize}
\end{def.}

For instance, the surface in Figure \ref{fig:example of Betti numbers}(D) is represented by a simplicial complex consisting of simplexes with maximal dimension 2, that is, triangles and their edges and vertices. 

In order to study the topological information of a given simplicial complex, we study relations among different dimensional simplexes.  It can be understood as a way to measure the complexity of simplicial complexes.  Homology theory is the main theory we count on to achieve this goal. In the end, we will define the Betti numbers.

\subsection{More on Betti numbers and Homology}\label{section:MoreBettiHomology}

\textit{Homology} is a classic subject { in algebraic topology} \cite{Munkres}, which captures ``holes'' of geometric objects in different dimensions. While we can discuss homology in more general geometric objects, in this work, we mainly consider simplicial complexes as our target object.  
We now discuss how to quantify $q$-dimensional holes.

In order to count $q$-dimensional holes in $\mathcal{K}$, we need to find the interactions among different simplexes.  To achieve it, { one adds} an algebraic structure to simplexes. % so that the algebraic tools can be utilized.  
Formally, given $q$-simplex $\sigma_1, \sigma_2, \ldots, \sigma_n \in \mathcal{K}$, one could write a formal sum as $c = \sum_{i = 1}^n \nu_i \sigma_i$, where $\nu_i\in \bbZ_2$.  This formal sum is commonly known as a $q$-{\it chain}. One could also define an addition as $\sum_{i = 1}^n \nu_i \sigma_i + \sum_{i = 1}^n \mu_i \sigma_i := \sum_{i = 1}^n (\nu_i+\mu_i) \sigma_i$.  We consider the collection of all $q$-chains, denoted by \begin{align}
C_q(\mathcal{K}) := \bigg\{ \sum_{i = 1}^n \nu_i \sigma_i~\Big|~ \nu_i \in \bbZ_2,~\sigma_i\in \mathcal{K},~\dim(\sigma)=q \bigg\}.
\end{align}
One could prove that $C_q(\mathcal{K})$ is actually a vector space over $\bbZ_2$ with the above addition.  For example, consider the simplicial complex $\mathcal{K}=\langle D,E,F\rangle$ in Figure \ref{fig:illustration of simplices}. $\langle D \rangle + \langle E \rangle$ is an element in $C_0(\mathcal{K})$, and $\langle D,E \rangle + \langle E,F \rangle$ is an element in $C_1(\mathcal{K})$.  Note that $\langle D,E \rangle + \langle E \rangle$ is not defined because they live in different spaces.  
There is a natural relation between $C_q(\mathcal{K})$ and $C_{q-1}(\mathcal{K})$, called the {\em boundary map}.

\begin{def.}
[\cite{Munkres} Sec. 1.5, p. 30]
Let $\sigma = \langle x_0, x_1, \cdots, x_q \rangle \in C_q(\mathcal{K})$. The $q^{\rm th}$ {\em boundary map} $\partial_q : C_q(\mathcal{K}) \rightarrow C_{q-1}(\mathcal{K})$ over $\bbZ_2$ is defined by
	\begin{equation}
	\partial_q(\langle x_0, x_1, \cdots, x_q \rangle) = \sum_{i = 0}^q \langle x_0, \cdots, \widehat{x_i}, \cdots x_q \rangle,
	\end{equation}
	where $\sigma = \langle x_0, x_1, \cdots, x_q \rangle$ being a $q$-simplex in $\mathcal{K}$ and the $\widehat{\bullet}$ denotes the drop-out operation. 
\end{def.}

For instance, $\partial_2( \langle D,E,F \rangle  ) = \langle E,F \rangle + \langle D,F \rangle + \langle D,E \rangle$. $\partial_q$ captures the boundary of a given simplex, which justifies the nomination. 
With the boundary maps, there is a nested relation among chains
\begin{equation}
\cdots\xrightarrow[]{\partial_{n+1}} C_n(\mathcal{K}) \xrightarrow[]{\partial_{n}} C_{n-1}(\mathcal{K}) \xrightarrow[]{\partial_{n-1}} \cdots C_1(\mathcal{K}) \xrightarrow[]{\partial_{1}} C_0(\mathcal{K}).
\end{equation}
This nested relation among chains is known as the {\em chain complex}, which is denoted as $\mathcal{C} = \{ C_q, \partial_q \}_{q \in \bbZ}$.

A fundamental result in homology theory (\cite{Munkres} Lemma 5.3 Sec. 1.5, p. 30) is that the composition of any two consecutive boundary maps is a trivial map, i.e.  $\partial_{q-1} \circ \partial_q = 0$. This result allows one to define the following quotient space.   We first denote \textit{cycles} and \textit{boundaries} by $Z_q$ and $B_q$, respectively, which are defined as
\begin{align}
Z_q &:= \ker(\partial_q) = \{ c \in C_q \ | \ \partial_q(c) = 0 \}, \\
B_q &:= {\rm im}(\partial_{q+1}) = \{ \partial_{q+1}(z) \in C_{q} \ | \ z \in C_{q+1} \}.\nonumber
\end{align}
Note that each $B_q$ is a subspace of $Z_q$. Therefore, we can define the $q^{\rm th}$ \textit{homology} to be the quotient space
\begin{equation}
\label{Equation : Quotient space as homology}
H_q(\mathcal{K}) := \frac{Z_q}{B_q} = \frac{\ker(\partial_q)}{{\rm im}(\partial_{q+1})},
\end{equation}
which is again a vector space.
Finally, the Betti number is defined to be the dimension of the homology.  More precisely, 
\begin{align}
\beta_q(\mathcal{K}) = \dim(H_q(\mathcal{K})). 
\end{align}

\begin{figure}[hbt!]
	\includegraphics[width = 0.95\textwidth]{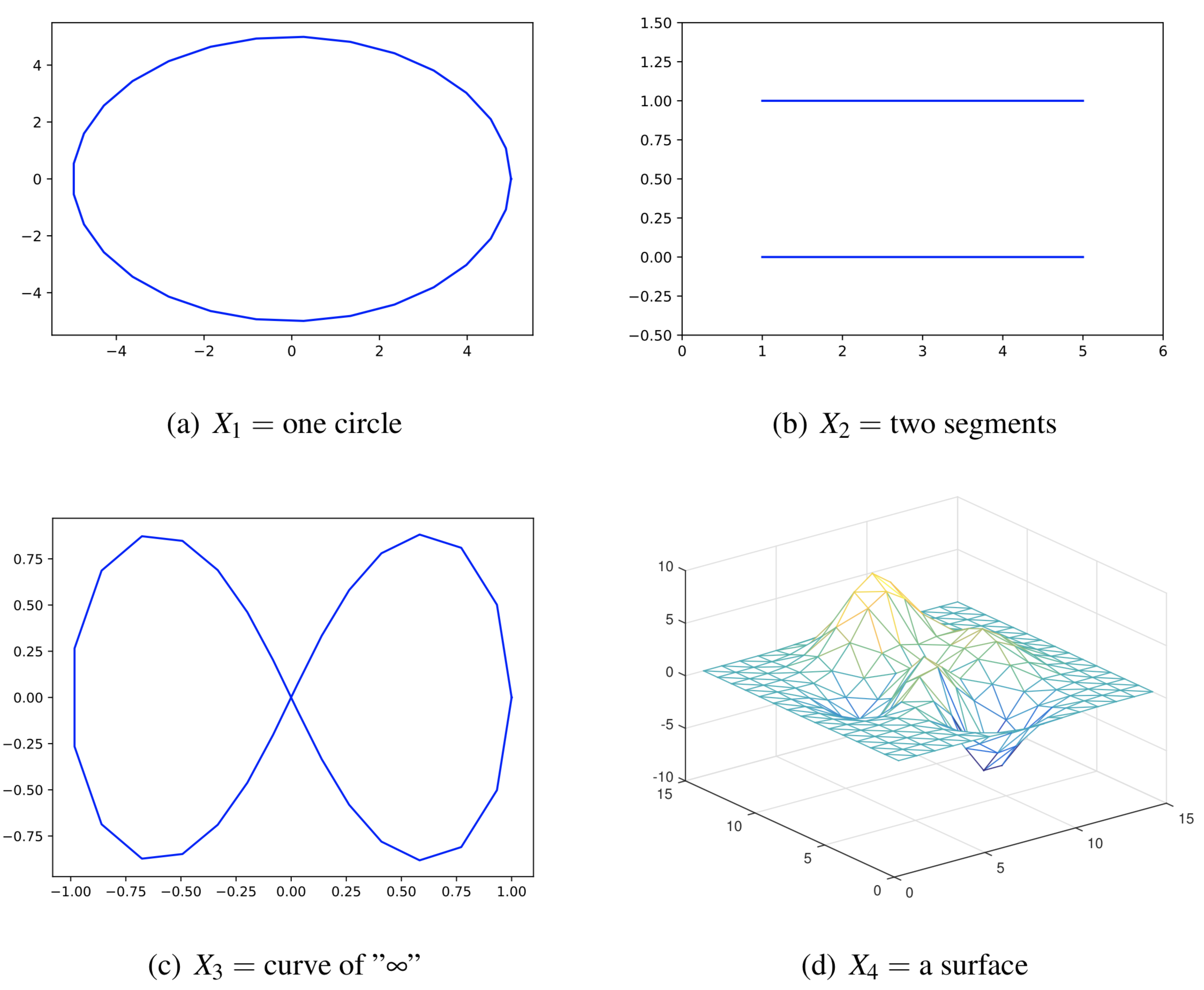}
	\includegraphics[width = 0.95\textwidth]{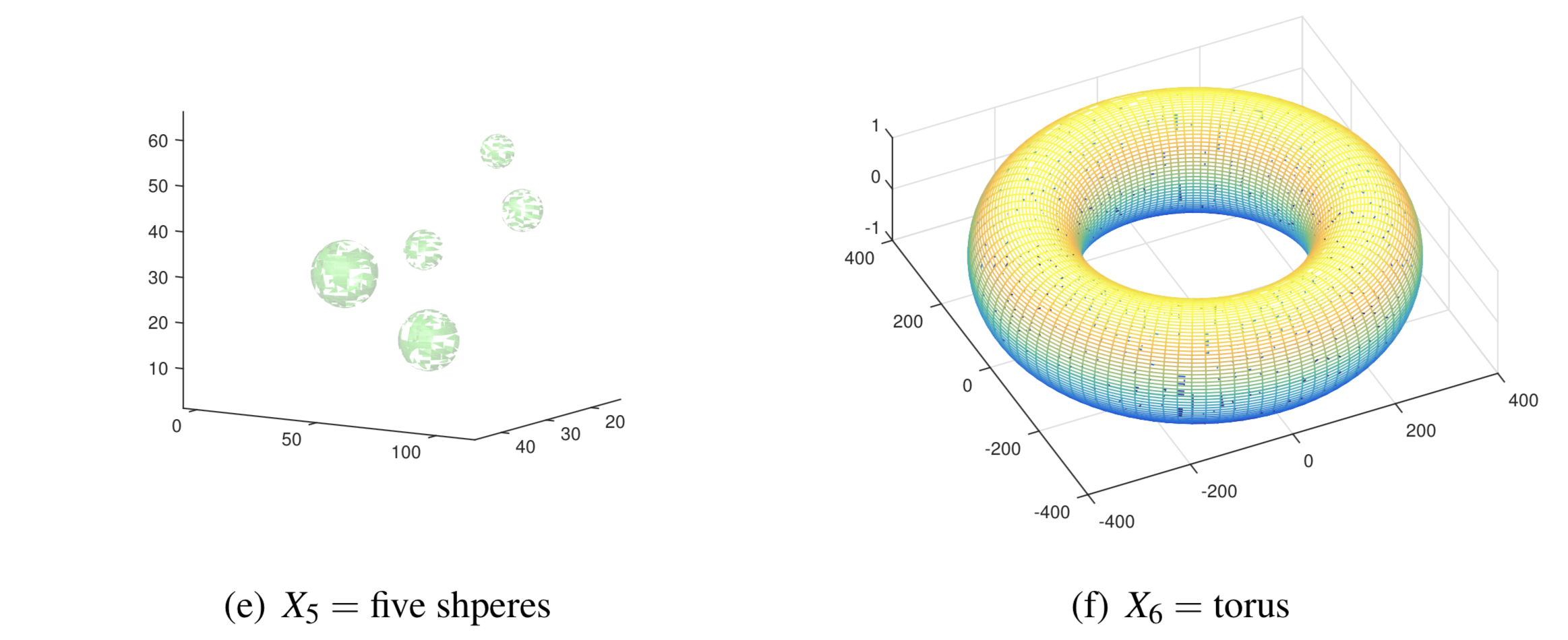}
		\caption{{ Some examples of geometric objects in $\bbR^2$ and $\bbR^3$. The Betti numbers are: $\beta(X_1) = (1,1)$, $\beta(X_2) = (2,0)$, $\beta(X_3) = (1,2)$, $\beta(X_4) = (1,0,0)$, $\beta(X_5) = (5,0,5)$ and $\beta(X_6) = (1,2,1)$}.}
	\label{fig:example of Betti numbers}
\end{figure}

As a result, given { a simplicial complex} $\mathcal{K}$, the {\em homology of $\mathcal{K}$} is a collection of vector spaces $\{ H_q(\mathcal{K}) \}_{q = 0}^{\infty}$, and its {\it Betti numbers} is denoted as $\beta(\mathcal{K}):=\{ \beta_q(\mathcal{K}) \}_{q = 0}^{\infty}$.  
Formally speaking, $\beta_q$ measures the number of $q$-dimensional holes.  For instance, $\beta_0$ counts the number of components, $\beta_1$ counts the number of loops, and $\beta_2$ counts the number of voids.  Using this intuition and by visual inspection, one may count the Betti numbers of those simplicial complexes appeared in Figure \ref{fig:example of Betti numbers}.  

{
Figure \ref{fig:example of Betti numbers} shows some examples of simplicial complexes in $\mathbb{R}^2$, $X_1$ to $X_3$, and $\mathbb{R}^3$, $X_4$ to $X_6$.  For instance, the direct computation shows that $H_0(X_1) = \mathbb{Z}_2$, $H_1(X_1) = \mathbb{Z}_2$, and $H_q(X_1)= \{ 0 \}$ for all $q>1$; hence $\beta(X_1) = \{1,1,0,\ldots\}$.  The only nontrivial homology group of $X_2$ is $H_0(X_2) = \mathbb{Z}_2^2$, and hence $\beta(X_2) = \{2,0,\ldots\}$. For objects $X_4$ to $X_6$, voids in $\bbR^3$ surrounded by spheres or tori are $2$-dimensional holes, so $H_2(X_5) = \bbZ_2^5$ and $H_2(X_6) = \bbZ_2$. Note that $H_1(X_6) = \bbZ_2^2$ because a torus has two $1$-dimensional holes (\cite{Munkres} Theorem 6.2 Sec. 1.6, p. 35).
}

\subsection{More about sub-level set and VR filtrations}\label{Section:MoresublevelVRfiltrations}

\begin{figure}[hbt!]
	\includegraphics[width=0.95\textwidth]{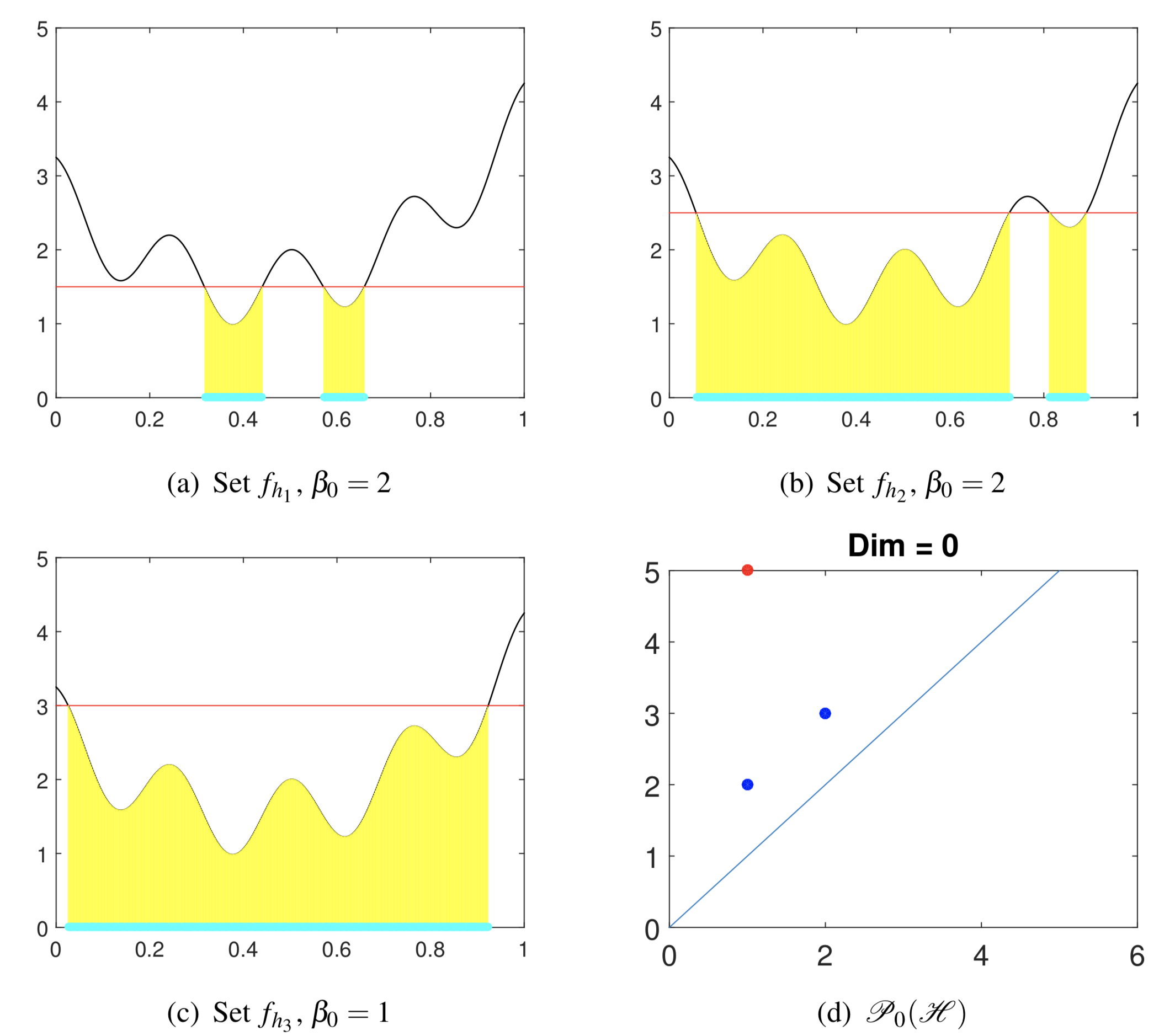} 
	\caption{An example of sub-level sets of signal $f(t) = 1 + t + 7(t - 0.5)^2 + \cos(8 \pi t)/2$. Unions of blue line segments lying on $x$-axis in terms $(a)$, $(b)$ and $(c)$ are sub-level sets $f_{h_1}, f_{h_2}$ and $f_{h_3}$ ($h_1 = 1.5$, $h_2 = 2.5$, $h_3 = 3$) with corresponding thresholds. We also use the convention $f_{h_1} = I_{1,\lleft} \cup I_{1,\rright}$ and $f_{h_2} = I_{2,\lleft} \cup I_{2,\rright}$ to decompose $f_{h_1}$ and $f_{h_2}$ by disjoint intervals as blue line segments in $(a)$ and $(b)$. Labels of $x, y$ axes in $(a)$, $(b)$ and $(c)$ { are arbitrary units.}}
	\label{fig:example of sub-level set filtration}
\end{figure}

First, we provide an example of the sublevel set filtration.
\begin{exam.}
Consider a simple filtration $\{f_{1.5}, f_{2.5}, f_{3}\}$ as shown in Figure \ref{fig:example of sub-level set filtration}. When $h = 1.5$, the sub-level set $f_{1.5}$ has $\beta_0 = 2$ since it contains two connected components i.e., disjoint intervals in $\bbR^1$. When $h = 2.5$, two connected components in $f_{1.5}$ merged to $f_{2.5}$. Moreover, there is a new interval ($\approx[0.8, 0.9]$) appeared in $f_{2.5}$, and hence $\beta_0(f_{2.5}) = 2$. Finally, when $h$ is lifted to $3$, previous intervals are merged to $f_{3}$ and we get $\beta_0(f_{3}) = 1$.  This filtering process can be depicted in the PD as shown in Figure \ref{fig:example of sub-level set filtration}(D), which is $\mathcal{P}_0(\{ f_{1.5}, f_{2.5}, f_3 \}) = \{ (1.5,\infty), (1.5,2.5), (2.5,3) \}$.
\end{exam.}

\begin{figure}[hbt!]
	\includegraphics[width=0.995\textwidth]{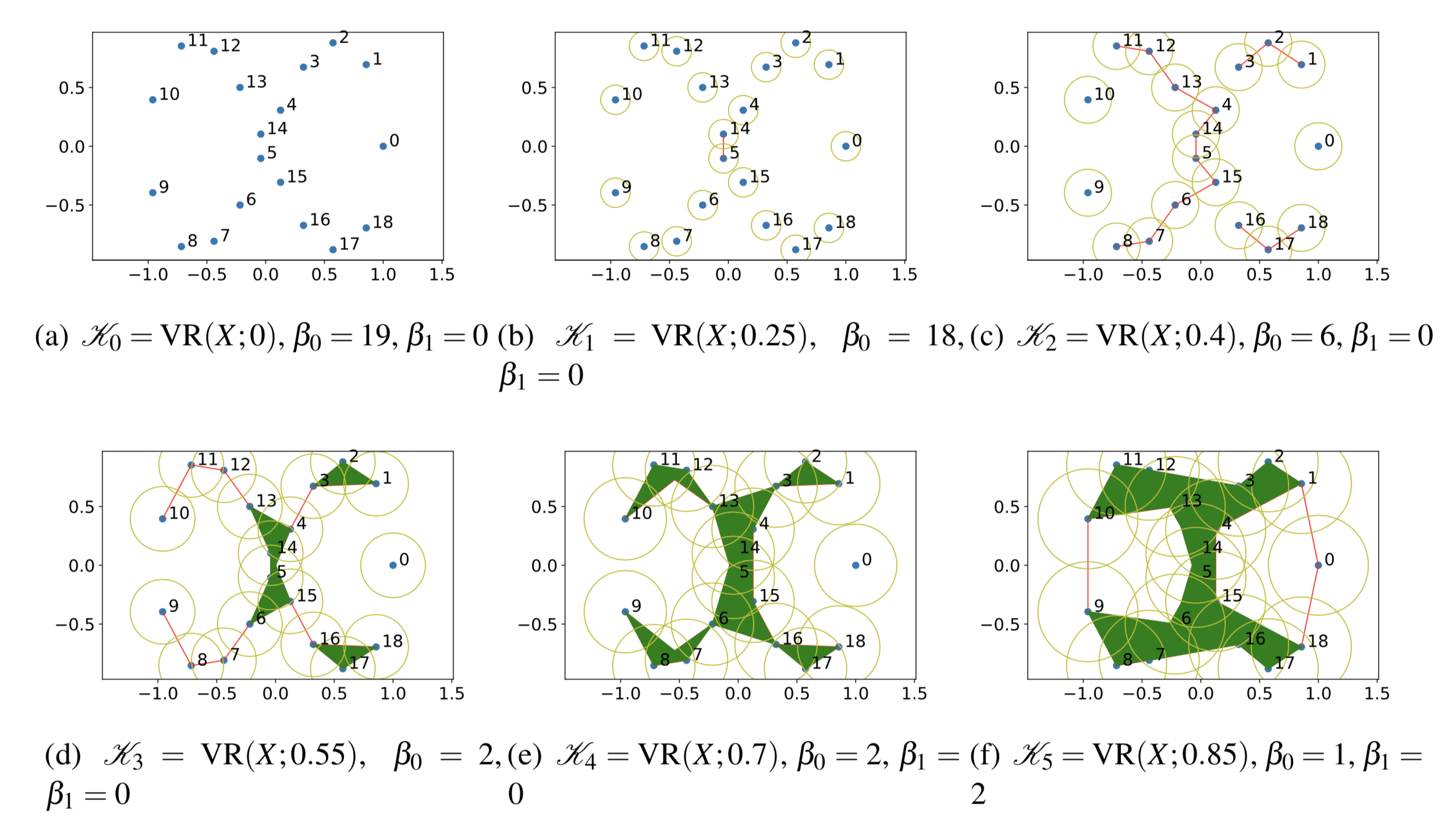}
	\includegraphics[width=0.995\textwidth]{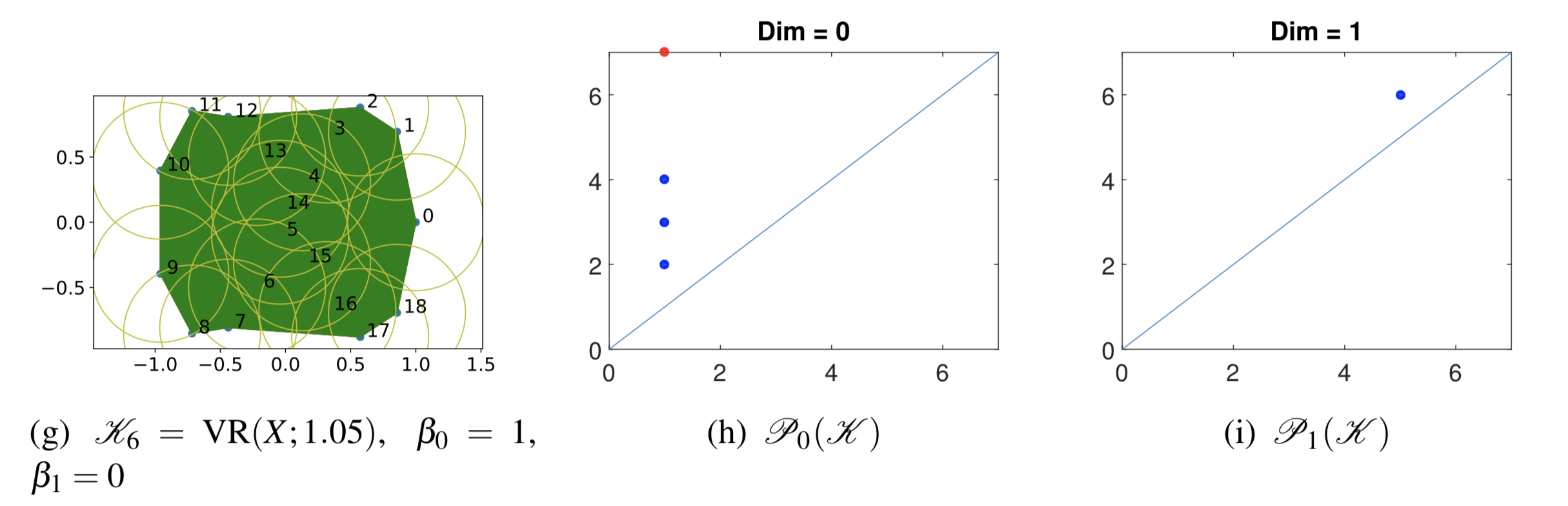}
	\caption{An example of filtration made by Vietoris-Rips complex. Labels of $x, y$ axes in $(a) \sim (c)$ are arbitrary units.
	\label{fig: VR filtration}}
\end{figure}

Next, we show an illustrative example of the VR filtration.
\begin{exam.}
Figure \ref{fig: VR filtration} shows an example of VR complex of point-cloud in $\bbR^2$, where points are sampled from curve $X_3$ in Figure \ref{fig:example of Betti numbers}. As in Figure \ref{fig: VR filtration}, when $\epsilon$ is too small, then the information of Betti numbers of { the simplicial complex $\VR(X;\epsilon)$} is nothing different to original point-cloud. On the other hand, for extremely large $\epsilon$, the only non-zero Betti number of $\VR(X;\epsilon)$ is $\beta_0 = 1$.   For example, the only two $1$-dimension holes in filtration $\mathcal{K}_1 \subseteq \mathcal{K}_2 \subseteq \cdots \subseteq \mathcal{K}_6$ of Figure \ref{fig: VR filtration} were born at time $5$ ($\mathcal{K}_5$) and died at time $6$ ($\mathcal{K}_6$), hence both of them have coordinate $(5,6)$. Moreover, if a $q$-dimensional hole is still alive in the final object of the filtration, we will use $\infty$ to denote its death value (which indicates the feature never dies).   We also use Figure \ref{fig: VR filtration} to explain the PD. First, in every $\mathcal{K}_i$, we encode each component to the smallest index of points { belonging to the components}. For example, because vertices $5$ and $14$ represent the same connected component $\langle 5,14 \rangle$ in $\mathcal{K}_1$, we use $0$-simplex $\langle 5\rangle$ to denote the component. 

With the VR complex, the lifespan of each hole can be computed. For example, both of components $\langle 5 \rangle$ and $\langle 11 \rangle$ in $\mathcal{K}_1$ has life span $(1,2)$ because they would be merged into the connected component represented by $\langle 4 \rangle$ in $\mathcal{K}_2$. On the other hand, because the connected component $\langle 0 \rangle$ is alive in whole filtration, it has lifespan $(1,\infty)$.
As a result, the PD of filtration in Figure \ref{fig: VR filtration} can be expressed as
\begin{equation}
\label{Equation : 0th Persistence diagram of VR example}
\begin{split}
\mathcal{P}_0(\mathcal{K}) = & \{ (1,\infty)_{\langle 0 \rangle}, (1,4)_{\langle 1\rangle}, (1,2)_{\langle 2\rangle}, (1,2)_{\langle 3\rangle}, (1,3)_{\langle 4\rangle}, (1,2)_{\langle 5\rangle}, \\ 
& \ \ (1,2)_{\langle 6\rangle}, (1,2)_{\langle 7\rangle}, (1,2)_{\langle 8\rangle}, (1,3)_{\langle 9\rangle}, (1,3)_{\langle 10\rangle}, (1,2)_{\langle 11\rangle}, \\
& \ \ (1,2)_{\langle 12\rangle}, (1,2)_{\langle 13\rangle}, (1,2)_{\langle 15\rangle}, (1,3)_{\langle 16\rangle}, (1,2)_{\langle 17\rangle}, (1,2)_{\langle 18\rangle}  \}
\end{split}
\end{equation}
and $\mathcal{P}_1(\mathcal{K}) = \{ (5,6), (5,6) \}$. The notations $\langle *\rangle$ of lifespans in \eqref{Equation : 0th Persistence diagram of VR example} are used for clarifying which component a lifespan belong to.  
\end{exam.}

\section{Takens' Lag Map}
\label{SI : Taken's Lag Map}

In this section, we provide a review of ``recovering the manifold'' underlying the one dimensional time series. Precisely, we discuss a set of theorems provided in~\cite{Takens:1981} and an associated embedding algorithm in the time series framework. The algorithm is well-known as the {\em Takens' lag map} or {\em lag map}, and has been extensively applied to many fields. In a nutshell, the recovered manifold is the phase space on which the dynamics lives that generates the time series. 
From now on, denote $M$ to be a $d$-dim compact manifold without boundary. For the sake of self-containedness, we recall the following definitions.

\begin{def.}[Discrete time dynamics]
By a discrete time dynamics, we mean a diffeomorhism $\varphi:\, M\rightarrow M$
with the time evolution $i\mapsto\varphi^{i}(x_{0})$, $i\in\mathbb{N}$, where
$x_{0}$ is the starting status.
\end{def.}

\begin{def.}[Continuous time dynamics]
By a continuous time dynamics, we mean a smooth vector field $X\in\Gamma(M)$ with
the time evolution $t\mapsto\gamma_{t}(x_{0})$, where $\gamma_{t}$ is the integral
curve with respect to $X$ via $x_{0}$.
\end{def.}

To simplify the discussion, in both cases, we denote $\Phi_{t}(x_{0})$ to be the
time evolution with time $t\in\mathbb{N}$ or $\mathbb{R}$ with the starting point
$x_0$.

\begin{def.}[Observed time series] Let $\Phi_{t}(x_{0})$ be a dynamics on $M$. The
observation is modeled as a function $f:\, M\rightarrow\mathbb{R}$ and the observed
time series is $f(\Phi_{t}(x_{0}))$.
\end{def.}

The question we have interest is that if we have an observed time series $f(\Phi_{t}(x_{0}))$, can we recover $M$ and/or the dynamics? 
The positive answer and the precise statements are provided in the following two theorems. The proof of these theorems can be found in~\cite{Takens:1981}. 

\begin{theorem}
[discrete time dynamics] 
For a pair $(\varphi,f)$, $\varphi:\, M^{d}\rightarrow M^{d}$ is the $C^{2}$-diffeomorphism and $f\in C^2$, it is generic that the map $\Psi:\, M\rightarrow\mathbb{R}^{2d+1}$ given by
$$
\Psi:\, x\mapsto(f(x),\, f(\varphi(x)),\, f(\varphi^{2}(x))\ldots
f(\varphi^{2d}(x)))^{T}\in\mathbb{R}^{2d+1}
$$
is an embedding.
\end{theorem}

By {\em generic}, we mean an open dense subset of all possible pairs $(\varphi,f)$. We mention that the theorems hold for non-compact manifolds if $f$ is proper. These technical assumptions are practically assumed held for real world datasets.

\begin{theorem}[Continuous time dynamics]
When $X\in C^{2}(\Gamma M)$ and $f\in C^{2}(M)$, it is generic that $\Psi:\,
M\rightarrow\mathbb{R}^{2d+1}$ given by
$$
\Psi:\, x\mapsto(f(x),\, f(\gamma_{1}(x)),\, f(\gamma_{2}(x))\ldots
f(\gamma_{2d}(x)))^{T}\in\mathbb{R}^{2d+1}
$$
is an embedding, where $\gamma_{t}(x)$ is the flow of $X$ of time $t$ via $x$.
\end{theorem}

These theorems tell us that we could embed the manifold into a $(2d+1)$ dimensional Euclidean space if we have access to all dynamical behaviors from all points on the manifold. 
However, in practice the above model and theorem cannot be applied directly. Indeed, for a given dynamical system, most of time we may only have one or few experiments that are sampled at discrete times; that is, we only have access to one or few $x\in M$. We thus ask the following question. Suppose we have the time series  
$$
\left\{ f(\Phi_{\ell \alpha}(x))\right\} _{\ell=0}^{N}
$$ 
as our dataset, where $x\in M$ is fixed and inaccessible to us, $\alpha>0$ is the sampling period, and $N\gg1$ is the number of samples, what can we do? We first give the following definition.

\begin{def.} 
The {\it positive limit set (PLS) of $x$} of a vector field $X\in C^{2}\left(\Gamma
M\right)$ is defined as
$$
L_{c}^{+}(x):=\left\{ x'\in M|\,\exists\, t_{i}\rightarrow\infty,\;
t_{i}\in\mathbb{R}\text{ such that }\gamma_{t_{i}}(x)\rightarrow x'\right\}
$$
and the PLS of $x$ of a diffeomorphism $\varphi:\, M\rightarrow M$ is defined as
$$
L_{d}^{+}(x):=\left\{ x'\in M|\,\exists\, n_{i}\in\mathbb{N}\rightarrow\infty,\text{
such that }\varphi^{n_{i}}\left(x\right)\rightarrow x'\right\}.
$$
\end{def.}

It turns out that in this case, we should know whether under generic assumptions the
topology and dynamics in the PLS of $x$ is determined by $\{ f(\Phi_{\ell
\alpha}(x))\} _{\ell=0}^{\infty}$. Precisely, we have the following theorem:

\begin{theorem}
[Continuous dynamics with 1 trajectory]
Fix $x\in M$. When $X\in C^{2}(\Gamma M)$ with flow $\gamma_{t}$ passing $x$, then
there exists a residual subset $C_{X,x}\subset\mathbb{R}^{+}$ such that for all
$\alpha\in C_{X,x}$ and diffeomorphism $\varphi:=\gamma_{\alpha}$, the PLS
$L_{c}^{+}(x)$ for flow $\gamma_{t}$ and $L_{d}^{+}(x)$ for $\varphi$ are the same;
that is, for all $\alpha\in C_{X,x}$ and for all $q\in L_{c}^{+}(p)$, there exists
$n_{i}\in\mathbb{N}\rightarrow\infty$ such that $\varphi^{n_{i}}(x)\rightarrow q$.
\end{theorem}

This theorem leads to the following corollary, which is what we need to analyze the time series. 

\begin{cor}
Take $x\in M$, a generic pair $X\in C^{2}(\Gamma M)$ and $f\in C^{2}(M)$, and
$a\in\mathbb{R}^{+}$ satisfying generic conditions depending on $X$ and $x$. Denote
the set
\[
\mathcal{P}:=\left\{f(\gamma_{k\alpha}(x)),\,f(\gamma_{k\alpha}(x)),\ldots,f(\gamma_{(k+2d)\alpha}(x))
\right\}_{k=0}^\infty.
\] 
Then there exists a smooth embedding of $M$ into $\mathbb{R}^{2d+1}$ mapping PLS
$L_{c}^{+}$ bijectively to the set $\mathcal{P}$.
\end{cor}

Last but not the least, we remark that the noise analysis of the Takens' lag map has been extensively studied, for example \cite{Stark_Broomhead_Davies_Huke:1997}.

%\clearpage
\section{More Automatic Annotation Results}
\label{sec:more automatic annotation results}
To further illustrate the robustness of our model and features, we perform more cross-database validation.

\subsection{Visualize features $\Phi^{(\texttt{PS})}(\mathcal{P}_i(R_{120,1}))$}

\begin{figure}[hbt!]
	\centering
	\subfigure[Boxplot. $\mathcal{P}_0$.]{
		\includegraphics[width = 0.48\textwidth]{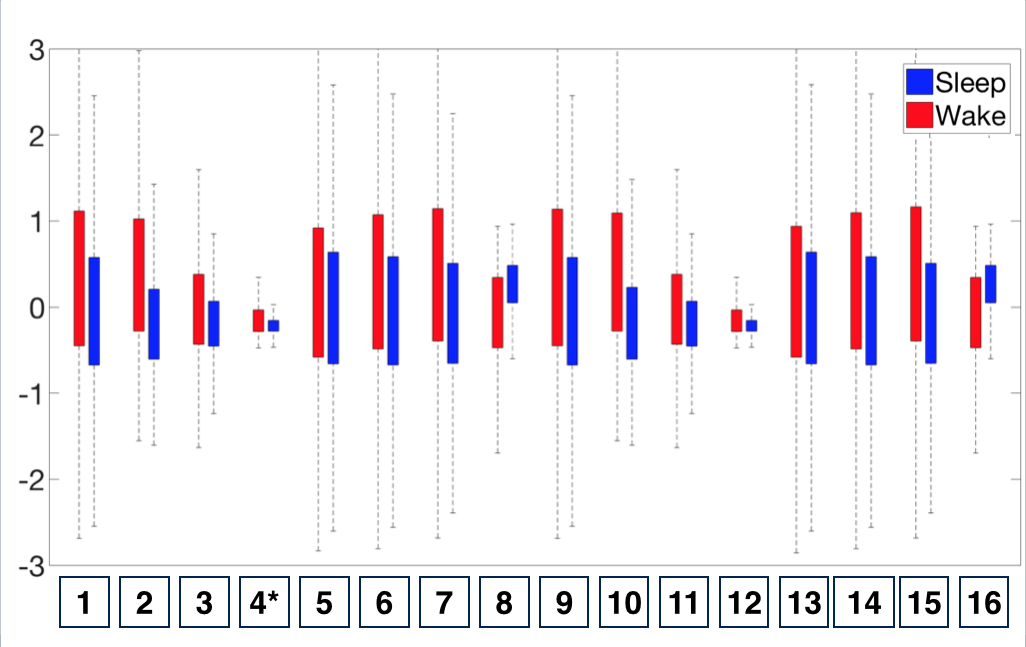}}
	\subfigure[Boxplot plot. $\mathcal{P}_1$.]{
		\includegraphics[width = 0.48\textwidth]{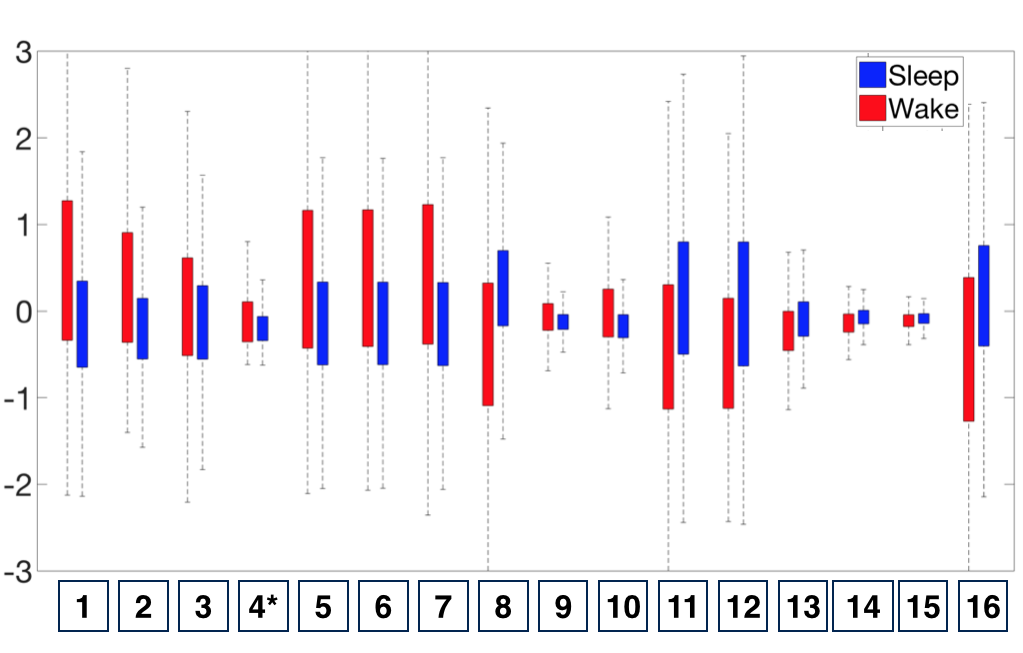}}
\subfigure[Scatter plot. $\mathcal{P}_0$.]{
	\includegraphics[width = 0.48\textwidth]{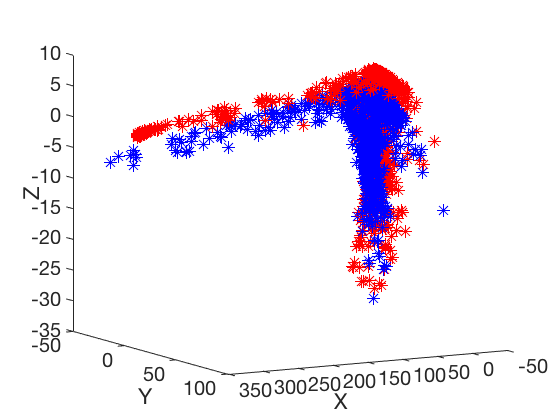}}
\subfigure[Scatter plot. $\mathcal{P}_1$.]{
	\includegraphics[width = 0.48\textwidth]{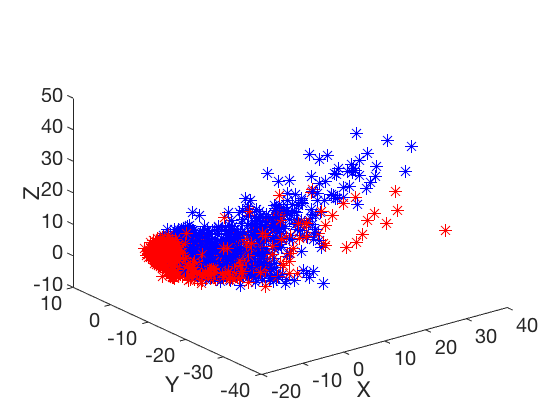}}	
	\vspace{-10pt}
	\caption{Distribution of normalized PS features, $\Phi^{(\texttt{PS})}(\mathcal{P}(R_{120,1}))$.  (a)-(b) Boxplot of the $\Phi^{(\texttt{PS})}(\mathcal{P}_i(R_{120,1}))$ where $i=0$, $1$, respectively.  The numbers listed on the horizontal axis indicates the number of PS. * indicates that the feature {\em fails} to reject the null hypothesis of the significance level of $0.05$ on the rank sum test with Bonferroni correction. (c)-(d) Visualization of $\cup_k\{\Phi^{(\texttt{PS})}(\mathcal{P}_i(R_{120,1}))\}_{j=1}^{n_k}$ by the first three { principal components}. }
	\label{fig:R120CGMH features}
\end{figure}

The visualization of $\Phi^{(\texttt{PS})}(\mathcal{P}_i(R_{120,1}))$, where $i=0,1$, is shown in Figure \ref{fig:R120CGMH features}.
In order to compare them on the same scale, we perform the standard $z$-score normalization for each parameter in $\Phi^{(\texttt{PS})}(\mathcal{P}_i(R_{120,1}))$. We abuse the notation and use the same notation $\Phi^{(\texttt{PS})}(\mathcal{P}_i(R_{120,1}))$ to denote the normalized parameters.
The boxplot of each normalized PS parameter, where blue (red) bars represent the PS associated with an IHR time series associated with the sleep (Wake) stage, is shown in Figure \ref{fig:R120CGMH features}.   
We performed a rank sum test with a significance level of $0.05$ with the Bonferroni correction. We found that there are significant differences between waking and sleeping features for all PS parameters, except two parameters in $\Phi^{(\texttt{PS})}(\mathcal{P}_0(R_{120,1}))$.
The first { principal components} of $\cup_k\{\Phi^{(\texttt{PS})}(\mathcal{P}_i(R_{120,1}))\}_{j=1}^{n_k}$ are shown in Figure~\ref{fig:R120CGMH features}(c) and (d).  We can observe a separation between sleep and Wake features.

\subsection{For the classification of sleep and Wake}
We consider the models trained on DREAMS and UCDSADB as shown in Table~\ref{tab:SW classification training on DREAMS (Subjects)} and Table~\ref{tab:SW classification training on UCDSB (Subjects)}, respectively.
In Table~\ref{tab:SW classification training on DREAMS (Subjects)}, we observe that the (SE, SP) pairs are consistent and { all are located} in the level of $(60, 70)$ except for the UCDSADB.  In Table~\ref{tab:SW classification training on UCDSB (Subjects)}, we also observe that the (SE, SP) pairs are all on the level of $(60,70)$. 
These consistent performances suggest that our proposed topological features, persistence statistics, are stable and well capture intrinsic characteristics about the sleep and Wake IHR signals.

\begin{table}
	\caption{SVM cross-database performance of subjects for Wake and Sleep classification with a single random seed. The training database is DREAMS. %{\color{red} Each feature vector used here is normalized via standard $z$-normalization among features of corresponding subject}. 
	For each database and each performance measurement, we report the mean $\pm$ standard deviation of all subjects.
	\label{tab:SW classification training on DREAMS (Subjects)}
}
	\centering
	\fbox{\footnotesize
		\begin{tabular}{c|cccc}
			& CGMH-training  & CGMH-validation & DREAMS & UCDSADB \\ \hline\hline
			TP & 74 $\pm$ 43 & 72 $\pm$ 40 & 112 $\pm$ 57 & 86 $\pm$ 47 \\
			FP & 166 $\pm$ 55 & 141 $\pm$ 54 & 176 $\pm$ 64 & 165 $\pm$ 63 \\
			TN & 447 $\pm$ 69 & 434 $\pm$ 97 & 592 $\pm$ 112 & 431 $\pm$ 114 \\
			FN & 29 $\pm$ 34 & 43 $\pm$ 44 & 45 $\pm$ 41 & 71 $\pm$ 54 \\
			\hline
			SE $(\%)$ & 75.9 $\pm$ 15.7 & 70.7 $\pm$ 17.0 & 73.7 $\pm$ 15.0 & 58.6 $\pm$ 16.9 \\ 
			SP $(\%)$ & 73.5 $\pm$ 7.1 & 76.5 $\pm$ 6.3 & 77.6 $\pm$ 6.0 & 72.5 $\pm$ 7.1 \\
			Acc $(\%)$ & 72.7 $\pm$ 6.7 & 73.6 $\pm$ 4.7 & 75.9 $\pm$ 5.3 & 68.7 $\pm$ 7.7\\
			\hline 
			PR $(\%)$ & 31.6 $\pm$ 18.3 & 35.9 $\pm$ 18.9 & 39.2 $\pm$ 18.1 & 33.6 $\pm$ 17.3 \\
			F1  & 0.411 $\pm$ 0.168 & 0.431 $\pm$ 0.129 & 0.481 $\pm$ 0.142 & 0.397 $\pm$ 0.149 \\
			AUC  & 0.810 $\pm$ 0.099 & 0.800 $\pm$ 0.101 & 0.825 $\pm$ 0.075 & 0.698 $\pm$ 0.125 \\
			Kappa & 0.285 $\pm$ 0.155 & 0.291 $\pm$ 0.110 & 0.351 $\pm$ 0.142 & 0.215 $\pm$ 0.166 \\ 
		\end{tabular}}
\end{table}

\begin{table}
	\caption{SVM cross-database performance for Wake and Sleep classification. The training database is UCDSB database with a single random seed. %{\color{red} Each feature vector used here is normalized via standard $z$-normalization among features of corresponding subject}. 
	For each database and each performance measurement, we report the mean $\pm$ standard deviation of all subjects.
	\label{tab:SW classification training on UCDSB (Subjects)}
}
	\centering
	\fbox{\footnotesize
		\begin{tabular}{c|cccc}
			& CGMH-training  & CGMH-validation & DREAMS & UCDSADB \\ \hline\hline
			TP & 77 $\pm$ 45 & 77 $\pm$ 44 & 110 $\pm$ 63 & 99 $\pm$ 55 \\
			FP & 192 $\pm$ 58 & 167 $\pm$ 60 & 232 $\pm$ 83 & 179 $\pm$ 67 \\
			TN & 421 $\pm$ 62 & 409 $\pm$ 92 & 535 $\pm$ 98 & 417 $\pm$ 105 \\
			FN & 25 $\pm$ 33 & 41 $\pm$ 44 & 47 $\pm$ 35 & 58 $\pm$ 44 \\
			\hline
			SE $(\%)$ & 79.6 $\pm$ 15.3 & 71.9 $\pm$ 17.8 & 70.4 $\pm$ 13.3 & 61.4 $\pm$ 18.8 \\ 
			SP $(\%)$ & 69.3 $\pm$ 7.0 & 72.0 $\pm$ 6.5 & 70.5 $\pm$ 7.7 & 70.1 $\pm$ 6.2 \\
			Acc $(\%)$ & 69.7 $\pm$ 6.8 & 70.2 $\pm$ 5.6 & 69.6 $\pm$ 6.7 & 68.2 $\pm$ 6.5\\
			\hline 
			PR $(\%)$ & 29.6 $\pm$ 17.7 & 32.8 $\pm$ 18.6 & 33.1 $\pm$ 19.0 & 34.4 $\pm$ 17.7 \\
			F1  & 0.398 $\pm$ 0.168 & 0.409 $\pm$ 0.149 & 0.420 $\pm$ 0.163 & 0.420 $\pm$ 0.170 \\
			AUC  & 0.815 $\pm$ 0.107 & 0.783 $\pm$ 0.120 & 0.771 $\pm$ 0.092 & 0.707 $\pm$ 0.113 \\
			Kappa & 0.263 $\pm$ 0.156 & 0.255 $\pm$ 0.136 & 0.266 $\pm$ 0.161 & 0.241 $\pm$ 0.157 \\ 
		\end{tabular}}
\end{table}

%%%%%%%%%

\subsection{For the classification of REM and NREM}

We consider the models trained on DREAMS and UCDSADB for the REM and NREM classification. The results are summarized in Tables~\ref{tab:REM/NREM classification training on DREAMS (Subjects)} and \ref{tab:SW classification training on UCDSADB (Subjects)} respectively. We observe that the overall performance is consistent. As expected, the performance of the model trained on DREAMS is degraded on UCDSADB, and the performance of the model trained on UCDSADB on other databases is slightly lower, particularly the SE's.

\begin{table}
	\caption{SVM cross-database performance for REM and NREM classification. The training database is DREAMS with a single random seed. %{\color{red} Each feature vector used here is normalized via standard $z$-normalization among features of corresponding subject}. 
	The subject $\#9$ in CGMH-validation and the subject $\# 24$ in UCDSADB were dropped because they {do not have REM epochs}. For each database and each performance measurement, we report the mean $\pm$ standard deviation of all subjects. 
	\label{tab:REM/NREM classification training on DREAMS (Subjects)}
}
	\centering
	\fbox{\footnotesize
		\begin{tabular}{c|cccc}
			& CGMH-training  & CGMH-validation & DREAMS & UCDSADB \\ \hline\hline
			TP & 75 $\pm$ 31 & 68 $\pm$ 30 & 102 $\pm$ 32 & 65 $\pm$ 35 \\
			FP & 130 $\pm$ 33 & 121 $\pm$ 47 & 155 $\pm$ 53 & 150 $\pm$ 34 \\
			TN & 382 $\pm$ 66 & 376 $\pm$ 90 & 472 $\pm$ 94 & 356 $\pm$ 70 \\
			FN & 25 $\pm$ 24 & 21 $\pm$ 16 & 37 $\pm$ 23 & 48 $\pm$ 36 \\
			\hline
			SE $(\%)$ & 76.8 $\pm$ 15.6 & 77.4 $\pm$ 16.7 & 74.8 $\pm$ 13.1 & 58.3 $\pm$ 18.0 \\ 
			SP $(\%)$ & 74.6 $\pm$ 4.9 & 76.4 $\pm$ 6.4 & 75.6 $\pm$ 5.2 & 70.2 $\pm$ 6.1 \\
			Acc $(\%)$ & 74.6 $\pm$ 6.0 & 75.3 $\pm$ 8.0 & 75.3 $\pm$ 5.9 & 67.8 $\pm$ 6.9\\
			\hline 
			PR $(\%)$ & 36.2 $\pm$ 12.4 & 38.1 $\pm$ 17.9 & 40.6 $\pm$ 12.2 & 29.5 $\pm$ 15.1 \\
			F1  & 0.479 $\pm$ 0.132 & 0.483 $\pm$ 0.155 & 0.519 $\pm$ 0.122 & 0.373 $\pm$ 0.155 \\
			AUC  & 0.823 $\pm$ 0.109 & 0.834 $\pm$ 0.109 & 0.819 $\pm$ 0.095 & 0.683 $\pm$ 0.131 \\
			Kappa & 0.343 $\pm$ 0.145 & 0.353 $\pm$ 0.166 & 0.375 $\pm$ 0.147 & 0.197 $\pm$ 0.161 \\
		\end{tabular}}
\end{table}

\begin{table}
	\caption{SVM cross-database performance for REM and NREM classification. The training database is UCDSADB with a single random seed. %{\color{red} Each feature vector used here is normalized via standard $z$-normalization among features of corresponding subject}. 
	The subject $\#9$ in CGMH-validation and the subject $\# 24$ in UCDSADB were dropped because they {do not have} REM epochs. For each database and each performance measurement, we report the mean $\pm$ standard deviation of all subjects.
	\label{tab:SW classification training on UCDSADB (Subjects)}
}
	\centering
	\fbox{\footnotesize
		\begin{tabular}{c|cccc}
			& CGMH-training  & CGMH-validation & DREAMS & UCDSADB \\ \hline\hline
			TP & 60 $\pm$ 29 & 52 $\pm$ 23 & 75 $\pm$ 26 & 73 $\pm$ 40 \\
			FP & 114 $\pm$ 32 & 108 $\pm$ 38 & 151 $\pm$ 45 & 117 $\pm$ 33 \\
			TN & 398 $\pm$ 69 & 389 $\pm$ 93 & 478 $\pm$ 103 & 390 $\pm$ 70 \\
			FN & 41 $\pm$ 25 & 36 $\pm$ 23 & 64 $\pm$ 32 & 41 $\pm$ 29 \\
			\hline
			SE $(\%)$ & 60.1 $\pm$ 17.9 & 60.3 $\pm$ 15.0 & 55.3 $\pm$ 15.5 & 65.3 $\pm$ 14.5 \\ 
			SP $(\%)$ & 77.6 $\pm$ 5.0 & 78.1 $\pm$ 5.4 & 76.0 $\pm$ 4.5 & 76.9 $\pm$ 5.4 \\
			Acc $(\%)$ & 74.6 $\pm$ 6.0 & 74.6 $\pm$ 9.0 & 72.0 $\pm$ 5.6 & 74.6 $\pm$ 5.9\\
			\hline 
			PR $(\%)$ & 33.8 $\pm$ 13.9 & 33.9 $\pm$ 14.5 & 33.6 $\pm$ 10.5 & 36.9 $\pm$ 16.2 \\
			F1  & 0.420 $\pm$ 0.143 & 0.412 $\pm$ 0.128 & 0.411 $\pm$ 0.115 & 0.452 $\pm$ 0.152 \\
			AUC  & 0.740 $\pm$ 0.135 & 0.742 $\pm$ 0.121 & 0.715 $\pm$ 0.113 & 0.773 $\pm$ 0.112 \\
			Kappa & 0.278 $\pm$ 0.158 & 0.271 $\pm$ 0.146 & 0.244 $\pm$ 0.142 & 0.308 $\pm$ 0.152 \\
		\end{tabular}}
\end{table}

%%%%%%%%
\subsection{For the classification of Wake, REM and NREM}

We consider the models trained on DREAMS and UCDSADB for the Wake, REM and NREM classification. The results are shown in Tables~\ref{tab:WRN classification training on DREAMS (Subjects)} and \ref{tab:WRN classification training on UCD (Subjects)} respectively. We observe that the overall performances are consistent, and +P of NREM is higher than other classes as expected.

%%%%%%%%%%%%%%%%%%%%%%%%%%%%%%%%%%%%%%%%%%%%%%%%%%%%%%%
%%%%%%%%%%%%%%%%%%%%%%%%%%%%%%%%%%%%%%%%%%%%%%%%%%%%%%%
\begin{table}
	\caption{SVM cross-database performance for Wake, REM and NREM classification. The training database is DREAMS with a single random seed. %{\color{red} Each feature vector used here is normalized via standard $z$-normalization among features of corresponding subject}. 
	For each database and each performance measurement, we report the mean $\pm$ standard deviation of all subjects.
	\label{tab:WRN classification training on DREAMS (Subjects)}
}
	\centering
	\fbox{\footnotesize 
		\begin{tabular}{c|cccc}
			& CGMH-training  & CGMH-validation & DREAMS & UCDSADB \\ \hline\hline
			SE $(\%)$ (Wake) & 63.3 $\pm$ 15.7 & 61.6 $\pm$ 17.7 & 64.8 $\pm$ 14.9 & 47.4 $\pm$ 14.7 \\					
			SE $(\%)$ (REM) & 54.7 $\pm$ 17.6 & 58.5 $\pm$ 19.0 & 59.4 $\pm$ 17.6 & 44.3 $\pm$ 17.6 \\
			SE $(\%)$ (NREM) & 67.9 $\pm$ 9.3 & 69.3 $\pm$ 7.4 & 70.6 $\pm$ 8.2 & 61.3 $\pm$ 9.4 \\
			+P $(\%)$ (Wake) & 34.7 $\pm$ 19.3 & 37.6 $\pm$ 16.4 & 44.6 $\pm$ 18.3 & 36.5 $\pm$ 17.5 \\
			+P $(\%)$ (REM) & 37.3 $\pm$ 15.6 & 37.4 $\pm$ 18.0 & 42.4 $\pm$ 14.0 & 29.0 $\pm$ 16.5 \\
			+P $(\%)$ (NREM) & 88.3 $\pm$ 8.7 & 85.5 $\pm$ 16.0 & 85.2 $\pm$ 7.8 & 75.9 $\pm$ 9.4 \\
			\hline
			Acc $(\%)$ & 64.5 $\pm$ 8.1 & 64.4 $\pm$ 8.3 & 66.9 $\pm$ 6.3 & 55.0 $\pm$ 8.5 \\
			Kappa & 0.351 $\pm$ 0.116 & 0.346 $\pm$ 0.106 & 0.398 $\pm$ 0.104  & 0.227 $\pm$ 0.127 \\
		\end{tabular}	
	}
\end{table}

\begin{table}
	\caption{SVM cross-database performance for Wake, REM and NREM classification. The training database is UCDSADB with a single random seed. %{\color{red} Each feature vector used here is normalized via standard $z$-normalization among features of corresponding subject}. 
	For each database and each performance measurement, we report the mean $\pm$ standard deviation of all subjects.
	\label{tab:WRN classification training on UCD (Subjects)}
}
	\centering
	\fbox{\footnotesize 
		\begin{tabular}{c|cccc}
			& CGMH-training  & CGMH-validation & DREAMS & UCDSADB \\ \hline\hline
			SE $(\%)$ (Wake) & 68.8 $\pm$ 17.0 & 61.3 $\pm$ 20.0 & 59.6 $\pm$ 15.8 & 53.6 $\pm$ 13.9 \\					
			SE $(\%)$ (REM) & 26.8 $\pm$ 13.4 & 27.8 $\pm$ 18.9 & 23.9 $\pm$ 14.0 & 39.6 $\pm$ 17.8 \\
			SE $(\%)$ (NREM) & 76.0 $\pm$ 7.5 & 76.7 $\pm$ 6.9 & 76.3 $\pm$ 6.9 & 72.6 $\pm$ 7.0 \\
			+P $(\%)$ (Wake) & 33.9 $\pm$ 18.6 & 34.1 $\pm$ 15.8 & 37.1 $\pm$ 20.7 & 39.3 $\pm$ 17.1 \\
			+P $(\%)$ (REM) & 36.7 $\pm$ 21.1 & 34.5 $\pm$ 16.6 & 34.7 $\pm$ 17.0 & 46.3 $\pm$ 22.0 \\
			+P $(\%)$ (NREM) & 86.4 $\pm$ 9.1 & 83.1 $\pm$ 16.0 & 81.2 $\pm$ 8.6 & 77.1 $\pm$ 8.4 \\
			\hline
			Acc $(\%)$ & 67.1 $\pm$ 7.2 & 65.5 $\pm$ 9.8 & 64.7 $\pm$ 6.2 & 63.6 $\pm$ 7.3 \\
			Kappa & 0.343 $\pm$ 11.2 & 0.313 $\pm$ 0.105 & 0.311 $\pm$ 0.105  & 0.314 $\pm$ 0.125 \\
		\end{tabular}	
	}
\end{table}

%%%

\subsection{Subsampling effect}
We further evaluate the impact of the subsampling process. We repeat all the above experiments with $20$ different random seeds. For each database, each performance measurement and each random seed, we record the mean of all subjects. Then we report the mean $\pm$ standard deviation of these $20$ results. Specifically, we take the random seeds $1\sim 20$ in MATLAB, which leads to $20$ different training datasets after the subsampling scheme.    
We observe that although subsampling process was utilized in the training procedure, the performance of each SVM model is stable in the sense that all performance measurements have small standard deviation below $1\%$. See Tables~\ref{tab:main results 20seeds} to \ref{tab:WRN classification training on UCDSADB 20 seeds}.

\begin{table}
\caption{SVM cross-database performance for Wake and Sleep classification. The training database is CGMH-training with 20 random seeds, $1\sim 20$; that is, we train 20 SVM models. %{\color{red} Each feature vector used here is normalized via standard $z$-normalization among features of corresponding subject}.
\label{tab:main results 20seeds}}
		\fbox{%
		\footnotesize
		\begin{tabular}{c|cccc}
			& CGMH-training  & CGMH-validation & DREAMS & UCDSADB \\ \hline\hline
			TP & 6728 $\pm$ 26& 2027 $\pm$ 14 & 2019 $\pm$ 10 & 2082 $\pm$ 17 \\
			FP & 13200 $\pm$ 143 & 3360 $\pm$ 39 & 3456 $\pm$ 41 & 3679 $\pm$ 37 \\
			TN & 41347 $\pm$ 143 & 12166 $\pm$ 39 & 11891 $\pm$ 41 & 11234 $\pm$ 37 \\
			FN & 2423 $\pm$ 26 & 1149 $\pm$ 14 & 1119 $\pm$ 10 & 1850 $\pm$ 17 \\
			\hline
			SE $(\%)$ & 73.5 $\pm$ 0.3 & 63.8 $\pm$ 0.4 & 64.3 $\pm$ 0.3 & 53.0 $\pm$ 0.4 \\ 
			SP $(\%)$ & 75.8 $\pm$ 0.3 & 78.4 $\pm$ 0.3 & 77.5 $\pm$ 0.3 & 75.3 $\pm$ 0.3 \\
			Acc $(\%)$ & 75.5 $\pm$ 0.2 & 75.9 $\pm$ 0.2 & 75.3 $\pm$ 0.2 & 70.7 $\pm$ 0.1\\
			\hline 
			PR $(\%)$ & 33.8 $\pm$ 0.2 & 37.6 $\pm$ 0.2 & 36.9 $\pm$ 0.2 & 36.1 $\pm$ 0.1 \\
			F1  & 0.463 $\pm$ 0.001 & 0.474 $\pm$ 0.002 & 0.469 $\pm$ 0.002 & 0.430 $\pm$ 0.002 \\
			AUC  & 0.809 $\pm$ 0.000 & 0.781 $\pm$ 0.001 & 0.777 $\pm$ 0.001 & 0.693 $\pm$ 0.001 \\
			Kappa & 0.331 $\pm$ 0.002 & 0.330 $\pm$ 0.002 & 0.323 $\pm$ 0.003 & 0.242 $\pm$ 0.002 \\ 
			\end{tabular}}
\end{table}

\begin{table}
\caption{SVM cross-database performance  for Wake and Sleep classification. The training database is DREAMS with 20 random seeds, $1\sim 20$; that is, we train 20 SVM models. %{\color{red} Each feature vector used here is normalized via standard $z$-normalization among features of corresponding subject}.  
}
	\label{tab:SW classification training on DREAMS 20 seeds}
	\centering
	\fbox{\footnotesize
		\begin{tabular}{c|cccc}
			& CGMH-training  & CGMH-validation & DREAMS & UCDSADB \\ \hline\hline
			TP & 6490 $\pm$ 49 & 1988 $\pm$ 16 & 2228 $\pm$ 13 & 2132 $\pm$ 23 \\
			FP & 14590 $\pm$ 181 & 3730 $\pm$ 50 & 3458 $\pm$ 65 & 4128 $\pm$ 52 \\
			TN & 39957 $\pm$ 181 & 11796 $\pm$ 50 & 11890 $\pm$ 65 & 10785 $\pm$ 52 \\
			FN & 2660 $\pm$ 49 & 1189 $\pm$ 16 & 910 $\pm$ 13 & 1800 $\pm$ 23 \\
			\hline
			SE $(\%)$ & 70.9 $\pm$ 0.5 & 62.6 $\pm$ 0.5 & 71.0 $\pm$ 0.4 & 54.2 $\pm$ 0.6 \\ 
			SP $(\%)$ & 73.3 $\pm$ 0.3 & 76.0 $\pm$ 0.3 & 77.5 $\pm$ 0.4 & 72.3 $\pm$ 0.4 \\
			Acc $(\%)$ & 72.9 $\pm$ 0.2 & 76.4 $\pm$ 0.3 & 73.6 $\pm$ 0.3 & 68.5 $\pm$ 0.2\\
			\hline 
			PR $(\%)$ & 30.8 $\pm$ 0.2 & 34.8 $\pm$ 0.3 & 39.2 $\pm$ 0.4 & 34.1 $\pm$ 0.3 \\
			F1  & 0.429 $\pm$ 0.002 & 0.447 $\pm$ 0.003 & 0.505 $\pm$ 0.003 & 0.418 $\pm$ 0.003 \\
			AUC  & 0.781 $\pm$ 0.002 & 0.747 $\pm$ 0.003 & 0.808 $\pm$ 0.001 & 0.679 $\pm$ 0.002 \\
			Kappa & 0.286 $\pm$ 0.002 & 0.293 $\pm$ 0.004 & 0.366 $\pm$ 0.004 & 0.218 $\pm$ 0.004 \\
		\end{tabular}}
\end{table}

\begin{table}
	\caption{SVM cross-database performance  for Wake and Sleep classification. The training database is UCDSADB database with 20 random seeds, $1\sim 20$; that is, we train 20 SVM models. %{\color{red} Each feature vector used here is normalized via standard $z$-normalization among features of corresponding subject}.
	\label{tab:SW classification training on UCDSB 20 seeds}}
\centering
\fbox{\footnotesize
	\begin{tabular}{c|cccc}
			& CGMH-training  & CGMH-validation & DREAMS & UCDSADB \\ \hline\hline
			TP & 6814 $\pm$ 51 & 2050 $\pm$ 27 & 2176 $\pm$ 21 & 2464 $\pm$ 16 \\
			FP & 16909 $\pm$ 219 & 4412 $\pm$ 67 & 4613 $\pm$ 62 & 4467 $\pm$ 54 \\
			TN & 37638 $\pm$ 219 & 11114 $\pm$ 67 & 10734 $\pm$ 62 & 10445 $\pm$ 54 \\
			FN & 2336 $\pm$ 51 & 1126 $\pm$ 27 & 962 $\pm$ 21 & 1468 $\pm$ 16 \\
			\hline
			SE $(\%)$ & 74.5 $\pm$ 0.6 & 64.6 $\pm$ 0.9 & 69.3 $\pm$ 0.7 & 62.7 $\pm$ 0.4 \\ 
			SP $(\%)$ & 69.0 $\pm$ 0.4 & 71.6 $\pm$ 0.4 & 69.9 $\pm$ 0.4 & 70.1 $\pm$ 0.4 \\
			Acc $(\%)$ & 69.8 $\pm$ 0.3 & 69.8 $\pm$ 0.3 & 68.7 $\pm$ 0.6 & 68.5 $\pm$ 0.2\\
			\hline 
			PR $(\%)$ & 28.7 $\pm$ 0.2 & 31.7 $\pm$ 0.3 & 32.1 $\pm$ 0.3 & 35.6 $\pm$ 0.2 \\
			F1  & 0.415 $\pm$ 0.002 & 0.425 $\pm$ 0.004 & 0.438 $\pm$ 0.003 & 0.454 $\pm$ 0.001 \\
			AUC   & 0.784 $\pm$ 0.002  & 0.733 $\pm$ 0.004 & 0.758 $\pm$ 0.003 & 0.716 $\pm$ 0.001 \\
			Kappa & 0.261 $\pm$ 0.003 & 0.256 $\pm$ 0.005 & 0.269 $\pm$ 0.005 & 0.255 $\pm$ 0.002 \\
	\end{tabular}}
\end{table}

\begin{table}
	\caption{\small SVM cross-database performance for REM and NREM classification. The training database is CGMH-training with 20 random seeds, $1\sim 20$; that is, we train 20 SVM models. %{\color{red} Each feature vector used here is normalized via standard $z$-normalization among features of corresponding subject}.
	\label{tab:main results for REM classification (SVM)}}
\centering
\fbox{\footnotesize
		\begin{tabular}{c|cccc}
			& CGMH-training  & CGMH-validation & DREAMS & UCDSADB \\ \hline\hline
			TP & 6731 $\pm$ 31 & 1776 $\pm$ 8 & 1877 $\pm$ 10 & 1558 $\pm$ 10 \\
			FP & 10168 $\pm$ 98 & 2890 $\pm$ 39 & 2817 $\pm$ 35 & 3272 $\pm$ 40 \\
			TN & 35446 $\pm$ 98 & 10338 $\pm$ 39 & 9749 $\pm$ 35 & 8921 $\pm$ 40 \\
			FN & 2202 $\pm$ 31 & 522 $\pm$ 8 & 904 $\pm$ 10 & 1161 $\pm$ 10 \\
			\hline
			SE $(\%)$ & 75.4 $\pm$ 0.4 & 77.3 $\pm$ 0.4 & 67.5 $\pm$ 0.4 & 57.3 $\pm$ 0.4 \\ 
			SP $(\%)$ & 77.7 $\pm$ 0.2 & 78.2 $\pm$ 0.3 & 77.6 $\pm$ 0.3 & 73.2 $\pm$ 0.3 \\
			Acc $(\%)$ & 77.3 $\pm$ 0.1 & 75.8 $\pm$ 0.2 & 71.3 $\pm$ 0.3 & 70.3 $\pm$ 0.2\\
			\hline 
			PR $(\%)$ & 39.8 $\pm$ 0.1 & 38.1 $\pm$ 0.3 & 40.0 $\pm$ 0.2 & 32.3 $\pm$ 0.2 \\
			F1  & 0.521 $\pm$ 0.001 & 0.510 $\pm$ 0.003 & 0.502 $\pm$ 0.002 & 0.413 $\pm$ 0.002 \\
			AUC  & 0.836 $\pm$ 0.000 &0.847 $\pm$ 0.001 & 0.796 $\pm$ 0.001 & 0.708 $\pm$ 0.001 \\
			Kappa & 0.391 $\pm$ 0.001 & 0.389 $\pm$ 0.004 & 0.356 $\pm$ 0.003 & 0.234 $\pm$ 0.003 \\
		\end{tabular}
		
	}
\end{table}

\begin{table}
	\caption{SVM cross-database performance for REM and NREM classification. The training database is DREAMS database with 20 random seeds, $1\sim 20$; that is, we train 20 SVM models. %{\color{red} Each feature vector used here is normalized via standard $z$-normalization among features of corresponding subject}.
	\label{tab:RN classification training on DREAMS 20 seeds}. 
	}
	\centering
	\fbox{\footnotesize
		\begin{tabular}{c|cccc}
			& CGMH-training  & CGMH-validation & DREAMS & UCDSADB \\ \hline\hline
			TP & 6608 $\pm$ 49 & 1747 $\pm$ 18 & 2020 $\pm$ 17 & 1533 $\pm$ 21 \\
			FP & 11235 $\pm$ 216 & 3171 $\pm$ 66 & 3032 $\pm$ 68 & 3502 $\pm$ 66 \\
			TN & 34379 $\pm$ 216 & 10057 $\pm$ 66 & 9534 $\pm$ 17 & 8692 $\pm$ 66 \\
			FN & 2325 $\pm$ 49 & 551 $\pm$ 18 & 761 $\pm$ 17 & 1186 $\pm$ 21 \\
			\hline
			SE $(\%)$ & 74.0 $\pm$ 0.6 & 76.0 $\pm$ 0.8 & 72.6 $\pm$ 0.6 & 56.4 $\pm$ 0.8 \\ 
			SP $(\%)$ & 75.4 $\pm$ 0.5 & 76.0 $\pm$ 0.5 & 75.9 $\pm$ 0.5 & 71.3 $\pm$ 0.5 \\
			Acc $(\%)$ & 75.1 $\pm$ 0.3 & 76.0 $\pm$ 0.4 & 75.3 $\pm$ 0.4 & 68.6 $\pm$ 0.4\\
			\hline 
			PR $(\%)$ & 37.0 $\pm$ 0.3 & 35.5 $\pm$ 0.4 & 40.0 $\pm$ 0.4 & 30.5 $\pm$ 0.3 \\
			F1  & 0.494 $\pm$ 0.002 & 0.484 $\pm$ 0.005 & 0.516 $\pm$ 0.003 & 0.395 $\pm$ 0.004 \\
			AUC  & 0.815 $\pm$ 0.002  & 0.829 $\pm$ 0.006 & 0.816 $\pm$ 0.001 & 0.684 $\pm$ 0.004 \\
			Kappa & 0.352 $\pm$ 0.004 & 0.354 $\pm$ 0.006 & 0.368 $\pm$ 0.004 & 0.208 $\pm$ 0.005 \\
		\end{tabular}
	}

\end{table}

\begin{table}
	\caption{SVM cross-database performance for REM and NREM classification. The training database is UCDSADB with 20 random seeds, $1\sim 20$; that is, we train 20 SVM models. %{\color{red} Each feature vector used here is normalized via standard $z$-normalization among features of corresponding subject}.
	\label{tab:RN classification training on UCDSADB 20 seeds}
	}
	\centering
	\fbox{\footnotesize
		\begin{tabular}{c|cccc}
			& CGMH-training  & CGMH-validation & DREAMS & UCDSADB \\ \hline\hline
			TP & 5340 $\pm$ 77 & 1351 $\pm$ 23 & 1492 $\pm$ 24 & 1758 $\pm$ 19 \\
			FP & 10351 $\pm$ 287 & 3044 $\pm$ 100 & 3368 $\pm$ 104 & 2869 $\pm$ 82 \\
			TN & 35263 $\pm$ 287 & 9522 $\pm$ 100 & 9198 $\pm$ 104 & 9324 $\pm$ 82 \\
			FN & 3593 $\pm$ 77 & 947 $\pm$ 23 & 1289 $\pm$ 24 & 961 $\pm$ 19 \\
			\hline
			SE $(\%)$ & 59.8 $\pm$ 0.9 & 58.8 $\pm$ 1.0 & 53.6 $\pm$ 0.9 & 64.7 $\pm$ 0.7 \\ 
			SP $(\%)$ & 77.3 $\pm$ 0.6 & 77.3 $\pm$ 0.7 & 75.8 $\pm$ 0.8 & 76.5 $\pm$ 0.7 \\
			Acc $(\%)$ & 74.4 $\pm$ 0.5 & 74.5 $\pm$ 0.6 & 71.8 $\pm$ 0.6 & 74.3 $\pm$ 0.4\\
			\hline 
			PR $(\%)$ & 34.0 $\pm$ 0.5 & 31.0 $\pm$ 0.7 & 40.8 $\pm$ 0.6 & 38.0 $\pm$ 0.5 \\
			F1  & 0.434 $\pm$ 0.004 & 0.406 $\pm$ 0.007 & 0.408 $\pm$ 0.005 & 0.479 $\pm$ 0.002 \\
			AUC  & 0.742 $\pm$ 0.005  & 0.733 $\pm$ 0.007 & 0.707 $\pm$ 0.006 & 0.771 $\pm$ 0.001\\
			Kappa & 0.284 $\pm$ 0.006 & 0.263 $\pm$ 0.009 & 0.236 $\pm$ 0.008 & 0.323 $\pm$ 0.004 \\
		\end{tabular}
	}

\end{table}

\begin{table}
	\caption{SVM cross-database performance for Wake, REM and NREM classification. The training database is CGMH-training with 20 random seeds, $1\sim 20$; that is, we train 20 SVM models. %{\color{red} Each feature vector used here is normalized via standard $z$-normalization among features of corresponding subject}.
	\label{tab:main results for 3 classes classification (SVM)}}
	\centering
	\fbox{	
	\small	
	\begin{tabular}{c|cccc}
			& CGMH-training  & CGMH-validation & DREAMS & UCDSADB \\ \hline\hline
			SE $(\%)$ (Wake) &  59.3 $\pm$ 0.2 & 53.4 $\pm$ 0.2 & 53.5 $\pm$ 0.2 & 37.3 $\pm$ 0.3 \\					
			SE $(\%)$ (REM) & 61.9 $\pm$ 0.2 & 65.2 $\pm$ 0.2 & 56.7 $\pm$ 0.4 & 47.7 $\pm$ 0.5 \\
			SE $(\%)$ (NREM) & 71.3 $\pm$ 0.2 & 72.3 $\pm$ 0.3 & 71.5 $\pm$ 0.3 & 65.5 $\pm$ 0.4 \\
			+P $(\%)$ (Wake) & 40.3 $\pm$ 0.1 & 45.3 $\pm$ 0.2 & 44.4 $\pm$ 0.1 & 38.2 $\pm$ 0.1 \\
			+P $(\%)$ (REM) & 39.8 $\pm$ 0.2 & 37.4 $\pm$ 0.3 & 39.7 $\pm$ 0.2 & 28.0 $\pm$ 0.2 \\
			+P $(\%)$ (NREM) & 89.5 $\pm$ 0.1 & 87.3 $\pm$ 0.1 & 83.7 $\pm$ 0.1 & 76.9 $\pm$ 0.1 \\
			\hline
			Acc $(\%)$ & 68.3 $\pm$ 0.1 & 68.2 $\pm$ 0.2 & 66.2 $\pm$ 0.1 & 57.0 $\pm$ 0.2 \\
			Kappa & 0.402 $\pm$ 0.001 & 0.395 $\pm$ 0.002 & 0.372 $\pm$ 0.002 & 0.240 $\pm$ 0.001 \\
		\end{tabular}		
	}
\end{table}

\begin{table}
	\caption{SVM cross-database performance for Wake, REM and NREM classification. The training database is DREAMS with 20 random seeds, $1\sim 20$; that is, we train 20 SVM models. %{\color{red} Each feature vector used here is normalized via standard $z$-normalization among features of corresponding subject}.
		\label{tab:WRN classification training on DREAMS 20 seeds}
}
	\centering
	\fbox{\footnotesize 
		\begin{tabular}{c|cccc}
			& CGMH-training  & CGMH-validation & DREAMS & UCDSADB \\ \hline\hline
			SE $(\%)$ (Wake) & 60.1 $\pm$ 0.3 & 53.9 $\pm$ 0.4 & 62.9 $\pm$ 0.4 & 52.4 $\pm$ 0.4 \\					
			SE $(\%)$ (REM) & 53.8 $\pm$ 0.4 & 56.8  $\pm$ 0.5 & 57.9 $\pm$ 0.5 & 40.7 $\pm$ 0.6 \\
			SE $(\%)$ (NREM) & 67.7 $\pm$ 0.3 & 69.6 $\pm$ 0.3 & 70.0 $\pm$ 0.3 & 72.5 $\pm$ 0.4 \\
			+P $(\%)$ (Wake) & 33.9 $\pm$ 0.2 & 38.8 $\pm$ 0.2 & 44.6 $\pm$ 0.3 & 40.2 $\pm$ 0.2 \\
			+P $(\%)$ (REM) & 37.8 $\pm$ 0.2 & 34.9 $\pm$ 0.3 & 42.8 $\pm$ 0.3 & 48.1 $\pm$ 0.6 \\
			+P $(\%)$ (NREM) & 88.7 $\pm$ 0.1 & 87.3 $\pm$ 0.1 & 85.4 $\pm$ 0.1 & 77.5 $\pm$ 0.1 \\
			\hline
			Acc $(\%)$ & 64.6 $\pm$ 0.2 & 65.4 $\pm$ 0.2 & 67.0 $\pm$ 0.2 & 63.8 $\pm$ 0.2 \\
			Kappa & 0.350 $\pm$ 0.002 & 0.354 $\pm$ 0.002 & 0.400 $\pm$ 0.002  & 0.321 $\pm$ 0.002 \\
		\end{tabular}	
	}
\end{table}

\begin{table}
	\caption{SVM cross-database performance for Wake, REM and NREM classification. The training database is UCDSADB with 20 random seeds, $1\sim 20$; that is, we train 20 SVM models. %{\color{red} Each feature vector used here is normalized via standard $z$-normalization among features of corresponding subject}.
	\label{tab:WRN classification training on UCDSADB 20 seeds}
}
\centering
	\fbox{\footnotesize 
		\begin{tabular}{c|cccc}
			& CGMH-training  & CGMH-validation & DREAMS & UCDSADB \\ \hline\hline
			SE $(\%)$ (Wake) & 63.7 $\pm$ 0.4 & 58.3 $\pm$ 0.5 & 47.1 $\pm$ 1.0  & 43.2 $\pm$ 0.9 \\
			SE $(\%)$ (REM) & 26.9 $\pm$ 0.4 & 26.0 $\pm$ 0.5 & 22.7 $\pm$ 0.4 & 36.7 $\pm$ 0.6 \\
			SE $(\%)$ (NREM) & 75.5 $\pm$ 0.3 & 77.0 $\pm$ 0.3 & 75.5 $\pm$ 0.3 & 70.8 $\pm$ 0.8 \\
			+P $(\%)$ (Wake) & 33.0 $\pm$ 0.2 & 35.0 $\pm$ 0.3 & 36.0 $\pm$ 0.2 & 35.3 $\pm$ 0.3 \\
			+P $(\%)$ (REM) & 38.5 $\pm$ 0.4 & 34.1 $\pm$ 0.5 & 35.6 $\pm$ 0.5 & 38.5 $\pm$ 0.2 \\
			+P $(\%)$ (NREM) & 86.6 $\pm$ 0.1 & 84.7 $\pm$ 0.2 & 81.6 $\pm$ 0.1 & 75.4 $\pm$ 0.2 \\
			\hline
			Acc $(\%)$ & 67.0 $\pm$ 0.1 & 66.9 $\pm$ 0.2 & 64.6 $\pm$ 0.2 & 60.1 $\pm$ 0.3 \\			Kappa & 0.339 $\pm$ 0.002 & 0.322 $\pm$ 0.003 & 0.308 $\pm$ 0.002 & 0.253 $\pm$ 0.002 \\
		\end{tabular}	
	}
\end{table}

\end{document}